\documentclass{article}

\usepackage{PRIMEarxiv}

\usepackage[utf8]{inputenc} % allow utf-8 input
\usepackage[T1]{fontenc}    % use 8-bit T1 fonts
\usepackage{url}            % simple URL typesetting
\usepackage{booktabs}       % professional-quality tables
\usepackage{amsfonts}       % blackboard math symbols
\usepackage{nicefrac}       % compact symbols for 1/2, etc.
\usepackage{microtype}      % microtypography
\usepackage{subcaption}
\usepackage{adjustbox}
\usepackage{cryptocode}
\graphicspath{{media/}}     % organize your images and other figures under media/ folder
\usepackage{graphicx}
\usepackage{easyReview}
\usepackage{fancyhdr}
\usepackage{natbib}
\usepackage{hyperref}       % hyperlinks

\begin{document}
\title{Broadband Ground Motion Synthesis via Generative Adversarial Neural Operators: Development and Validation
\thanks{Accepted by \textit{Bulletin of the Seismological Society of America}}
}

\author{ 
    %first author
    \href{https://orcid.org/0000-0001-8863-2026}{\includegraphics[scale=0.06]{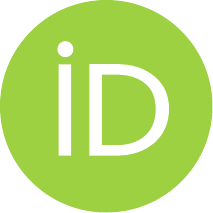}\hspace{1mm}Yaozhong Shi} \\
    Division of Engineering and Applied Sciences\\
    California Institute of Technology\\
    Pasadena, California, U.S.A. \\
	\texttt{yshi5@caltech.edu} \\
	%second author
	\And
    \href{https://orcid.org/0000-0001-6546-1340}{\includegraphics[scale=0.06]{orcid.pdf}\hspace{1mm}Grigorios Lavrentiadis} \\
    Division of Engineering and Applied Sciences\\
    California Institute of Technology\\
    Pasadena, California, U.S.A. \\
	\texttt{glavrent@caltech.edu} \\
    %third author
	\And
    \href{https://orcid.org/0000-0002-3008-8088}{\includegraphics[scale=0.06]{orcid.pdf}\hspace{1mm}Domniki Asimaki} \\
    Division of Engineering and Applied Sciences\\
    California Institute of Technology\\
    Pasadena, California, U.S.A. \\
	\texttt{domniki@caltech.edu} \\   
    %fourth author
	\And
    \href{https://orcid.org/0000-0002-3008-8088}{\includegraphics[scale=0.06]{orcid.pdf}\hspace{1mm}Zachary E. Ross} \\
    Seismological Laboratory\\
    California Institute of Technology\\
    Pasadena, California, U.S.A. \\
	\texttt{zross@caltech.edu} \\ 
    %fourth author
	\And
    \href{https://orcid.org/0000-0001-8507-1868}{\includegraphics[scale=0.06]{orcid.pdf}\hspace{1mm}Kamyar Azizzadenesheli} \\
    NVIDIA Corporation\\
    Santa Clara, California, U.S.A.\\
	\texttt{kamyara@nvidia.com} \\  
}

\maketitle
\begin{abstract}
We present a data-driven framework for ground-motion synthesis that generates three-component acceleration time histories conditioned on moment magnitude ($\mathbf{M}$), rupture distance ($R_{rup}$), time-average shear-wave velocity at the top $30m$ ($V_{S30}$), and style of faulting. We use a Generative Adversarial Neural Operator (GANO), a resolution invariant architecture that guarantees model training independent of the data sampling frequency. We first present the conditional ground-motion synthesis algorithm (cGM-GANO) and discuss its advantages compared to previous work. We next train cGM-GANO on simulated ground motions generated by the Southern California Earthquake Center Broadband Platform (BBP) and on recorded KiK-net data and show that the model can learn the overall magnitude, distance, and $V_{S30}$ scaling of effective amplitude spectra (EAS) ordinates and pseudo-spectral accelerations (PSA). Results specifically show that cGM-GANO produces consistent median scaling with the training data for the corresponding tectonic environments over a wide range of frequencies for scenarios with sufficient data coverage. For the BBP dataset, cGM-GANO cannot learn the ground motion scaling of the stochastic frequency components ($f>1Hz$); for the KiK-net dataset, the largest misfit is observed at short distances ($R_{rup}<50km$) and for soft soil conditions ($V_{S30}<200m/s$) due to the scarcity of such data. Except for these conditions, the aleatory variability of EAS and PSA are captured reasonably well. Lastly, cGM-GANO produces similar median scaling to traditional GMMs for frequencies greater than 1Hz for both PSA and EAS but underestimates the aleatory variability of EAS. Discrepancies in the comparisons between the synthetic ground motions and GMMs are attributed to inconsistencies between the training dataset and the datasets used in GMM development. Our pilot study demonstrates GANO's potential for efficient synthesis of broad-band ground motions. 

\end{abstract}

\section{Introduction}

The demand for reliable ground motion time histories for seismic analysis is growing, yet there are still not enough records of large earthquakes covering the desired range of source, path, and site characteristics \citep{katsanos_selection_2010}, in particular at near-source ground motions or on very soft site conditions. 
In order to resolve the limitations caused by data scarcity, ground motion records are usually selected and scaled in the time domain or adjusted in the frequency domain to satisfy specified intensity measurements or frequency content \citep{goulet_assessment_2008,rezaeian_simulation_2012}. However, scaling or modification of records that is not based on physics may only scale some intensities or frequency measures while leaving other elements unaltered, leading to unrealistic ground motions \citep{naeim_use_1995,luco_does_2007, baker_seismic_2021}. Additionally, most of the scaling and spectrum matching techniques only modify the two horizontal components making them unsuitable for structural models that are sensitive to all three components \citep{bhosale_vertical_2017}.

Methods for synthesizing ground motions offer an alternative solution by directly generating ground motions for the scenarios of interest. These methods can be broadly grouped into “physics-based” numerical simulations and “stochastic-based” approaches, depending on whether the process of generating synthetic waveforms involves physics \citep{douglas_survey_2008},
In general, stochastic-based methods for ground motion generation involve modifying white noise to emulate specific ground motion characteristics
\citep{dabaghi_stochastic_2017, baglio_generating_2023, sabetta_simulation_2021, burks_validation_2014}, while physics-based numerical simulations rely on modeling the physics that govern earthquakes, such as wave propagation and seismic source radiation to generate ground motions for a scenario of interest \citep{mccallen_eqsimmultidisciplinary_2021, paolucci_bbspeedset_2021, touhami_sem3d_2022}. 

Physics-based methods have the advantage of generating complex path and site effects, but they are still far from being incorporated in routine engineering applications due to the following reasons: (1) Such approaches demand in-depth knowledge of sources, paths, and site effects, which are usually unavailable for most engineering purposes. \citep{causse_calibrating_2008,frankel_constant_2009,rezaeian_simulation_2012}. (2) they require expensive computations to provide sufficient high-resolution frequency. Nowadays, with the assistance of supercomputers, it is possible to synthesize broadband ground motions up to $10 Hz$ \citep{rodgers_regional_2020}; however, such computational resources are not available for most engineering projects. 
(3) Even with earthquake simulations capable of generating high-frequency ground motions, the accuracy and spatiotemporal variability will continue to lag behind the input model accuracy for source physics, crustal complexity, nonlinear site response, and other constituent elements of simulating earthquakes end-to-end. 
(4) To achieve the desired frequency resolution, the generated ground motions may be applicable only to a limited range of site conditions due to the application of minimum $V_S$ threshold for wave propagation or the omission of non-linear effects.  For instance, to achieve computational efficiency, some physics-based methods impose a $V_{S}$ lower limit at 500 m/s, which means such methods cannot accurately simulate strong motions at soil sites \citep{pitarka_broadband_2013}. 

Stochastic methods take a different approach to generating ground motions for engineering applications.
The main advantages are that: (1) the source, site, and path scaling terms can be determined from empirical data and can simply be combined to generate the desired ground motions \citep{hanks_character_1981, boore_stochastic_1983,boore_simulation_2003}, and (2) they are computationally efficient. At the same time, the main limitations of stochastic methods are (1) Such methods are based on the reconstruction of particular features, such as pseudo-spectral accelerations (PSA) \citep{tavakoli_empirical-stochastic_2005} or pseudo-spectral velocity (PSV) \citep{boore_simulation_2003} without taking physics into account to fully characterize the ground motion.
(2) Due to their inability to properly represent P, S, surface-wave modes, and polarization, stochastic methods cannot be used to produce coherent 3-component waveforms with multiple arrivals and coda that are representative of real ground motions \citep{graves_broadband_2010}. (3) They tend to underestimate the correlation between amplitudes at different frequencies compared to observed motions, which underestimates the variability in the building response and can have an impact on the calculated risk \citep{stafford_interfrequency_2017,bayless_empirical_2019}. 

In both stochastic and physics-based models, a source of model uncertainty arises from the approximate representation of unresolved processes through parameterized functional forms. 
Model errors can be reduced to a certain degree through targeted data collection, leading --for example-- to an improved spatial resolution of 3D velocity models in physics-based simulations, but they cannot be eliminated because there is no resolution separation between resolved and unresolved processes. 
At the same time, functional parameterization is often necessary for robust model extrapolation, especially outside the range of available data. 
However, for applications with dense data coverage and emphasis on interpolation (i.e., surrogate modeling of hybrid empirical-numerical strong motion datasets), machine learning models offer a fundamentally different approach to reduce model uncertainty: they parameterize the complex, nonlinear sub-resolution processes directly from the data that describe those processes using well‐defined optimization rules \citep{gagne2020machine, kong2019machine}. With the growth of available high-quality strong ground motion data and the explosion of modern machine learning algorithms, generating realistic ground motions that span the broadband range of engineering interest ($\sim 0.1-25$Hz) is no longer a far-reaching goal. 

The types of stochastic methods mentioned previously are not data-driven, (i.e., they are not generative models learned from data). However, major advances in machine learning have been made in the last decade on the topic of deep generative models, in which the goal is to learn an unknown data distribution given a large dataset for training. Generative adversarial networks, or GANs \citep{goodfellow2014generative}, are a class of generative models well suited for time-series synthesis. GANs draw samples from random distributions to generate new data that mimic an existing sample. Carefully designed GANs have uncovered and replicated the properties of time series in various fields such as finance, medicine, and seismology, including ground motion synthesis for earthquake early warning. Among others, \cite{wang2019earthquakegen} and \citet{wang2021seismogen} generated realistic seismic signals for training of early warning systems; \citep{li2020seismic} also proposed a GAN-based method to artificially generate time series that resemble realistic seismic signals for early warning system training; \cite{gatti2020towards} used GANs to extract high frequency features from ground motion records and use them to enhance ground motion simulated time series; \cite{matinfar2023deep} used GANs to synthesize ground motion acceleration time series compatible with a given target design spectrum; \cite{florez_data-driven_2022} used GAN to synthesize realistic three-component accelerograms conditioned on a set of continuous physical variables (magnitude, distance, and shear wave velocity) and statistically evaluated the quality of the signals using common engineering ground-motion intensity measures; while \cite{esfahani2023tfcgan}, inspired by \cite{florez_data-driven_2022}, simulated non-stationary ground-motion recordings leveraging information contained in the frequency domain and incorporating physics-based knowledge such magnitude and distance. For a detailed overview of the use of GANs in earthquake engineering fields, the reader is referred to \cite{marano2023generative}.

In this paper, we use a new class of generative models -- a Generative Adversarial Neural Operator (GANO) \citep{rahman2022generative} with the ultimate goal of building a framework that synthesizes ground motions for engineering applications. We first present the ground-motion synthesis algorithm and discuss its advantages compared to previous work. We next evaluate the GANO framework using noiseless ground motions simulated by the Southern California Earthquake Center (SCEC) Broadband Platform (BBP) \citep{maechling_scec_2014}, and finally validate it by training it on three-component acceleration time histories from a strong-motion dataset processed from the Kik-Net network in Japan.
In both applications, the generated ground motions are conditioned on moment magnitude ($\mathbf{M}$), closest-point-on-the-rupture to site distance ($R_{rup}$). 
Additionally, for the BBP dataset the style of faulting (strike-slip or reverse event) was included as a conditional parameter, while for the Kik-Net dataset, the time average shear wave velocity at the top $30 m$ ($V_{S30}$) and tectonic environment type (subduction or shallow crustal event) were the conditional variables.
Lastly, we assess the performance of our model through residual analysis and through a comparison with published GMMs. 
While still in progress, quantitative analyses show that our GANO-generated ground motion model can overall reproduce realistic ground motion time histories with a broad range of frequencies (0.1$\sim$30Hz) for horizontal and vertical components and can approximate the variability in the empirical Kik-net dataset.

\section{Methods}
\phantomsection

\subsection{Neural Operators}
The classical development of neural networks has primarily focused on learning mappings between finite-dimensional Euclidean spaces or finite sets. Neural operators were introduced as a generalization of neural networks, intended to learn mappings between continuous function spaces from finite input-output pairs \citep{li_neural_2020_a,li_fourier_2021}. Learning mappings between function spaces has shown great success at efficiently solving complex partial differential equations (PDEs). 
For instance, \cite{pathak_fourcastnet_2022, wen_u-fno_2022, yang_seismic_2021}, have shown that neural operators can efficiently reproduce simulations governed by PDEs in problems of weather forecasting, CO2 injection, and wave propagation in solids to name a few.

Neural operators can be understood as mappings between infinite-dimensional function spaces. Suppose, for instance, that there is a non-linear mapping $\mathcal{F}$:$\mathcal{A}\rightarrow\mathcal{U}$, where $\mathcal{A}$ is the input and $\mathcal{U}$ is the output function space. The neural operator approximates the solution operator $\mathcal{F}$ using a fixed set of parameters ($\mathcal{F}_\theta\approx\mathcal{F}$) that provide a continuous representation of $\mathcal{U}$ in the function space. \cite{kovachki2023neural} formally established that neural operator models with a fixed number of parameters satisfy discretization invariance; and further showed that they are universal approximators of continuous operators, namely, can uniformly approximate any continuous operator defined on a compact set of normed vector spaces. According to \cite{kovachki2023neural}, neural operators are the only known class of models that guarantee both discretization-invariance and universal approximation.

A computationally efficient variant of Neural Operators particularly suited for wave propagation (and synthesis) are Fourier Neural Operators (FNO) \citep{li_fourier_2021}, whereby the representation and parameterization of the neural operator kernel are formulated in the Fourier space. Despite the advantages in terms of efficiency and accuracy, FNOs, which according to \cite{kovachki2023neural} \emph{achieved the best numerical performance of all the numerical experiments}, are fully connected networks, and as such, the memory requirements associated with their training can be significant. Inspired by the U-net architecture \citep{ronneberger_u-net_2015}, \cite{rahman_u-no_2022} proposed an architecture referred to as the U-shaped Neural Operator (U-NO), which relies on gradually compressing and decoding information to alleviate issues associated with excess memory requirements. U-NO has the advantage of faster training and less memory requirements compared to standard FNO. 

\subsection{Generative Adversarial Neural Operators (GANO)}
Generative Adversarial Neural Operators (GANO) generalize the Generative Adversarial Network (GAN) framework \citep{rahman2022generative}, which is one of the most prominent paradigms in machine learning for analyzing unsupervised data. While GANs are successful generative models with a wide range of theoretical developments \citep{arjovsky_wasserstein_2017}, their empirical success has been only in finite-dimensional data regimes. In fact, there has been little progress in developing generative models for infinite-dimensional spaces and function spaces, which limited the use of GANs for learning functional data (i.e., data on irregular grids in space and time), which is often the case in the fields of science and engineering.

A GANO framework consists of a neural operator generator and a discriminator functional (i.e., a neural operator followed by an integral function). The generator uses Gaussian Random Fields (GRFs) or any predefined probability function as input and outputs a function sample.
The discriminator accepts as input real or synthetic data and returns a scalar (score).
The objective function, namely the score generated by the discriminator, is used for the loss function. \cite{rahman2022generative} devised the GANO
framework by i) generalizing the Wasserstein GAN setting \citep{arjovsky_wasserstein_2017} to infinite dimension space, and ii) using the generalized Wasserstein distance and a gradient penalty as regularization terms to make the models more stable and less sensitive to the tuning of their hyperparameters. 

\subsection{Conditional Ground Motion GANO (cGM-GANO)}

\begin{figure*}
\centering
\includegraphics[scale=0.5]{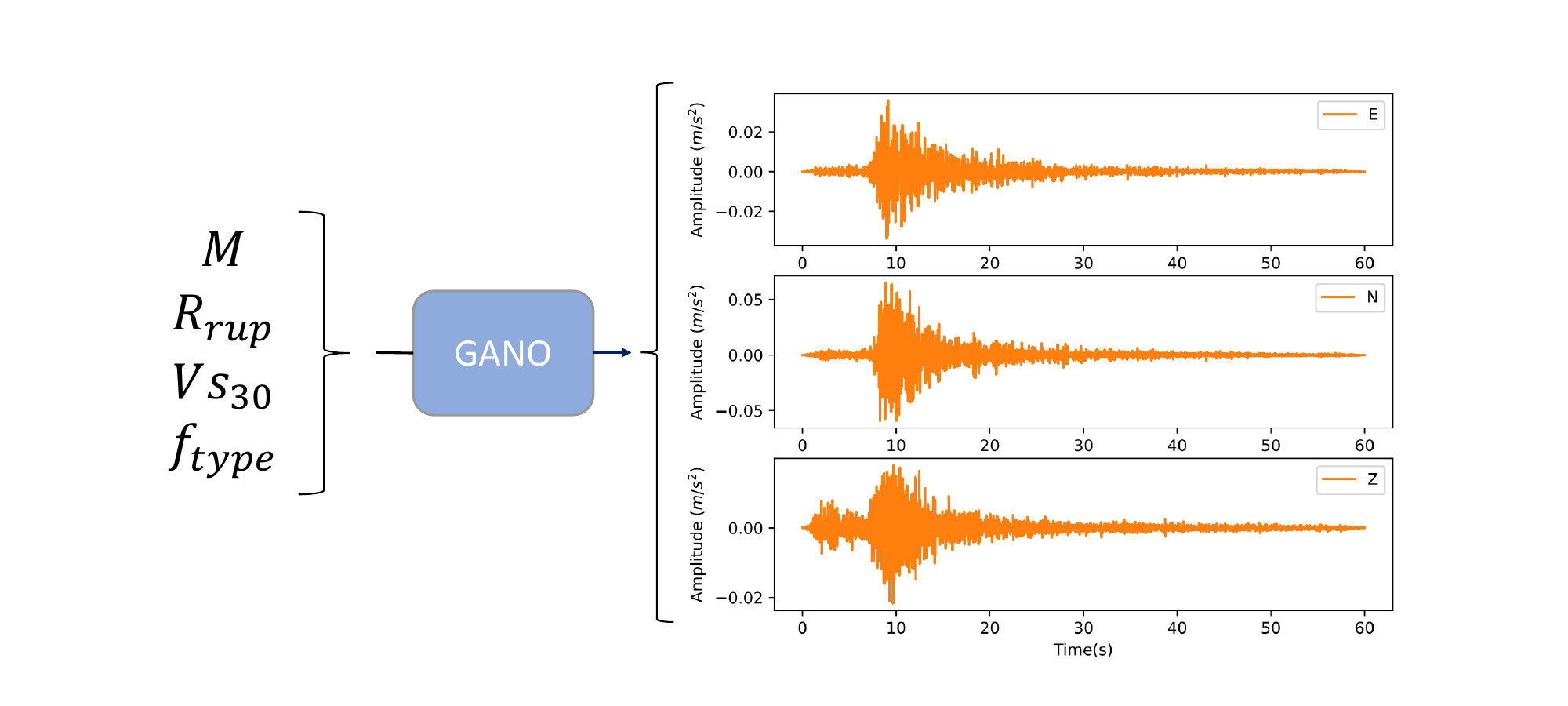}
\caption{Schematic of GANO input-output structure used in this work: given a set of conditional variables (left), the trained GANO (middle) produces three-component acceleration time series (right). }
\label{fig:model_example}
\end{figure*}

In this paper, we present a conditional GANO architecture designed to generate realistic 3-component acceleration time histories; we will refer to the architecture in the following sections as cGM-GANO. The architecture presented here has been trained on four conditional variables: the moment magnitude ($\mathbf{M}$), the closest-point-on-the-rupture to site distance ($R_{rup}$), the time-averaged shear wave velocity in the top $30 m$ ($V_{S30}$), 
and the style of faulting or tectonic environment ($f_{type}$). $f_{type}$ is used to differentiate between strike-slip ($f_{type}=1$) and reverse  ($f_{type}=0$) events for the BPP data set, or shallow crustal ($f_{type}=1$) and subduction ($f_{type}=0$) events for the Kik-net data set. 
Given a vector of the aforementioned parameters, it generates 3-component acceleration time series (Fig \ref{fig:model_example}). We should note here that the selection of four variables is a modeling decision and could be replaced with any number of conditional variables, provided that there is a sufficiently large dataset for GANO to learn the effect of the different parameters. 

\begin{figure*}
\centering
\includegraphics[scale=0.33]{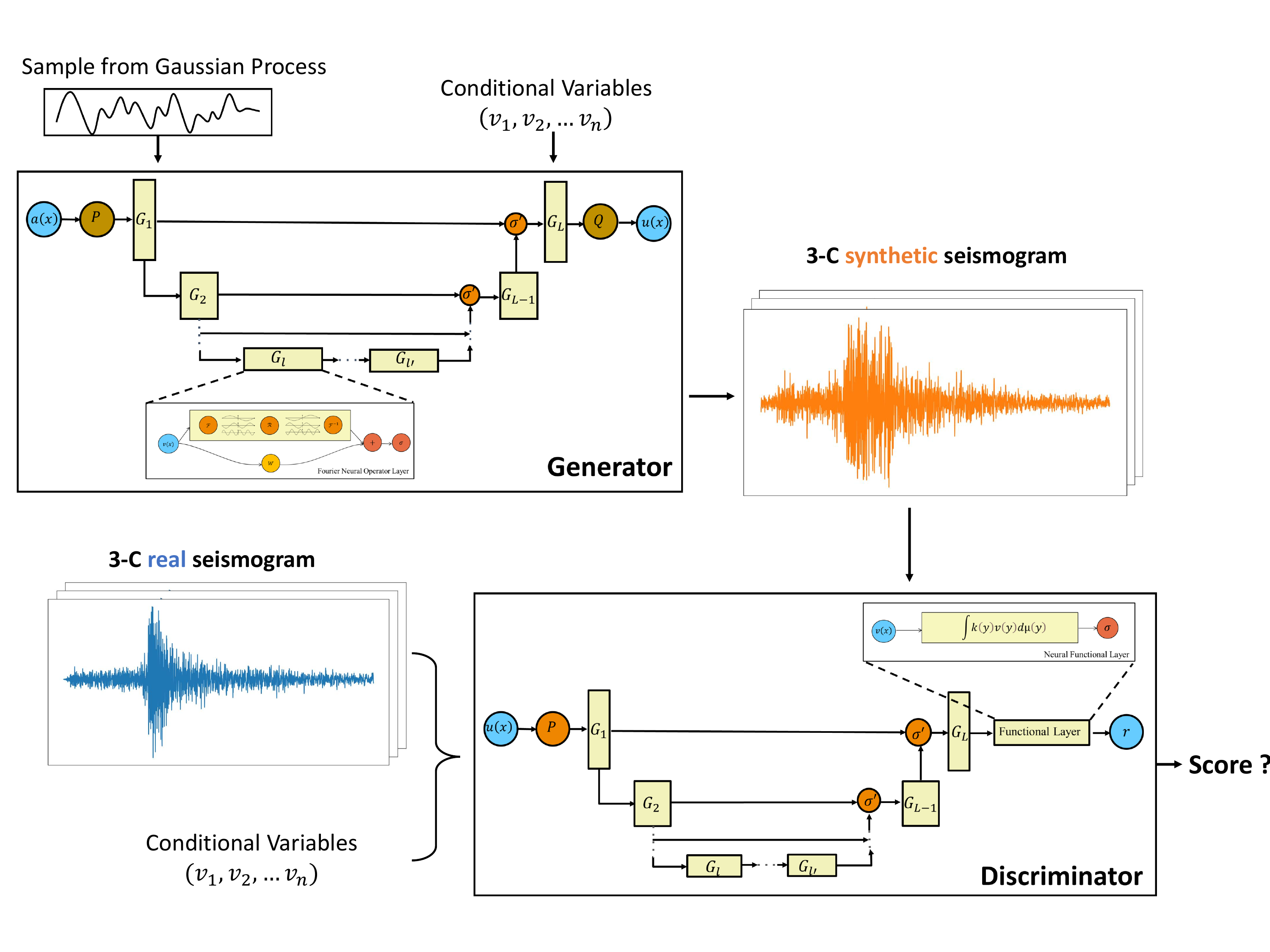}
\caption{Schematic of cGM-GANO architecture, and generator - discriminator interaction. The input conditional variables are $\mathbf{M}$, $R_{rup}$, $V_{S30}$ and $f_{type}$. The input function for the generator is finite duration Gaussian random field sample while the output is the normalized three-component acceleration time series and three PGAs. The input function of the discriminator are the real or synthetic three-component acceleration time series while the output is score evaluating the realism of the input ground motions. }
\label{fig:model_architecture}
\end{figure*}

To design cGM-GANO, we implemented the U-NO architecture in a GANO framework, as outlined in the previous section. 
cGM-GANO consists of a generator and discriminator trained in a min-max adversarial game (Fig. \ref{fig:model_architecture}). We should note here that the training database of ground motions spans 3-4 orders of magnitude of PGA amplitude; as such, acceleration time series needed to be normalized to a common amplitude scale to ensure that there was no loss of information \citep{goodfellow_deep_2016,wu_group_2018} so that small events are not interpreted as noise. The normalization constants for each channel of the three component time series are also \emph{learned} and recovered when the trained generator is called to synthesize a conditional ground motion scenario. The input to the trained generator consists of two sets of one-dimensional (1D) discretized functions: (i) a 1D Gaussian Random Field (Gaussian Process), which is stochastic (ii) and a set of constant functions that are used to convert the input parameters ($\mathbf{M}$, $R_{rup}$, $V_{S30}$, $f_{type}$) to a measurable function. The output includes the 3-component normalized ground motion waveforms and three normalization constants, in this case, the peak amplitudes associated with the normalized waveforms. 

Compared to the previous work by the authors \citep{florez_data-driven_2022}, the important advantage of GANOs that pertains to the problem of ground motion synthesis is their resolution and discretization invariance \citep{kovachki_neural_2021}: the input function to the generator can be expressed with an arbitrary discretization (i.e., a dataset of time series sampled at various frequencies), yet the generated output is a function, which can be queried at any (i.e., frequency) resolution. These properties follow the recent advancements in operator learning that generalize neural networks (and thus GANs) that only operate on a fixed resolution \citep{kovachki_neural_2021}; specifically in \cite{florez_data-driven_2022}, the input to the GAN was a sample from a finite-dimensional multivariate random variable and the output was a finite-dimensional object that could be queried only at a predetermined time-frequency resolution. 

Additionally, GANOs can learn probability measures on function spaces, in contrast to GANs, a difference that has been shown to avoid the mode collapse issues seen in GANs when the training data is generated from a mixture of GRFs (GANO reliably recovers the modes included in the
mixture). For more information related to mode collapse and the strategies of hyperparameter tuning that have been developed to avoid it, as well as issues associated with training stability and mode entanglement of GANs, the reader is referred to \cite{rahman2022generative, wang_faces_2021, abdal_styleflow_2020}. 

In what follows, we present the cGM-GANO training strategy, as well as the verification framework using simulated (noise-free) ground motions, and the validation framework using strong motion recordings from the Japanese strong motion network Kik-Net and empirical ground motion models. By contrast to \cite{florez_data-driven_2022}, we here train GANO using velocity time series and then differentiate the generated ground motions to compute the target 3-component acceleration time series. Learning the acceleration time integral is equivalent to learning a linear filter of the target time series that balances the low and high frequencies. This effect is better described with \eqref{1} where for non-zero frequencies, the integral amplifies low-frequency components (for $\omega \le 1$) and suppresses the high-frequency components (for $\omega \ge 1$). 
An example of the frequency content comparison between acceleration, velocity, and displacement is presented in the electronic supplement Fig S1. 

\begin{equation}
    \int_{0}^{t} f(t) \overset{FT}\leftrightarrow \frac{F(\omega)}{j\omega}
    \label{1}
\end{equation}

\subsection{Residual Analysis and Comparison to Empirical Ground Motion Models}  \label{sec:processing_reseval}

The statistical evaluation of the generated ground motions was performed (i) by comparing the PSA and FAS, for BBP dataset, or EAS, for Kik-net dataset, ordinates of the observed and synthetically generated ground motions for specific scenario bins and (ii) by performing residual analysis of the PSA and FAS or EAS of the training and testing datasets, for both the horizontal and vertical directions.
For the horizontal component, PSA was calculated as the rotation-independent 5\%-damped response spectra (RotD50) for periods in the $0.01 - 10 sec$ range \citep{boore_orientation-independent_2006, boore_orientation-independent_2010} using a modified GPU parallel computation version of the Pyrotd \citep{kottke_arkottkepyrotd_2018} python library, while in the vertical direction PSA was calculated as 5\%-damped response spectrum of the vertical component. 
The horizontal FAS was calculated as the power mean spectrum of the two horizontal components:
\begin{equation}
A_h(f) = \sqrt{ \frac{1}{2} \left(A_{h1}(f)^2+A_{h2}(f)^2\right)}
\end{equation}
\noindent where, $A_{h1}(f)$ and $A_{h2}(f)$ are the FAS of the two horizontal components at frequency $f$. 
The calculation of the horizontal EAS followed the recommendations of \cite{Kottke2021EAS}, smoothing the power-mean FAS through a \cite{Konno1998} smoothing window (Equation \ref{eq:ko}) with bandwidth, $b_w = 0.0333$.
The vertical FAS was calculated directly from the vertical component, and the vertical EAS was smoothed with the Konno-Ohmachi filter. 

\begin{align} \label{eq:ko}
    b &= 2 \pi/b_w \\
    W(f) &= \left( \frac{sin(b \ln(f/f_c))}{b \ln(f/f_c)} \right)^4
\end{align}

%For the comparison in the vertical direction, both the 5\%-damped PSA and FAS were calculated using the standard processes. 

The residual analysis was performed by comparing the observed intensity measures ($IM$) with the distribution of the synthetic ground motion $IM$, within the useable frequency range of the observed ground motions.
The comparison is performed by: (i) generating 100 synthetic 3-component realizations with the same ($\mathbf{M}$, $R_{rup}$, $V_{S30}$, $f_{type}$) conditional variable values of the observed ground motion, (ii) computing the relevant $IM$ from the observed and synthetic ground motions (${IM}^{\text{observed}}$, $IM^{\text{cGM-GANO}}$, respectively), (iii) calculating the mean and standard deviation of $IM^{\text{cGM-GANO}}$ in log space, and (iv) computing the normalized model residual, $\epsilon$, for each observation as:
\begin{equation}
    \epsilon = \frac{ln({IM}^{\text{observed}}) - mean(ln(IM^{\text{cGM-GANO}}))}{std(ln(IM^{\text{cGM-GANO}}))}
\end{equation}
this way the mean of $\epsilon$ quantifies bias and the standard deviation of $\epsilon$ the variability of the predictive model with respect to the data. A zero mean $\epsilon$ indicates that the ensemble of synthetic ground motions is unbiased with respect to the observations, and a unit $\epsilon$ standard deviation, or a two-unit width between the $16^{th}$ and $84^{th}$ $\epsilon$  percentile, suggests that the variability of the synthetic ground motion ensemble is consistent with the variability of the empirical dataset.

Since the BBP waveforms are considered noiseless, the entire frequency content of the ground motion was part of the useable frequency range.
Kik-net ground motions, on the other hand, were filtered based on the noise-to-signal ratio \citep{bahrampouri_updated_2021}.
In this case, the useable frequency range is defined as:
\begin{align}
    \text{Lowest usable frequency }(LUF)  &= 1.25 max(HPF_{H1},HPF_{H2},HPF_{V}) \\
    \text{Highest usable frequency }(HUF) &= 0.8 min(LPF_{H1},LPF_{H2},LPF_{V})
\end{align}
\noindent where $HPF_{H1}$, $HPF_{H2}$, and $HPF_{V}$ are the corner frequencies of the high pass filter for the two horizontal and vertical component, while $LPF_{H1}$, $LPF_{H2}$, $LPF_{V}$ are the corner frequencies for the low pass filter of the same components. 
The calculation of residuals was performed only within the usable frequency range to remove the effect of filtering  

% In each of the cases presented below, a zero mean $\epsilon$ indicates that the ensemble of synthetic ground motions is unbiased with respect to the observations, and a unit $\epsilon$ standard deviation suggests that the variability of the synthetic ground motion ensemble is consistent with the variability of the empirical dataset.

Lastly, we compared the scaling predicted by cGM-GANO against representative GMMs for the KiK-net dataset as part of our ongoing effort to develop a tool for efficient ground motion synthesis for engineering applications. Comparisons were performed against published PSA and EAS models such as the NGA-West2 GMMs \citep{abrahamson_summary_2014} and BA18 \citep{bayless_empirical_2019} for the shallow crustal events and BCHydro \citep{BCHydro016} for the subduction events.

\section{Training with Simulated Ground Motions: Verification}

We first verified our cGM-GANO architecture by training it on simulated ground motion time series generated using the SCEC BBP, a collection of open-source scientific software modules that can simulate broadband ($0\sim20 Hz$) ground motions for earthquakes at regional scales \citep{maechling_scec_2014}.

The goal of this exercise was to verify that cGM-GANO is able to assimilate noiseless ground motion training data to reproduce the median scaling and aleatory variability of the simulations. Verifying the reasonableness of the simulated ground motions, however, was outside the scope of this study. 
This exercise was conducted first, before testing cGM-GANO on observed ground motions, as it offered a simpler learning environment. 
The BBP-generated ground motions do not require filtering and baseline correction, reducing the number of factors affecting the waveforms, and the uniform site conditions across the ensemble of simulated scenarios reduce the number of conditional variables that cGM-GANO was called to learn.

\begin{figure*}
    \centering
    \begin{subfigure}[b]{0.45\textwidth}
        \caption{}
        \includegraphics[height=.8\linewidth]{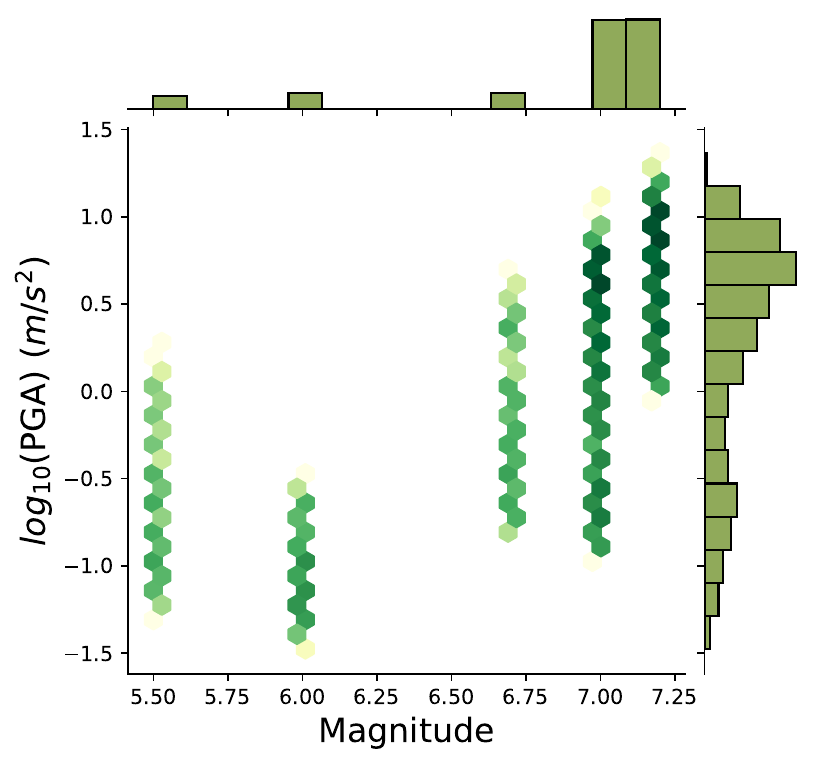}
    \end{subfigure}
    \quad
    \begin{subfigure}[b]{0.45\textwidth}
        \caption{}
        \includegraphics[height=.8\linewidth]{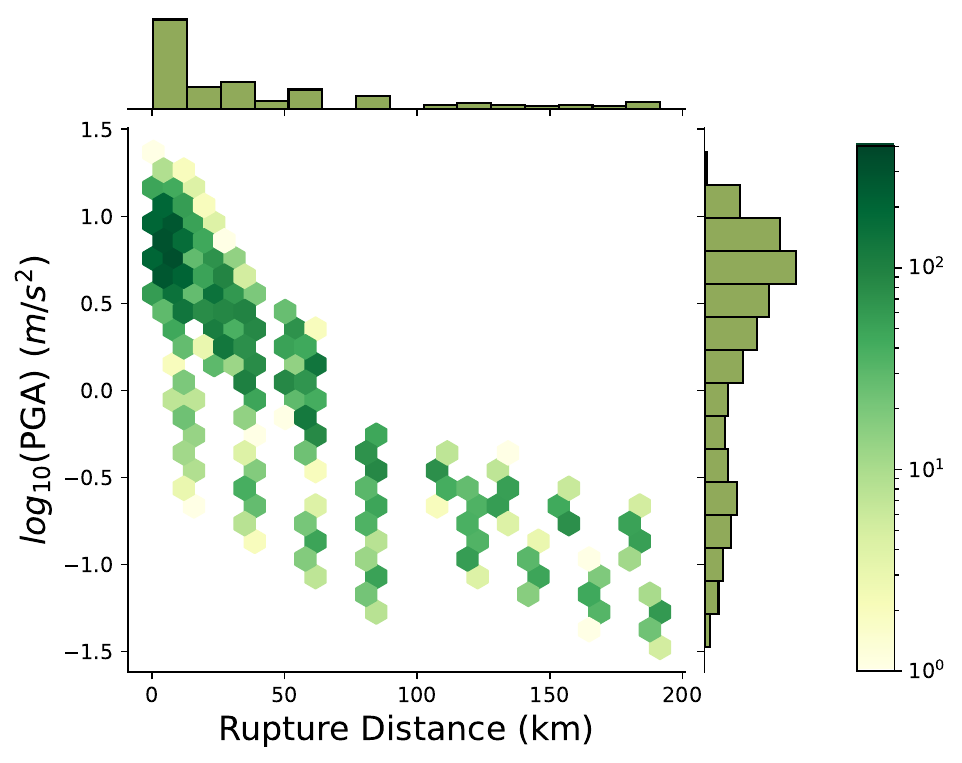}
    \end{subfigure}
    \caption{Data distribution of BBP dataset. Subfigures (a) and (b) show crossplot between $log_{10}$(PGA) and magnitude and rupture distance, respectively}
    \label{fig:bbp_dataset}
\end{figure*}

For this study, we used a dataset developed on BBP v17.3.0. following the \cite{graves2015refinements} simulation module (BBP-GP). This module is a hybrid approach that combines low-frequency (often referred to as deterministic) and high-frequency (often referred to as stochastic) motions computed with different methods into a single broadband response. The database was developed by \cite{shi2019improving} in collaboration with the BBP-GP developer, Robert Graves (pers. communication), and it contains 113 events ranging from $\mathbf{M}$ $5.5$ to $7.2$ on strike-slip and from $\mathbf{M}$ $7.0$ to $7.2$ on reverse faults, recorded at stations on reference site conditions  ($V_{s30} = 760$ m/s), at rupture distances between 0 to 200 km. The database contains 6704 acceleration time histories: each record has 3-component 40s waveforms sampled at 100Hz.
The magnitude-distance-PGA distribution of the data is shown in Fig \ref{fig:bbp_dataset}. A representative illustration of the fault and stations' locations is shown in Fig S2 in the electronic supplement.

In this case, the cGM-GANO was trained conditioned on the moment magnitude ($\mathbf{M}$), on the closest-point-on-the-rupture to site distance ($R_{rup}$), and on the fault type ($f_{type}$, strike-slip or reverse). 
% All SCEC BBP simulations of the \cite{shi2019improving} database were developed for reference ($V_{s30} = 760$ m/s) site conditions, so this cGM-GANO training was not conditioned on $V_{s30}$.

\begin{figure*}
    \centering
    \begin{subfigure}{0.23\textwidth}
        \caption{}
        \includegraphics[width=1.1\textwidth]{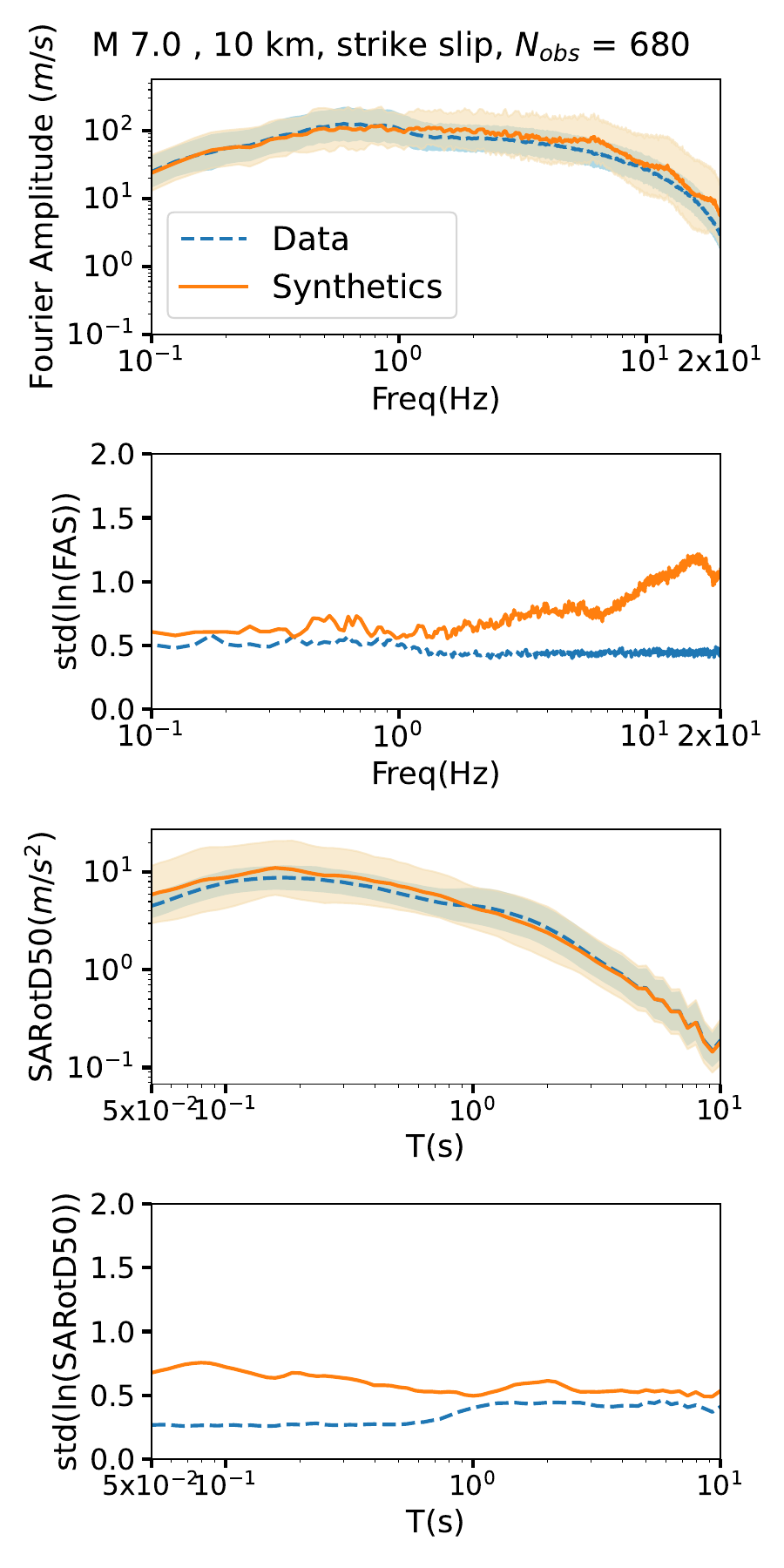}
    \end{subfigure}
    \quad
    \begin{subfigure}{0.23\textwidth}
        \caption{}
        \includegraphics[width=1.1\textwidth]{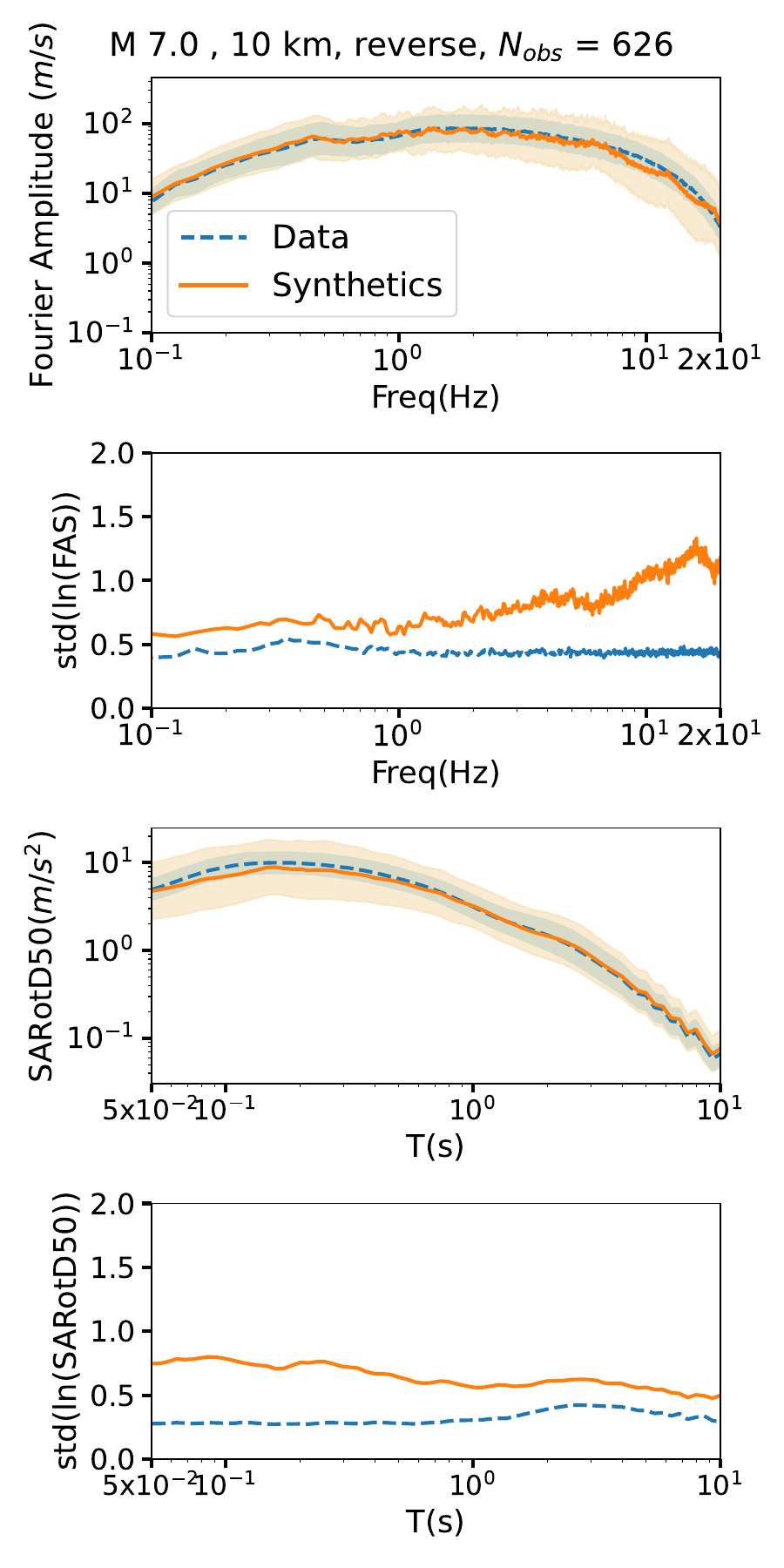}
    \end{subfigure}
    \quad
    \begin{subfigure}{0.23\textwidth}
        \caption{}
        \includegraphics[width=1.1\textwidth]{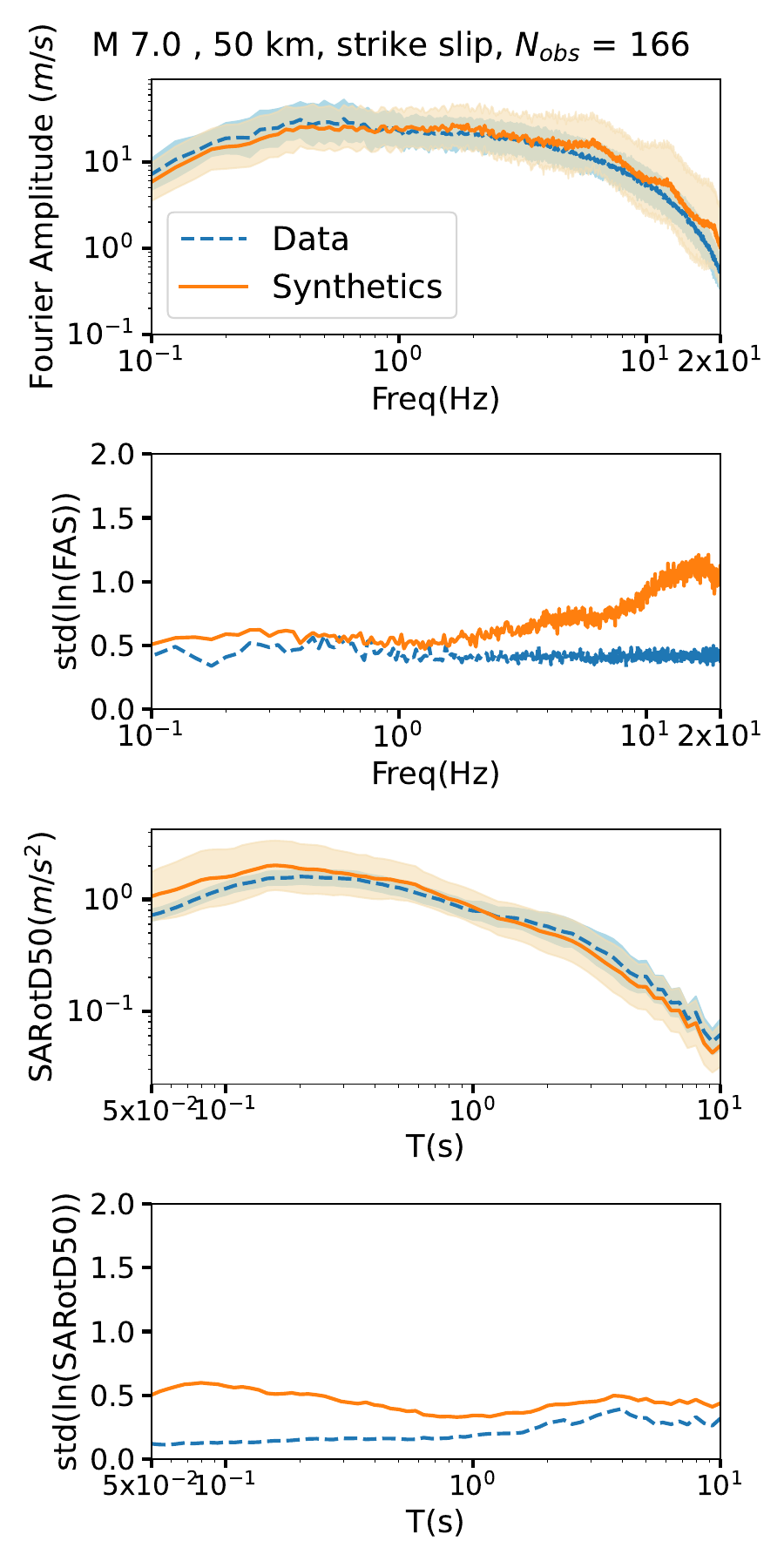}
    \end{subfigure}
    \quad
    \begin{subfigure}{0.23\textwidth}
        \caption{}
        \includegraphics[width=1.1\textwidth]{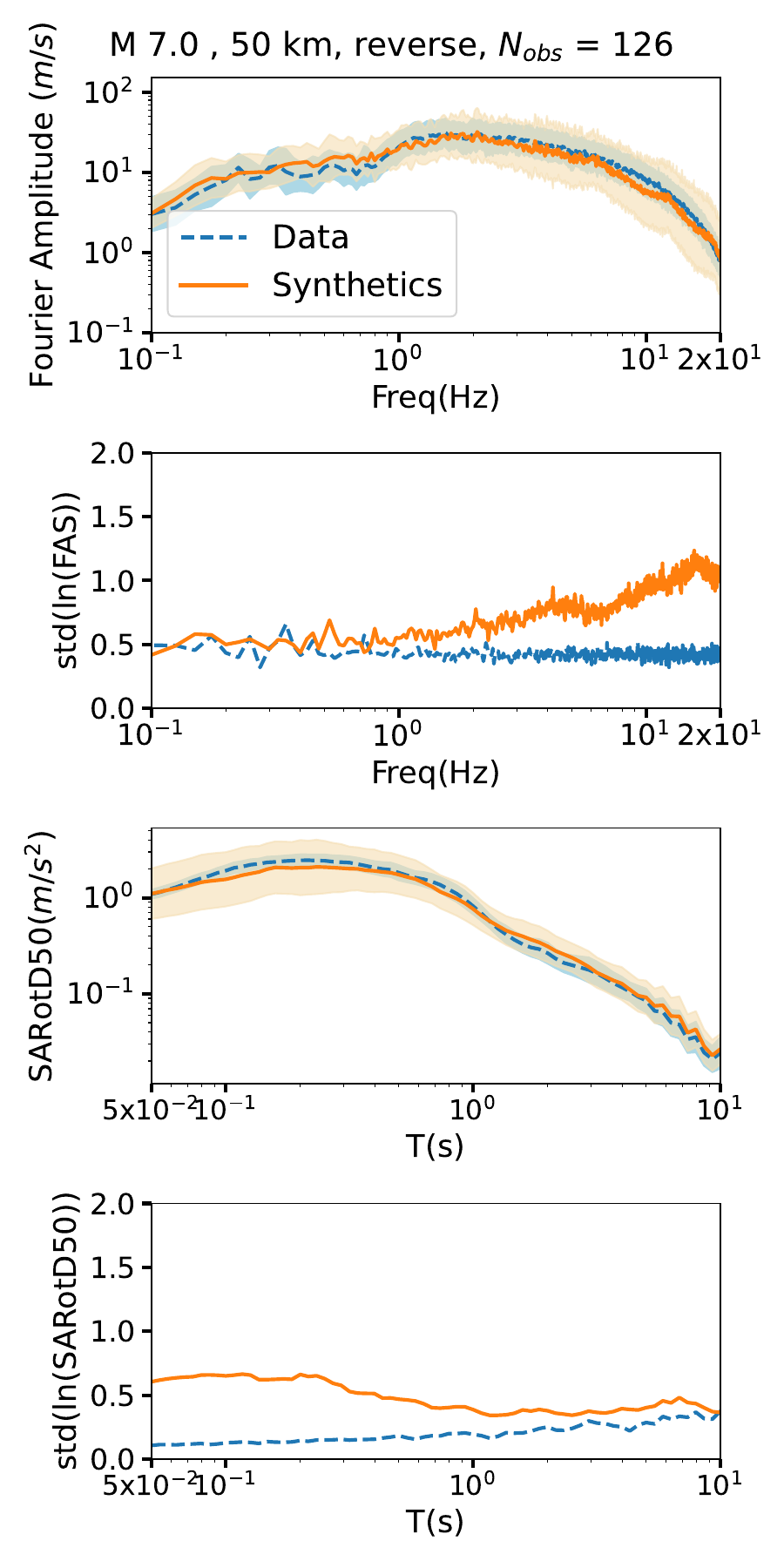}
    \end{subfigure}
    \caption{Scenario event comparisons for cGM-GANO trained on the BBP dataset. Each vertical sub-panel depicts the Fourier Amplitude Spectrum (FAS), aleatory standard deviation of FAS, Rotd50 Pseudo-spectral acceleration (PSA) spectrum, and aleatory standard deviation of Rotd50 PSA.
     The title of each plot corresponds to the midpoint of each bin, and $N_{obs}$ is the number of observed records within each bin.
    The lines represent the mean in log space of the horizontal components of the synthetic (solid line) or BBP (dashed line) data, while the shaded zone represents the $16^{th}$ to $84^{th}$ percentile range.}
    \label{fig:bbp_fas_sa_scen}
\end{figure*}

Fig.~\ref{fig:bbp_fas_sa_scen} compares the FAS and PSA spectra of BBP simulated and GANO-generated 3-component waveforms for $\mathbf{M}~7.0$ strike-slip and reverse events at  $10~km$ rupture distance. 
The comparison of the FAS and PSA ordinates is encouraging, especially in the long period, short frequency regime, where both the mean and standard deviation are in agreement.
In particular, one can observe that the standard deviation of the two data sets converges for frequencies smaller than $f < 1 Hz$ ($T > 1 sec$), while for higher frequencies, the cGM-GANO-generated time series have systematically higher variability. 
Our working hypothesis is that the discontinuity occurs at $f = 1 Hz$, the cross-over frequency between the 
low- and high-frequency methods in the BBP-GP hybrid module, because cGM-GANO cannot learn the stochastic high-frequency method and resorts to carrying over the learned physics-based solution of $f < 1 Hz$ into the higher frequency range. One reason for this might be that GANO can only learn distributions over functions that are square-integrable, and the stochastic generation method used in the BBP does not necessarily satisfy this property.
Representative waveforms from both the BPP and cGM-GANO are also depicted in the electronic supplement (Fig S3), showing that the synthetic time series manifest realistic features, such as P-,  S-, and coda wave arrivals, as well as time history envelopes consistent with the simulated ground motions. 

\begin{figure*}
    \centering
    \begin{subfigure}{0.4\textwidth}
        \caption{}
        \includegraphics[width=\textwidth]{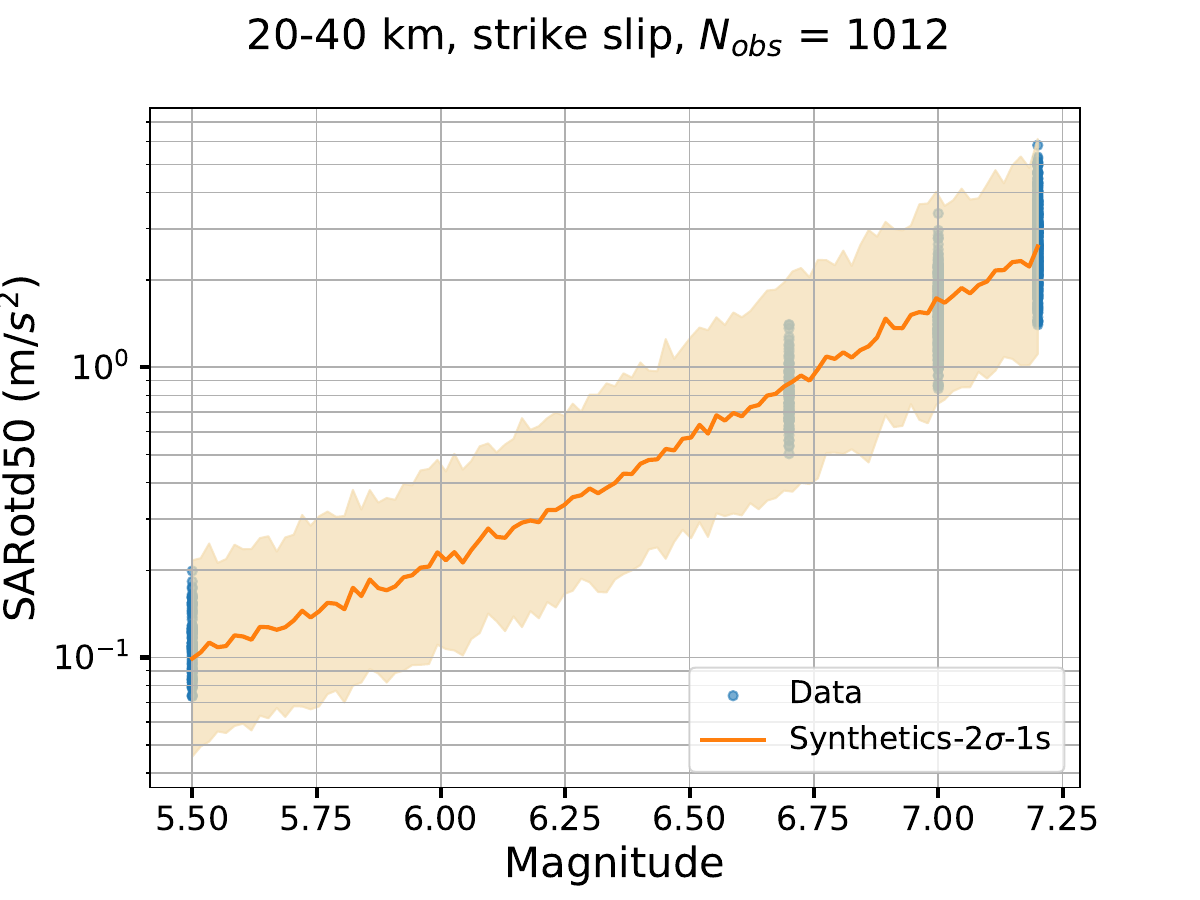}\quad
    \end{subfigure}
    \begin{subfigure}{0.40\textwidth}
        \caption{}
        \includegraphics[width=\textwidth]{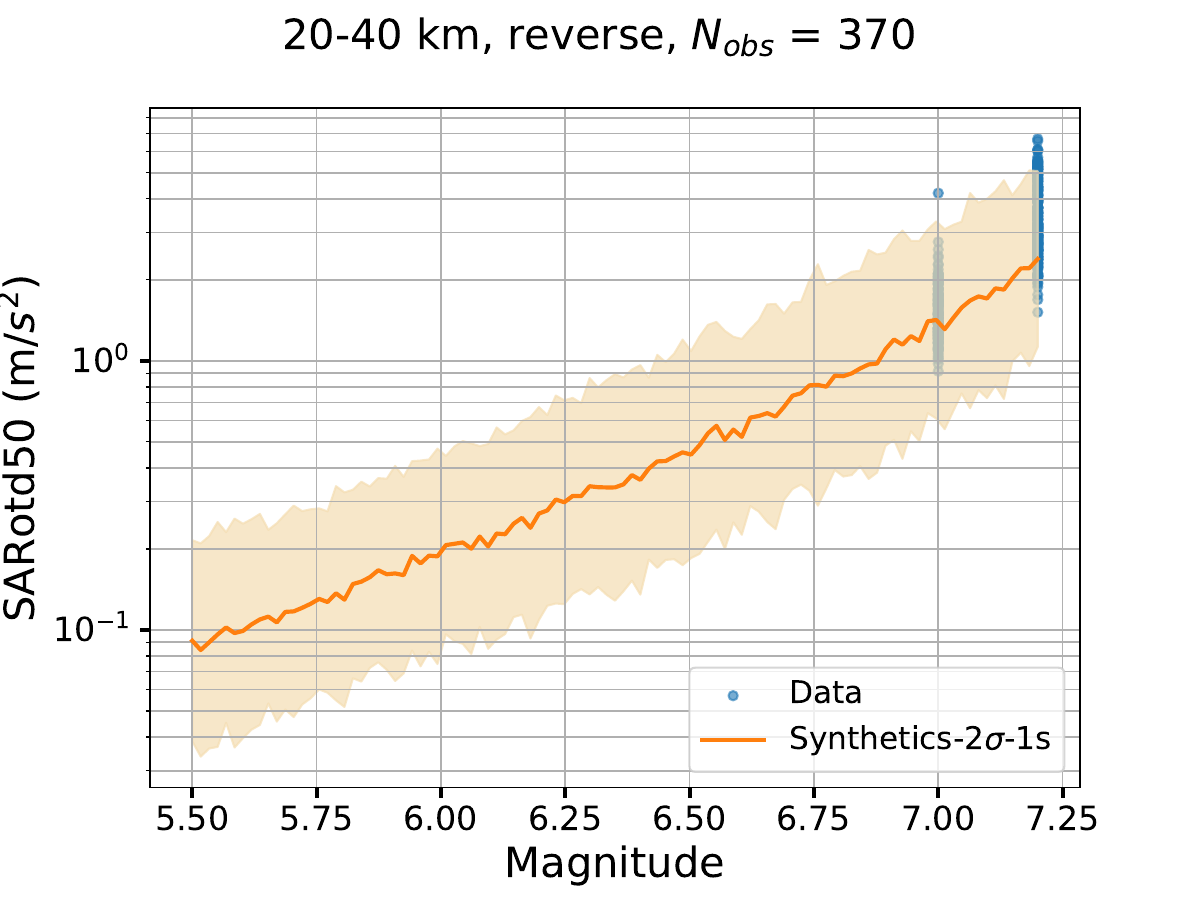}
    \end{subfigure}
    \begin{subfigure}{0.40\textwidth}
        \caption{}
        \includegraphics[width=\textwidth]{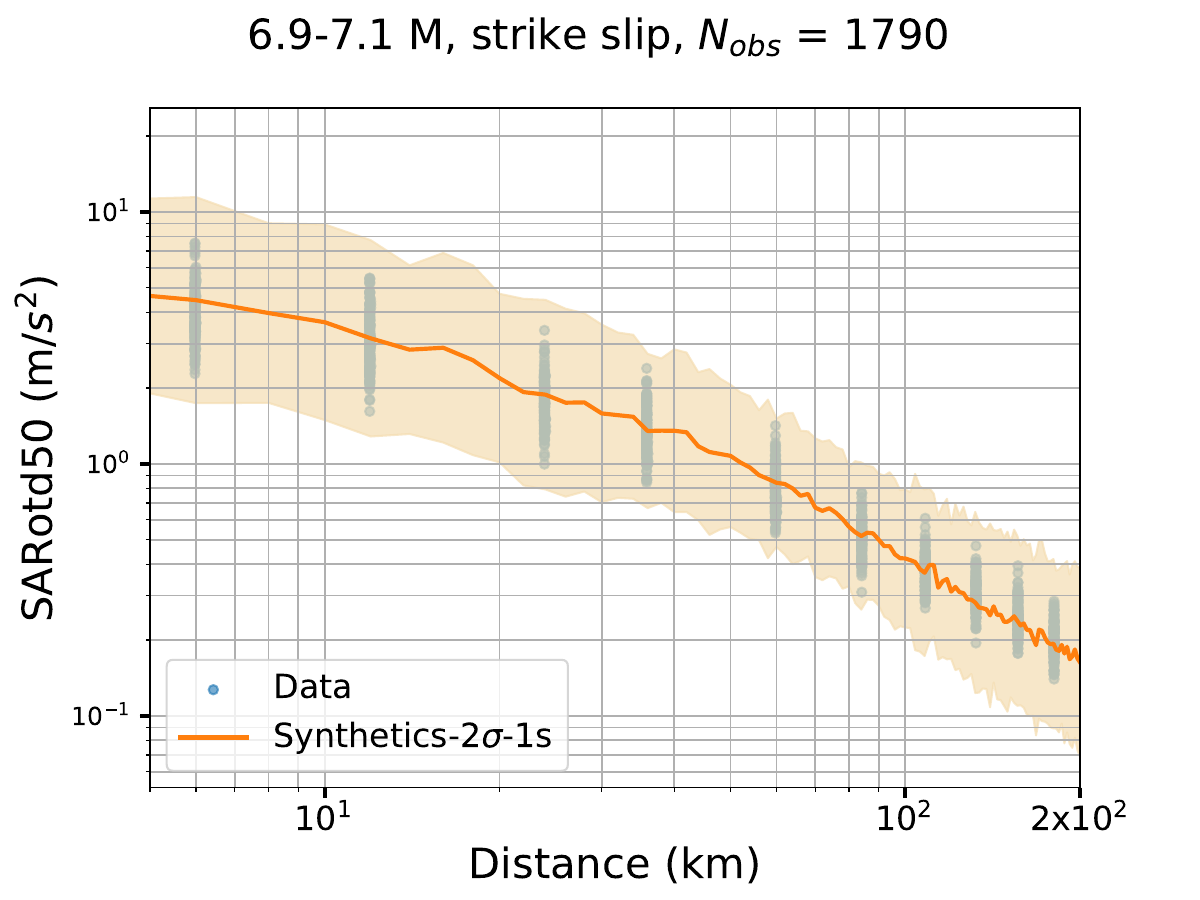}\quad
    \end{subfigure}
    \begin{subfigure}{0.40\textwidth}
        \caption{}
        \includegraphics[width=\textwidth]{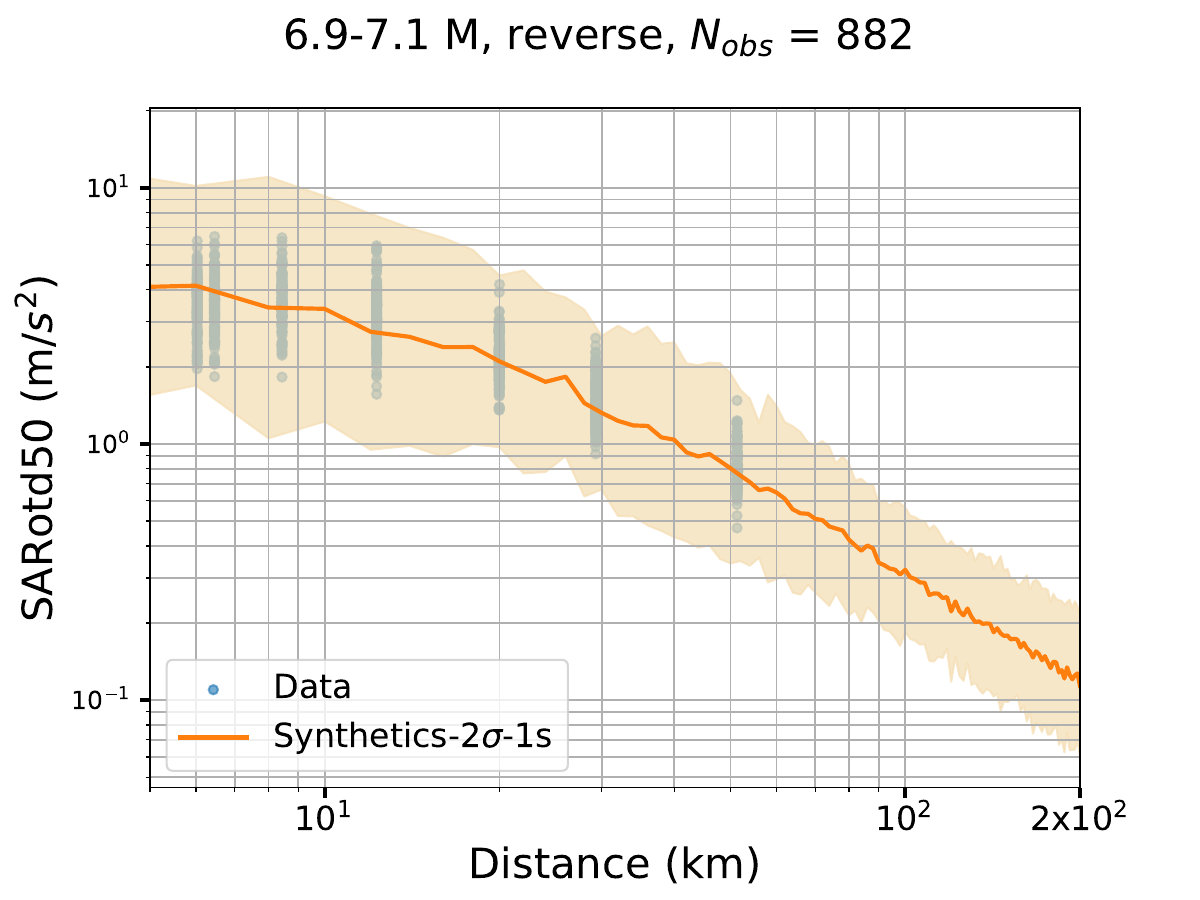}\quad
    \end{subfigure}
    \caption{Comparison of cGM-GANO magnitude and distance scaling with BBP dataset for PSA RotD50(T=1 s).
    Solid lines represent the mean scaling in log space. The shaded region shows the $2^{nd}$ to $98^{th}$ percentile of aleatory variability. 
    Solid dots correspond to the training data. }
    \label{fig:BBP_fill}
\end{figure*}

Next, we test the capabilities of cGM-GANO to learn the magnitude and distance scaling in sparsely sampled regions of the training dataset. Fig.~\ref{fig:BBP_fill} shows that the BBP training database by \cite{shi2019improving} has no reverse event records (here, depicted by the dot markers) for small magnitude at distances $20-40 km$, or for magnitudes $M6.9-7.1$ at large distances ($\ge 50 km$). Still, the cGM-GANO was trained using data from both reverse and strike-slip events, each appropriately tagged.
The example in the figure suggests that the model has, in part, inferred the median magnitude and distance scaling of the reverse style faulting from the data ensemble training, transferring the information across faulting styles. At the same time, however, the variability of cGM-GANO is larger than the BBP data in the densely sampled regions of the training dataset (for example, $M6.9-7.1$ far-field strike-slip as well as near-field reverse) where one would expect the variability to be best captured.

\begin{figure*}
    \centering
    \begin{subfigure}{0.4\textwidth}
        \caption{}
        \includegraphics[width=\textwidth]{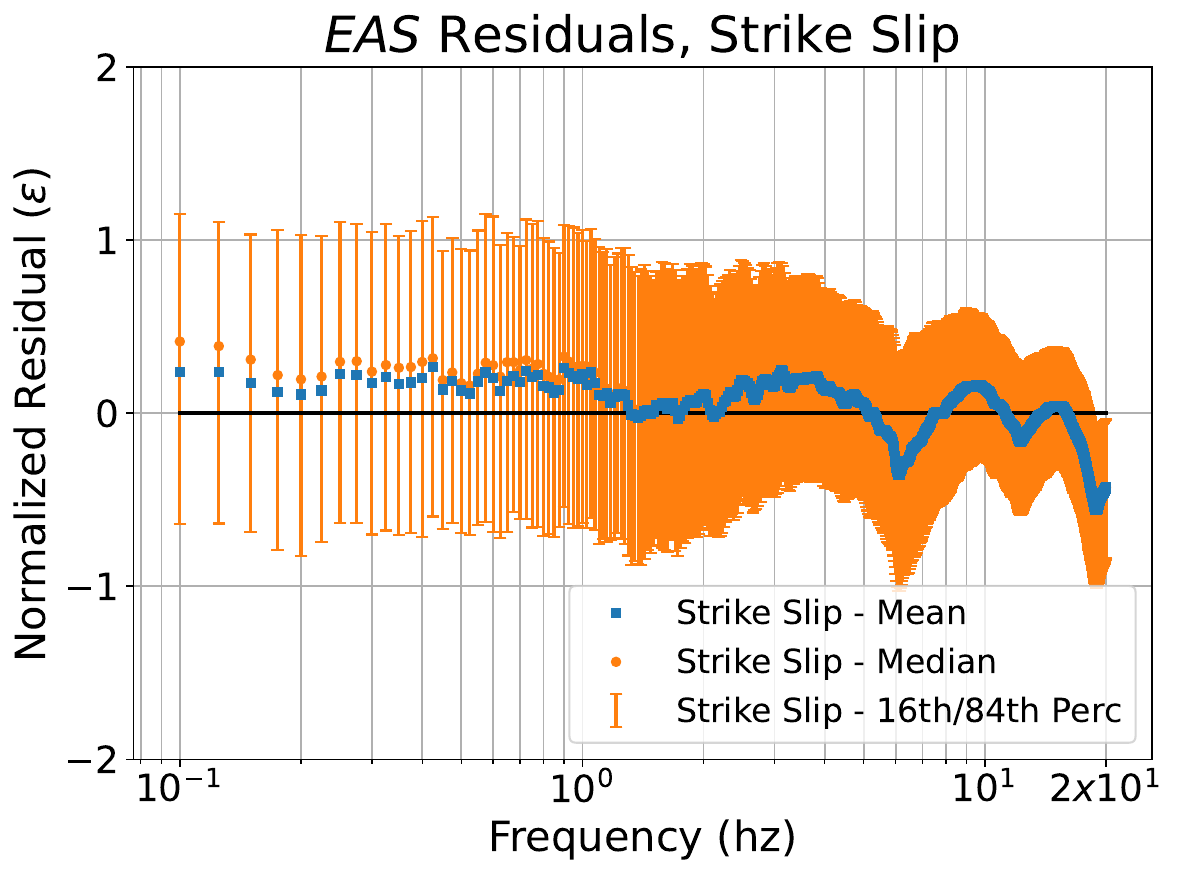}
    \end{subfigure}
    \quad
    \begin{subfigure}{0.4\textwidth}
        \caption{}
        \includegraphics[width=\textwidth]{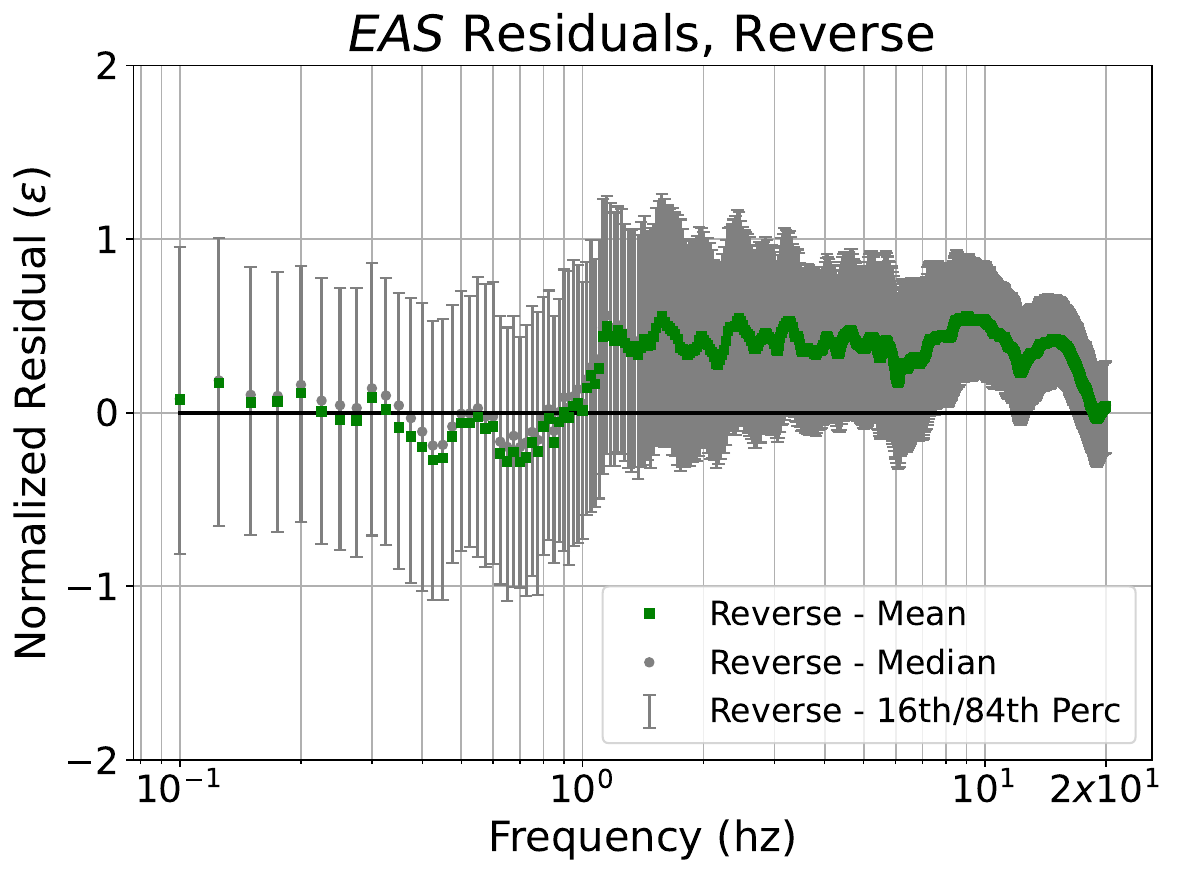}
    \end{subfigure}
    \quad
    \begin{subfigure}{0.4\textwidth}
        \caption{}
        \includegraphics[width=\textwidth]{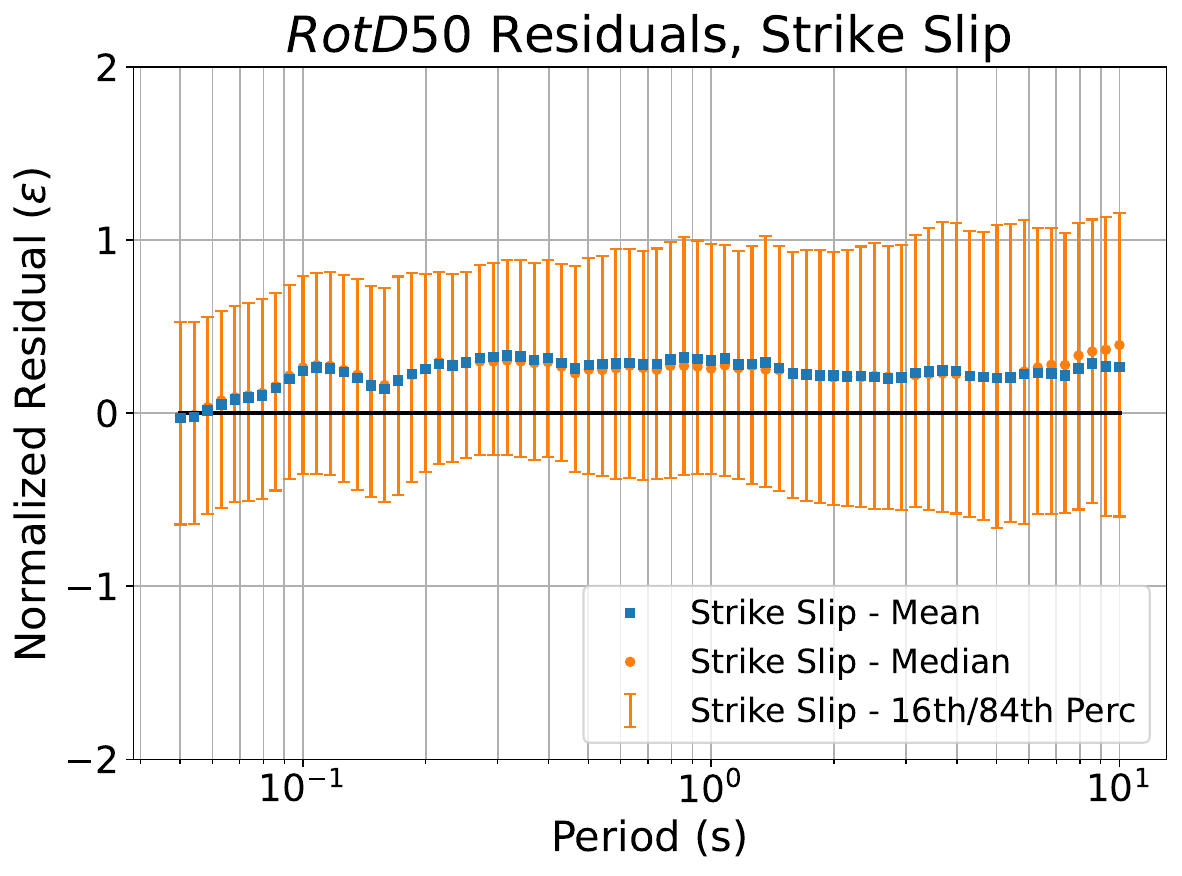}
    \end{subfigure}
    \quad
    \begin{subfigure}{0.4\textwidth}
        \caption{}
        \includegraphics[width=\textwidth]{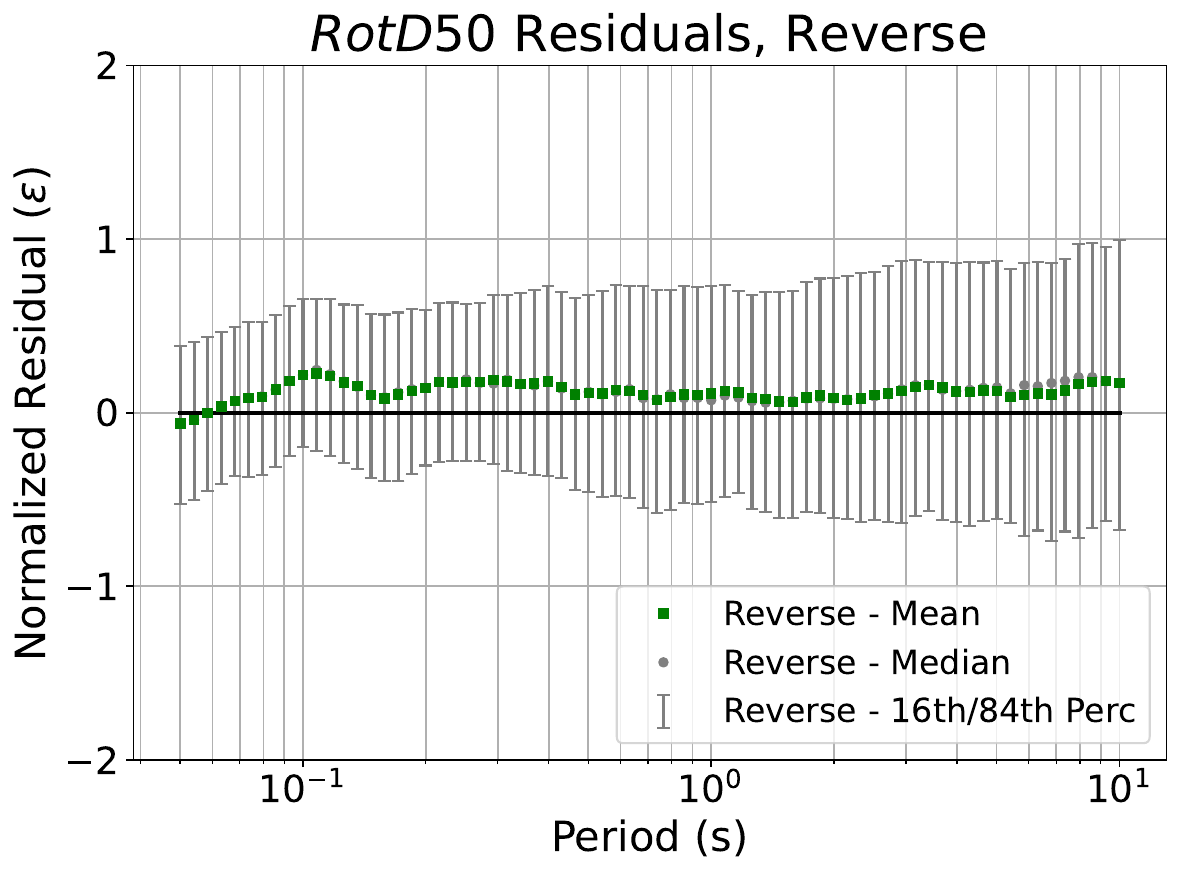}
    \end{subfigure}
    \caption{Residual plots of the BBP trained model. 
    Subfigures (a) and (b) show the EAS residuals for strike-slip and reverse faults, respectively.
    Subfigures (c) and (d) show the RotD50 PSA residuals for strike-slip and reverse faults, respectively.}
    \label{fig:BBP_res}
\end{figure*}

Lastly, we examined the degree to which cGM-GANO can recover the frequency domain characteristics of the BBP simulated ground motions through a residual analysis. Results for the horizontal and vertical components of the EAS and RotD50 residuals are shown in Fig \ref{fig:BBP_res}, separately for the strike-slip and the reverse simulated events. 
The cGM-GANO results for the horizontal EAS components for both types of ruptures show discontinuities at $1$Hz.
The mean and aleatory variability are captured reasonably well for both strike-slip and reverse events in the low-frequency range ($f \le 1$Hz).
In the high-frequency range ($f > 1$Hz), the mean is captured reasonably well, but the aleatory variability is underestimated for strike-slip events, and there is a noticeable bias in the mean and underestimation of the variability for reverse events. 
The RotD50 spectral accelerations show a slight constant positive bias across the period range for both styles of faulting. 
The difference between EAS and RotD50 residual distributions is attributed to the unsuitable inter-frequency correlation of the synthetic ground motions that is influenced by the stochastic component of the training dataset (Electronic Supplement Fig S4); further investigation is needed, nonetheless, to find the exact cause of the mismatch between the bias and variability of EAS and PSA in the various frequency-period ranges.

Our working hypothesis is that verification of cGM-GANO using BBP simulations, which are synthesized by two separate processes, is compromised when the neural operator attempts to recover the entire frequency range as if it were the solution of the wave equation for propagation in a continuum.
%the neural operator, by learning a continuous broadband set of ground motion characteristics, is suspected to be compromised in recovering the entire frequency range. 
Whether the identified biases in the aforementioned figures stem from the hybrid BBP-GP method and the difficulties associated with training neural operators on stochastic data; or it is evidence that more training data is necessary to reliably recover the median and standard deviation scaling relationships across the entire dataset, is still being debated. More extensive validation is currently being carried out by the authors to confirm this claim, including the simulation of large-magnitude, near-field reverse faulting events to confirm the aforementioned recovered scaling. 
These results, along with the foregoing sections, however, point to the possibility of using cGM-GANO in the future to fill sparsely sampled regions of observational datasets used to develop GMMs with defensible, realistic, 3-component ground motion time series appropriate for engineering applications.

\section{Training with Recorded Ground Motions: Verification and Validation}

In this section, we evaluated the performance of cGM-GANO by training it on a ground motion dataset compiled from the Japanese Strong Motion Network KiK-net \citep{nied_nied_2019}. The compiled KiK-net dataset included 42481 strong ground motions sampled at 100Hz from 1985 events, 620 shallow crustal and 1365 subduction events, collected over a period of 20 years (from 1997 to 2017) at 613 stations. For the data preprocessing, we used the method described in \cite{bahrampouri_updated_2021}, which comprises of the following steps: (i) identify the filter corner frequencies for each record based on noise windows, and (ii) detect the P- and S-wave arrivals. We ensured that each time history contained the ground motion from a single event, and we used a fixed duration of $60 sec$ for training, starting from the onset of the P wave. The choice of a fixed duration window stems from FNO requirements that we used in our architecture. Prior to training, we detrended and baseline-corrected the KiK-net ground motions to ensure that the displacement, velocity, and acceleration are zero at the end of the $60 sec$. The same baseline correction was applied on the cGM-GANO generated ground motion time series before the validation tests described below. 

\begin{figure*}
    \centering
    \begin{subfigure}{0.28\textwidth}
        \includegraphics[height=\textwidth]{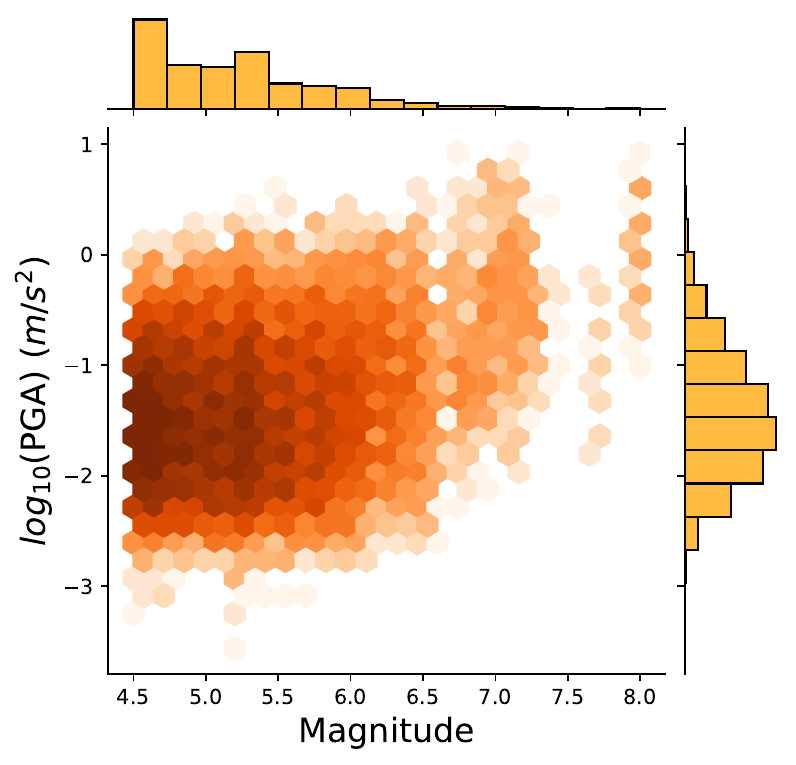}
    \end{subfigure}
    \quad
    \begin{subfigure}{0.28\textwidth}
        \includegraphics[height=\textwidth]{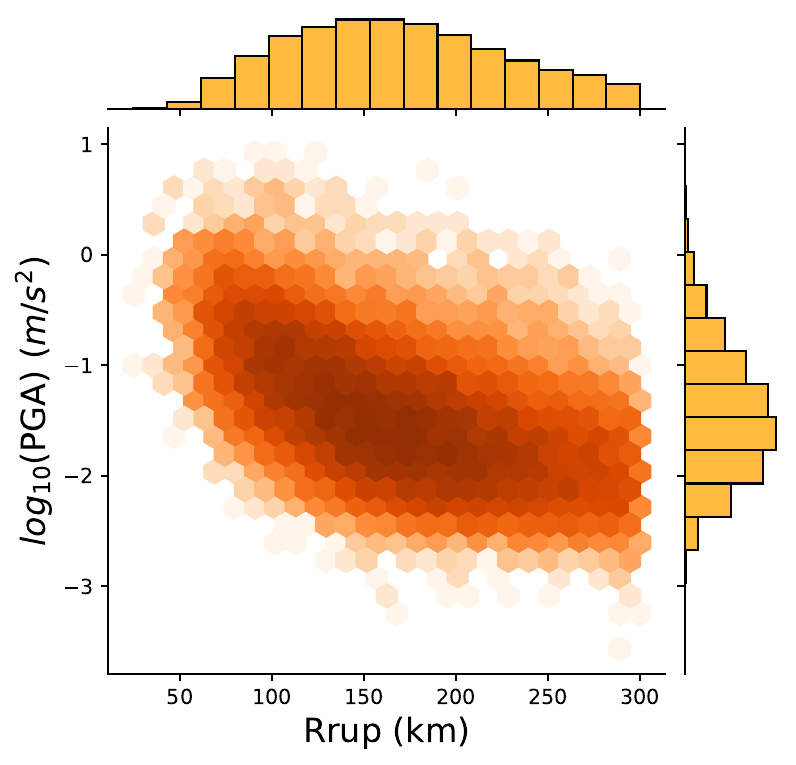}
    \end{subfigure}
    \quad    
    \begin{subfigure}{0.28\textwidth}
        \includegraphics[height=\textwidth]{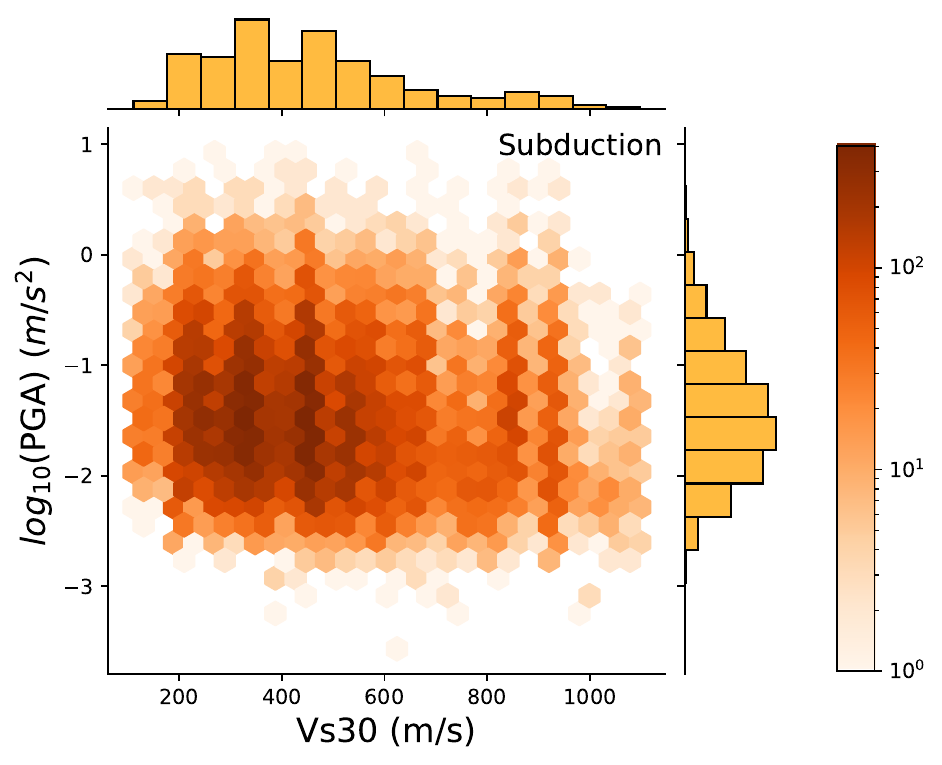}
    \end{subfigure}
    \quad  

    \centering
    \begin{subfigure}{0.28\textwidth}
        \includegraphics[height=\textwidth]{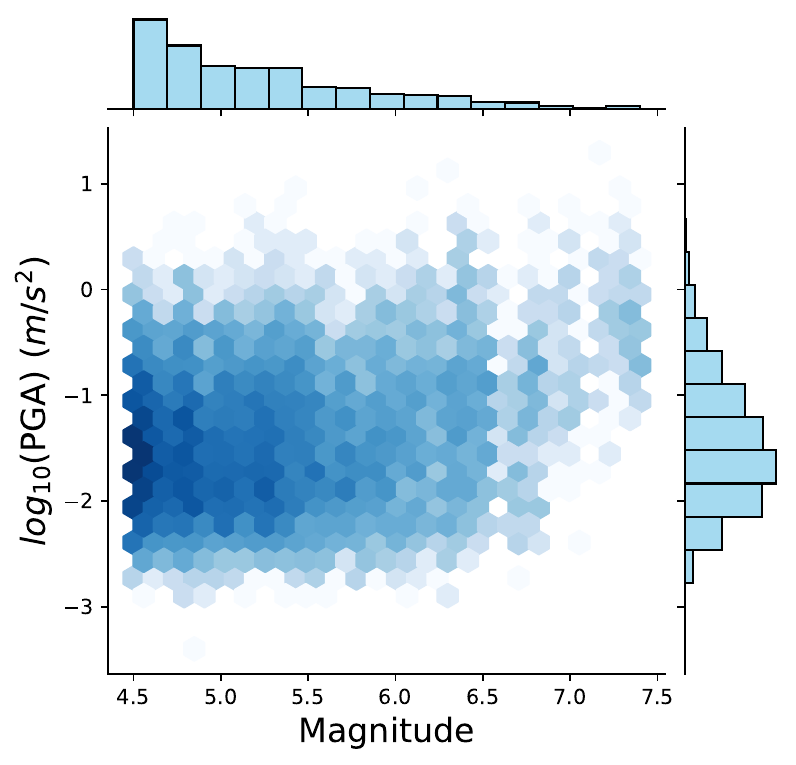}
    \end{subfigure}
    \quad
    \begin{subfigure}{0.28\textwidth}
        \includegraphics[height=\textwidth]{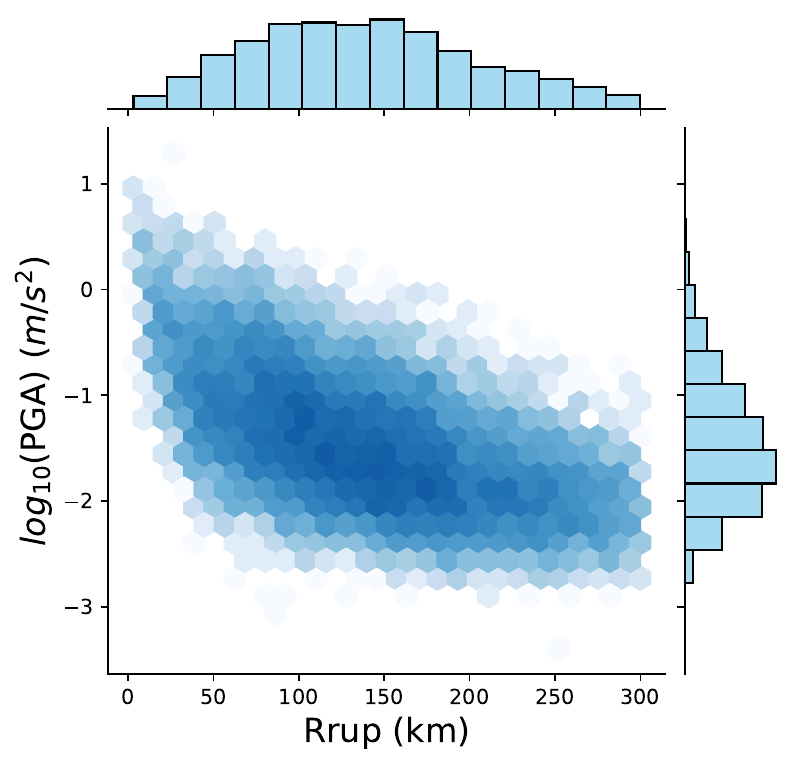}
    \end{subfigure}
    \quad    
    \begin{subfigure}{0.28\textwidth}
        \includegraphics[height=\textwidth]{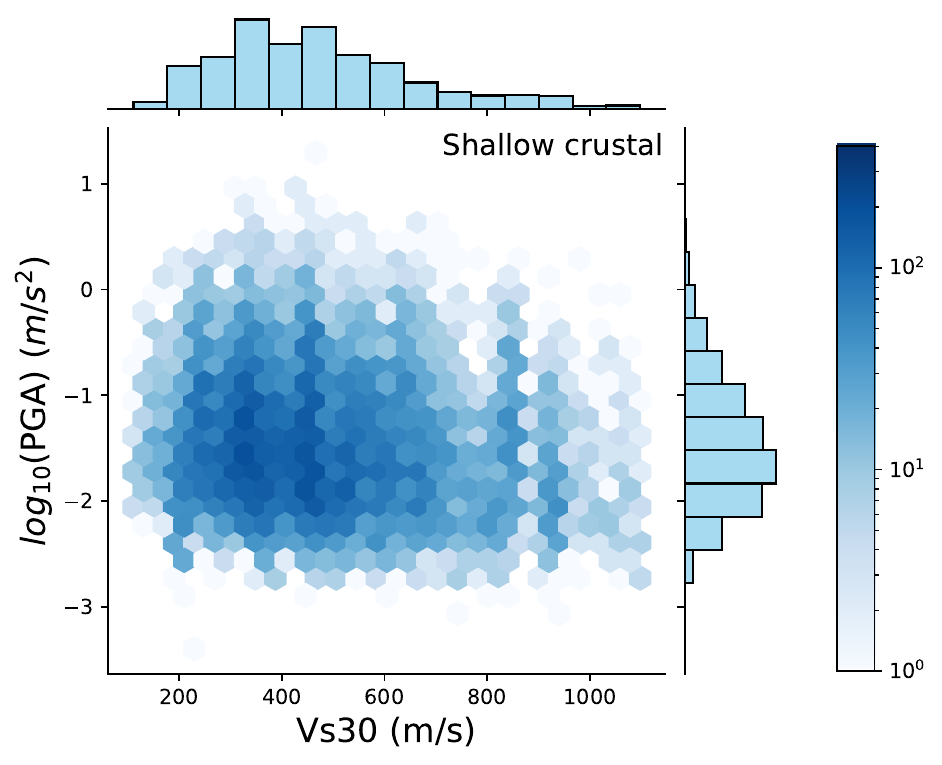}
    \end{subfigure}
    \quad 

\caption{
Distribution of earthquake magnitude, rupture distance, and $V_{S30}$ versus  $log_{10}$(PGA) for the subduction (first row) and shallow crustal (second row) events. }
\label{fig:kik_dataset}
\end{figure*}

The conditional parameters for the KiK-net dataset ($\mathbf{M}$, $R_{rup}$, $V_{S30}$, and $f_{type}$) were obtained from \cite{bahrampouri_updated_2021}; $M$ ranges from 4.5 to 8.0, $R_{rup}$ from 0 to 300 km, $V_{S30}$ from 100 to 1100 m/s, and the tectonic environment is subduction or shallow crustal. 
Figure \ref{fig:kik_dataset} shows the magnitude, distance, and $V_{S30}$ distribution of the compiled data, which is primarily comprised of small-to-moderate events recorded at distances $R_{rup} \ge 50$km, on moderate stiffness sites. Only 1.6\% of the dataset ground motions corresponded to events with $\mathbf{M} \ge 7.0$, 2.8\% of records have rupture distances less than $R_{rup} \le 50$km, and  2.7\% of the ground round motions were recorded on stations with $V_{S30} < 200 m/sec$.

For training purposes, the KiK-net dataset was randomly divided into 90\% training data and 10\% testing data, not used in training. In what follows, we present verification results against the entire database, while validation results based on the testing dataset alone are provided in the electronic supplement (Fig S5).  
Both the entire and testing datasets show similar trends, indicating no overfitting on the training dataset. 
It was decided to include the comparison with the entire dataset in the manuscript for more robust magnitude, rupture distance, $V_{S30}$ scaling, and individual scenario comparisons. 

\begin{figure*}
\centering
    \begin{subfigure}{0.48\textwidth}
        \caption{}
        {\includegraphics[width=\textwidth]{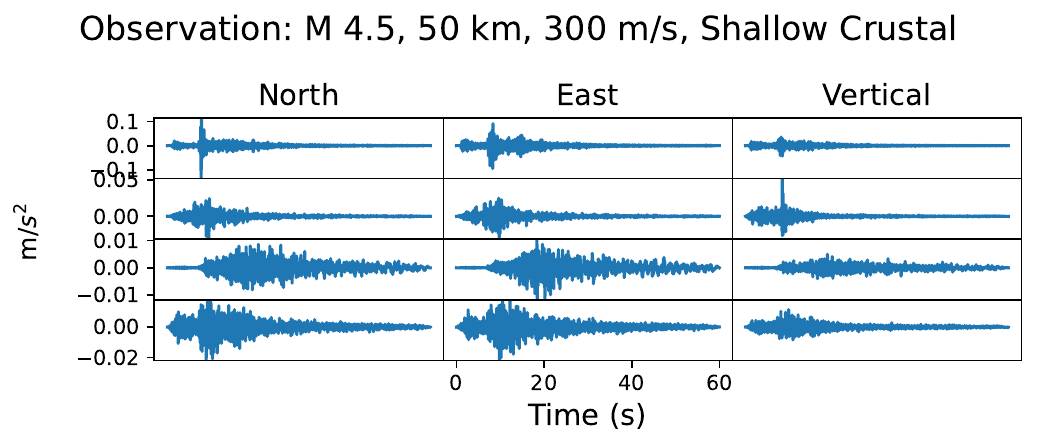}}
    \end{subfigure}
    \quad
    \begin{subfigure}{0.48\textwidth}
        \caption{}
        {\includegraphics[width=\textwidth]{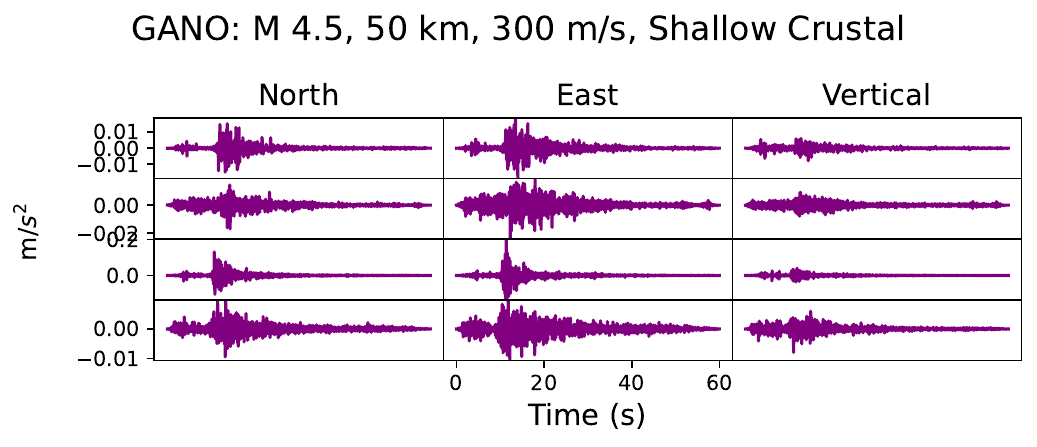}}
    \end{subfigure} 
    \medskip
    \begin{subfigure}{0.48\textwidth}
        \caption{}
        {\includegraphics[width=\textwidth]{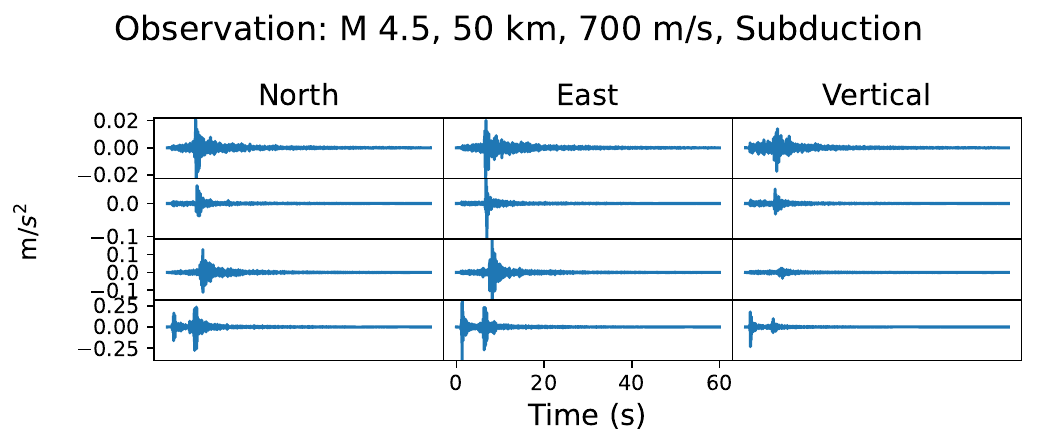}}
    \end{subfigure}
    \quad
    \begin{subfigure}{0.48\textwidth}
        \caption{}
        {\includegraphics[width=\textwidth]{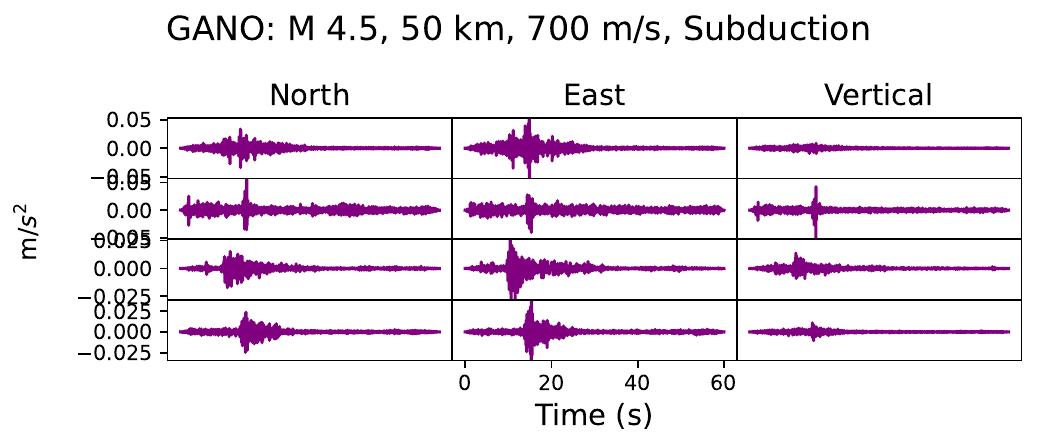}}
    \end{subfigure}    
    \medskip
    \begin{subfigure}{0.48\textwidth}
        \caption{}
        {\includegraphics[width=\textwidth]{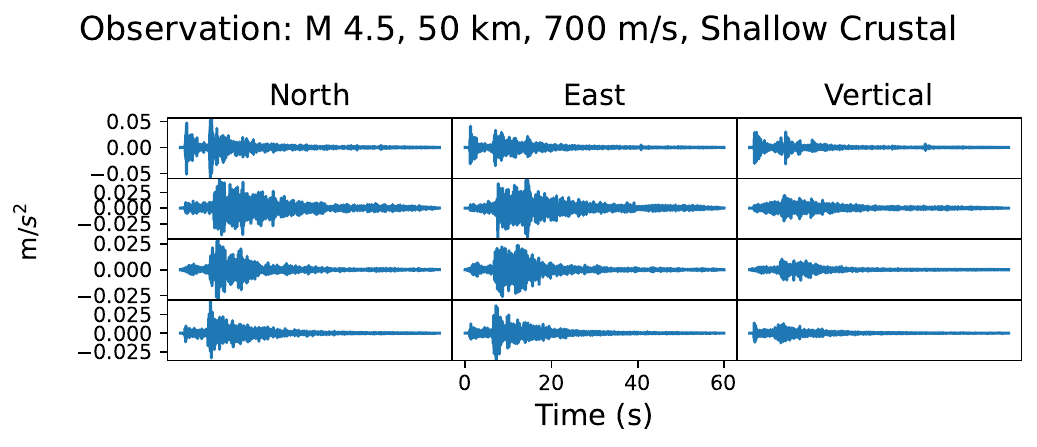}}
    \end{subfigure}
    \quad
    \begin{subfigure}{0.48\textwidth}
        \caption{}
        {\includegraphics[width=\textwidth]{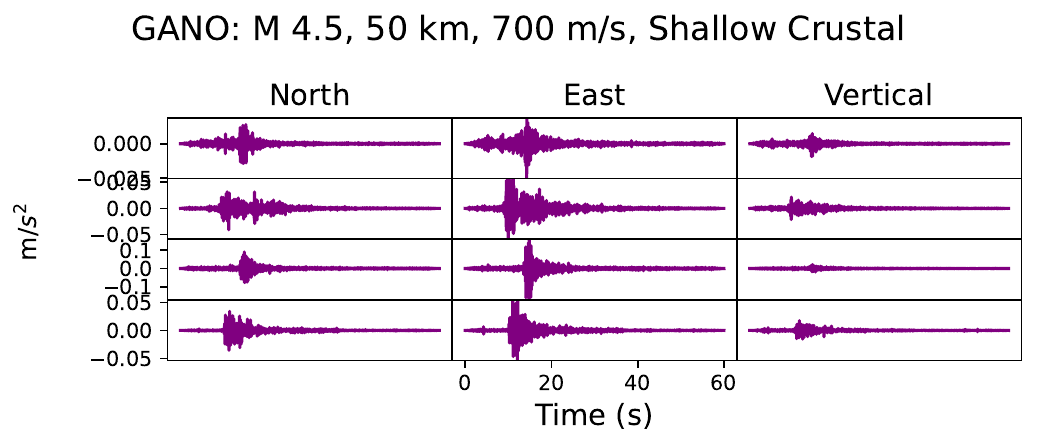}}
    \end{subfigure} 
    \medskip
    \begin{subfigure}{0.48\textwidth}
        \caption{}
        {\includegraphics[width=\textwidth]{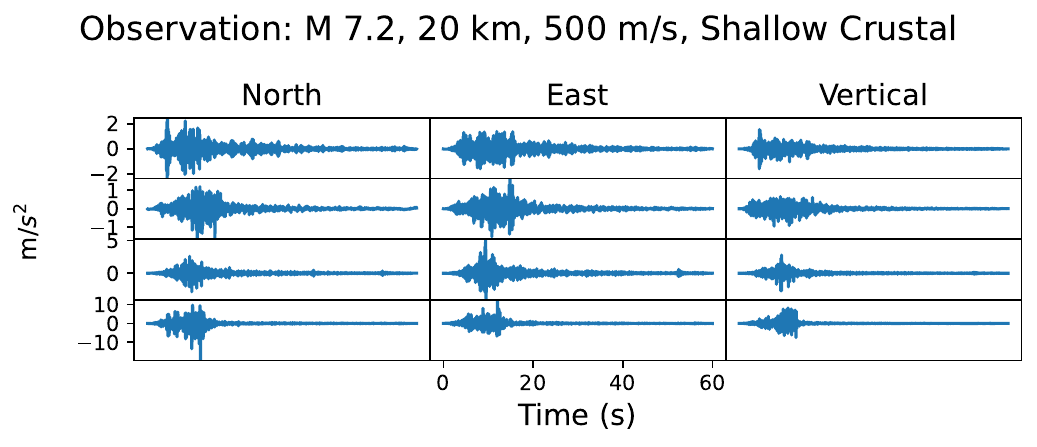}}
    \end{subfigure}
    \quad
    \begin{subfigure}{0.48\textwidth}
        \caption{}
        {\includegraphics[width=\textwidth]{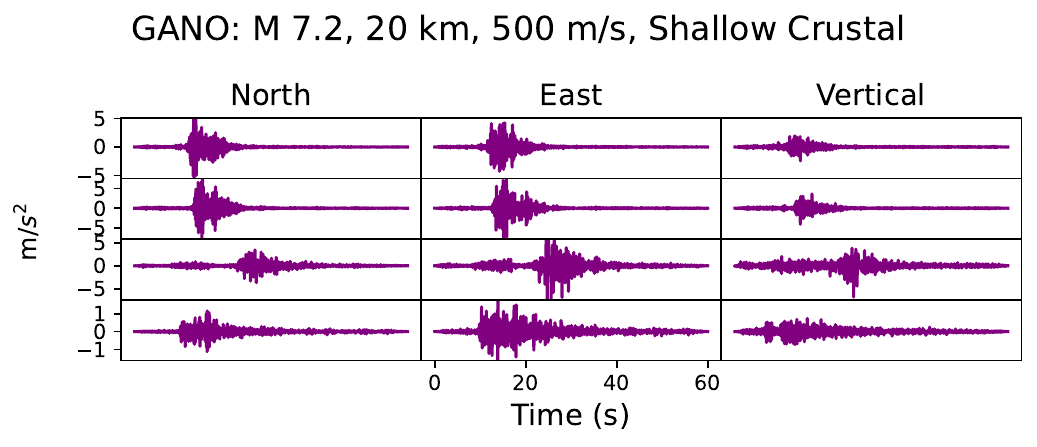}}
    \end{subfigure}  
    \caption{Accelerograms comparisons real and synthetically generated ground motions for  cGM-GANO trained on the Kik-net dataset.}
    \label{fig:visual_check}
\end{figure*}

Fig \ref{fig:visual_check} shows a visual comparison between randomly sampled observed time series for different $\mathbf{M}$, $R_{rup}$, $V_{S30}$ combinations and the corresponding cGM-GANO random realizations for the same scenarios. 
This comparison is intended to show that the generated waveforms are comparable in amplitude, duration and phase arrival with the recorded ground motions.

\begin{figure*}
    \begin{subfigure}{0.32\textwidth}
        \caption{}
        {\includegraphics[width=\textwidth]{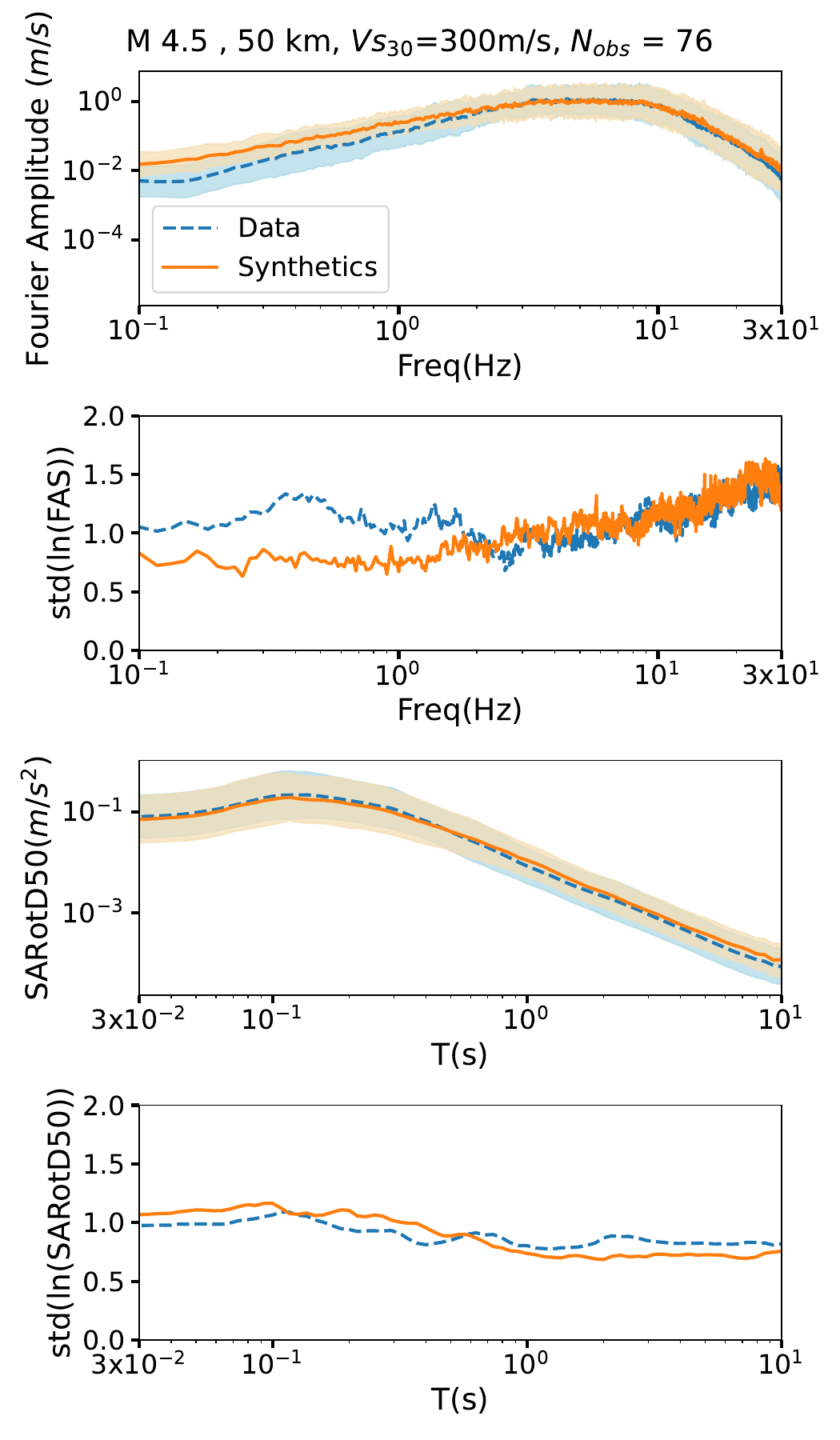}}
    \end{subfigure}
    \quad
    \begin{subfigure}{0.32\textwidth}
        \caption{}
        {\includegraphics[width=\textwidth]{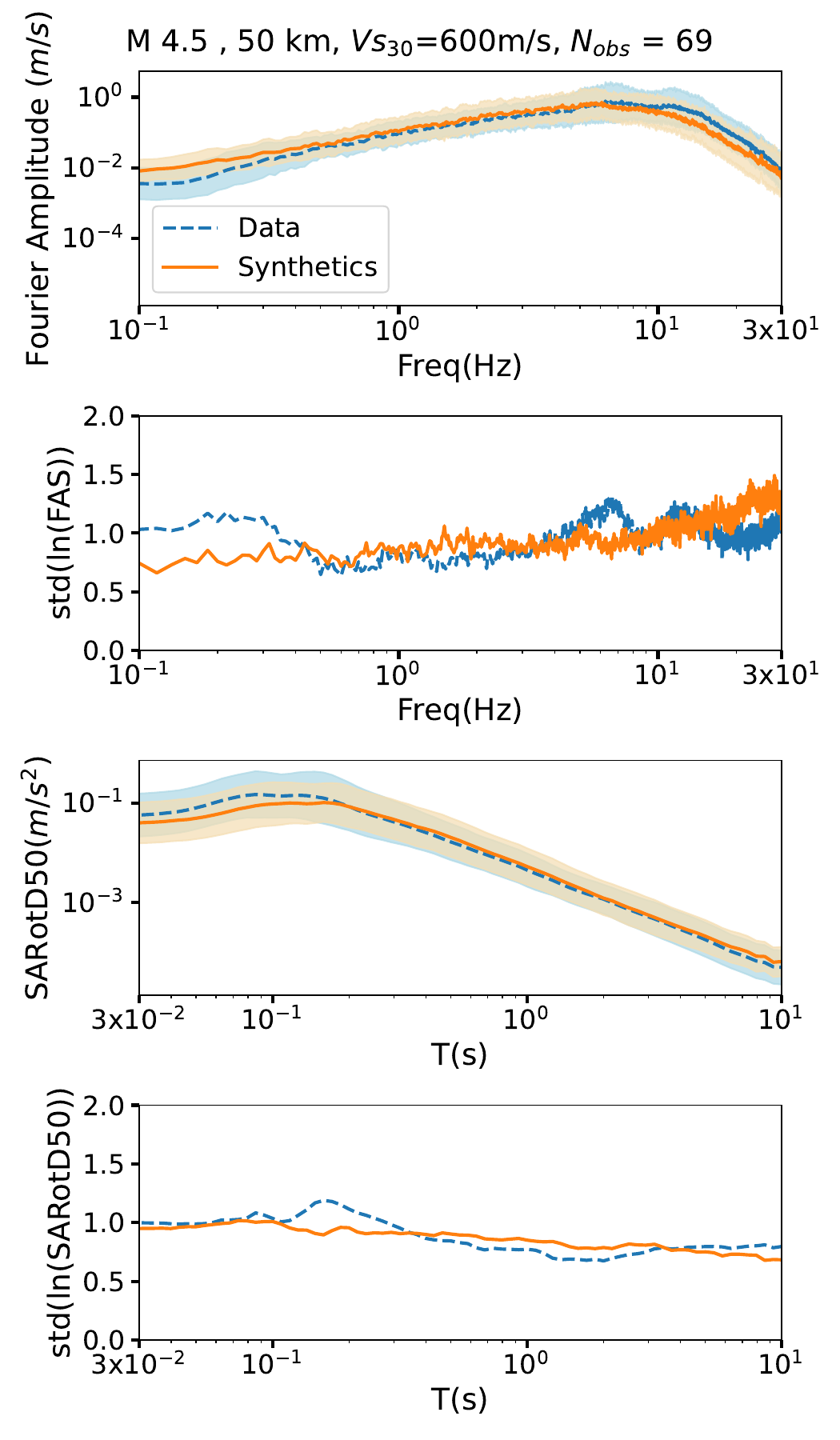}}
    \end{subfigure}
    \quad
    \begin{subfigure}{0.32\textwidth}
        \caption{}
        {\includegraphics[width=\textwidth]{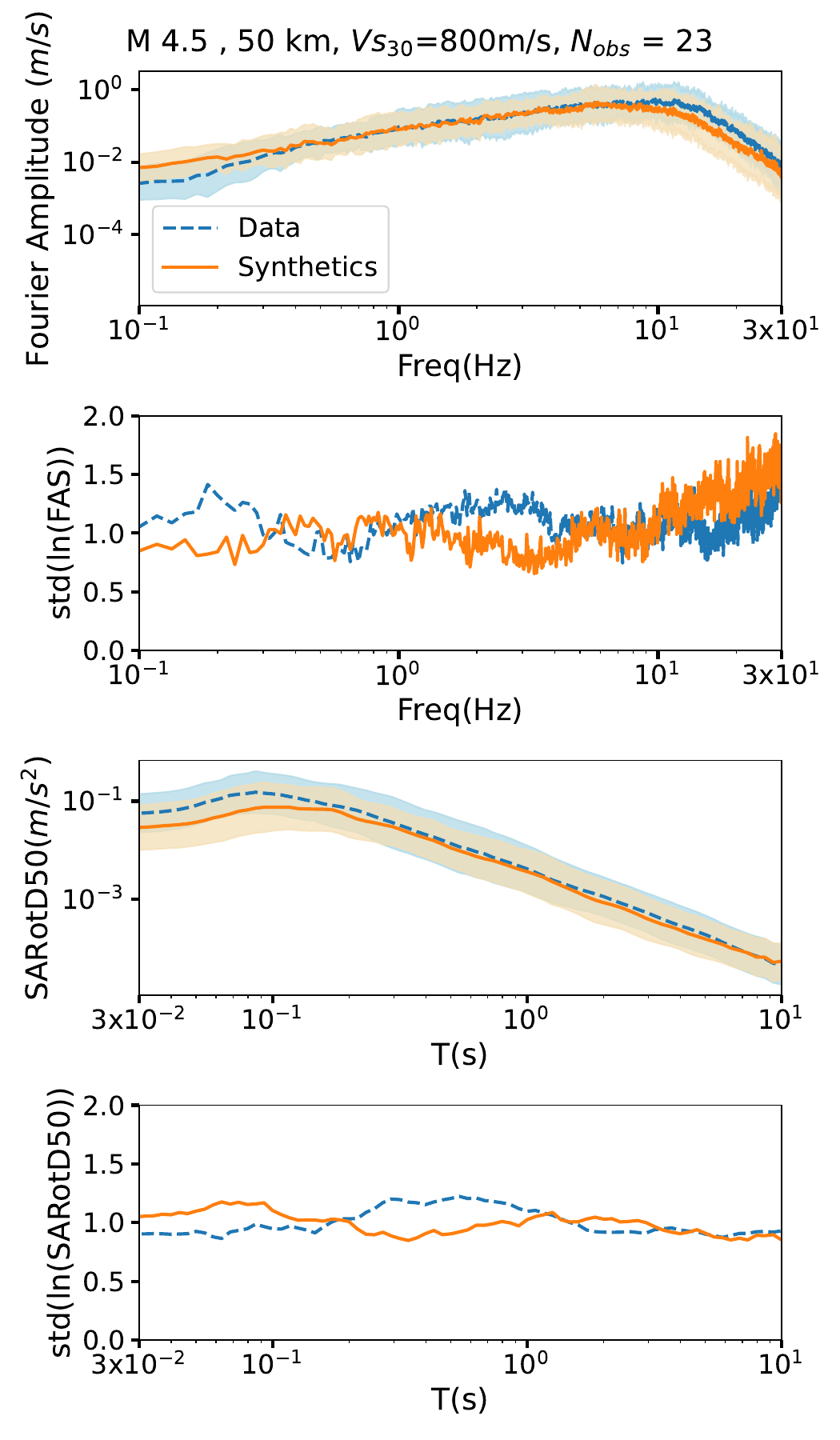}}
    \end{subfigure}
    
    \begin{subfigure}{0.32\textwidth}
        \caption{}
        {\includegraphics[width=\textwidth]{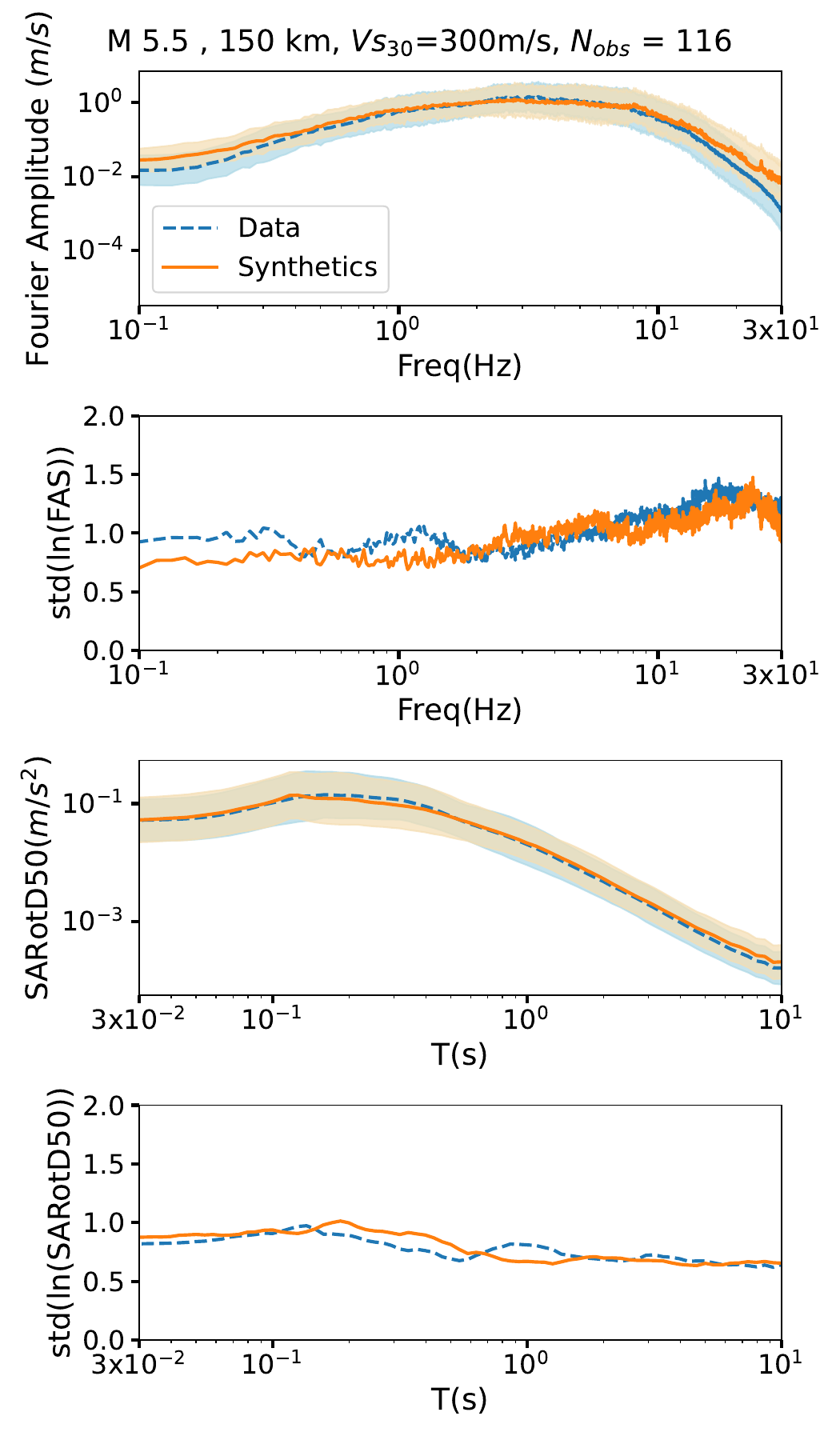}}
    \end{subfigure}
    \quad
    \begin{subfigure}{0.32\textwidth}
        \caption{}
        {\includegraphics[width=\textwidth]{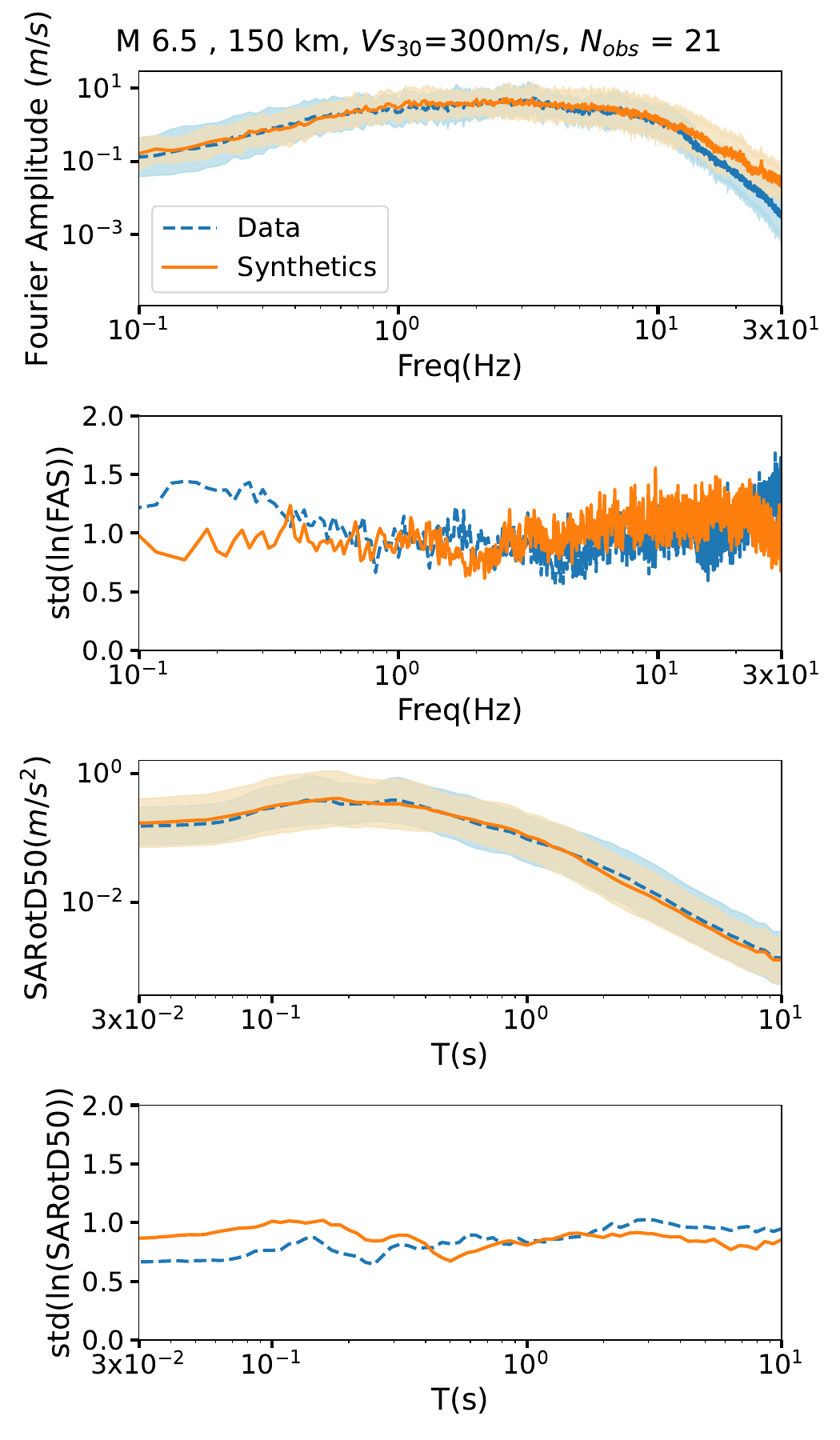}}
    \end{subfigure}
    \quad
    \begin{subfigure}{0.32\textwidth}
        \caption{}
        {\includegraphics[width=\textwidth]{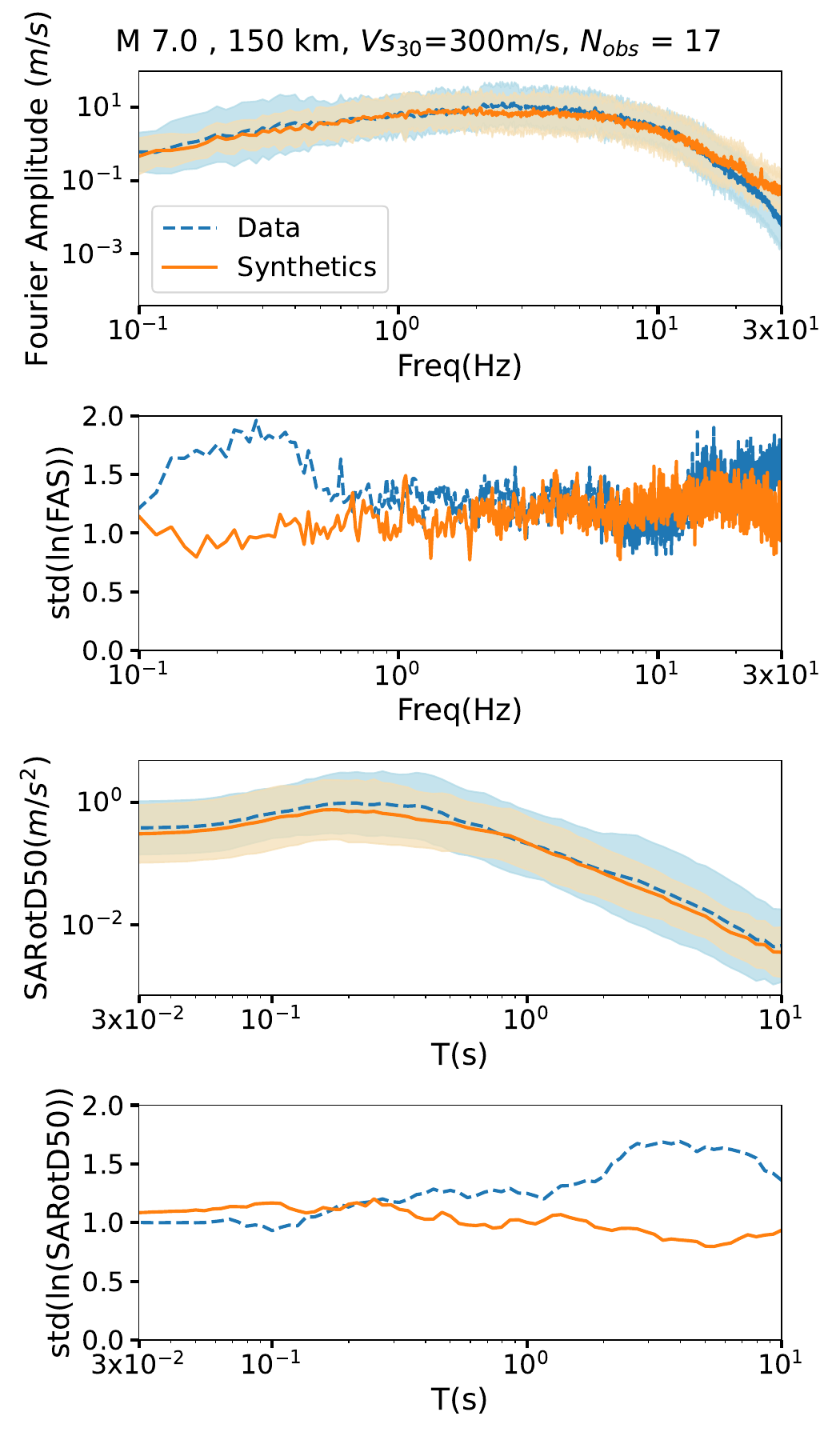}}
    \end{subfigure}
    \caption{Scenario event comparisons for cGM-GANO trained on the kik-net dataset. Each sub-panel depicts the Fourier Amplitude Spectrum (FAS), aleatory standard deviation of FAS, Rotd50 Pseudo-spectral acceleration (PSA) spectrum, and aleatory standard deviation of Rotd50 PSA.
    The title of each plot corresponds to the midpoint of each bin, and $N_{obs}$ is the number of observed records within each bin.
    The lines represent the mean in log space of the horizontal components of the synthetic (solid line) or kik-net (dashed line) data, while the shaded zone represents the $16^{th}$ to $84^{th}$ percentile range.}
    \label{fig:scenario_kik}
\end{figure*}

Fig.~\ref{fig:scenario_kik} compares the frequency domain statistics of the observed and cGM-GANO generated time-series belonging to the same $\mathbf{M}$, $R_{rup}$, $V_{S30}$ bin. Although neural operators map function spaces and by definition have continuous input spaces (i.e., conditional variables), for the validation we binned the data in windows of increment $\mathbf{M}=0.2$ , $R_{rup} = 20$km and $V_{S30}=100$ m/s to ensure that the number of records in each bin would provide statistically significant results for comparison purposes. We successively generated cGM-GANO time-series for each scenario in each bin to create a database of synthetic ground motions for the same scenarios as the observations in each bin. The statistics of the two datasets are compared for representative bins in Fig.~\ref{fig:scenario_kik}. For the examples shown, cGM-GANO recovers the mean and aleatory variability of the FAS in the frequency range between $\sim 0.5$ Hz and 30 Hz. Similarly, cGM-GANO recovers the mean and aleatory variability of PSA RotD50 in most cases; exceptions such as scenario (e) for periods shorter than $T=0.2$sec, or scenario (f) for periods longer than $T=1$sec which may stem from the small number of observations available in those bins (21 and 17 observations, respectively) and will be further investigated by the authors. 

\begin{figure*}
    \centering
    \begin{subfigure}{0.35\textwidth}
        \caption{}
        {\includegraphics[width=\textwidth]{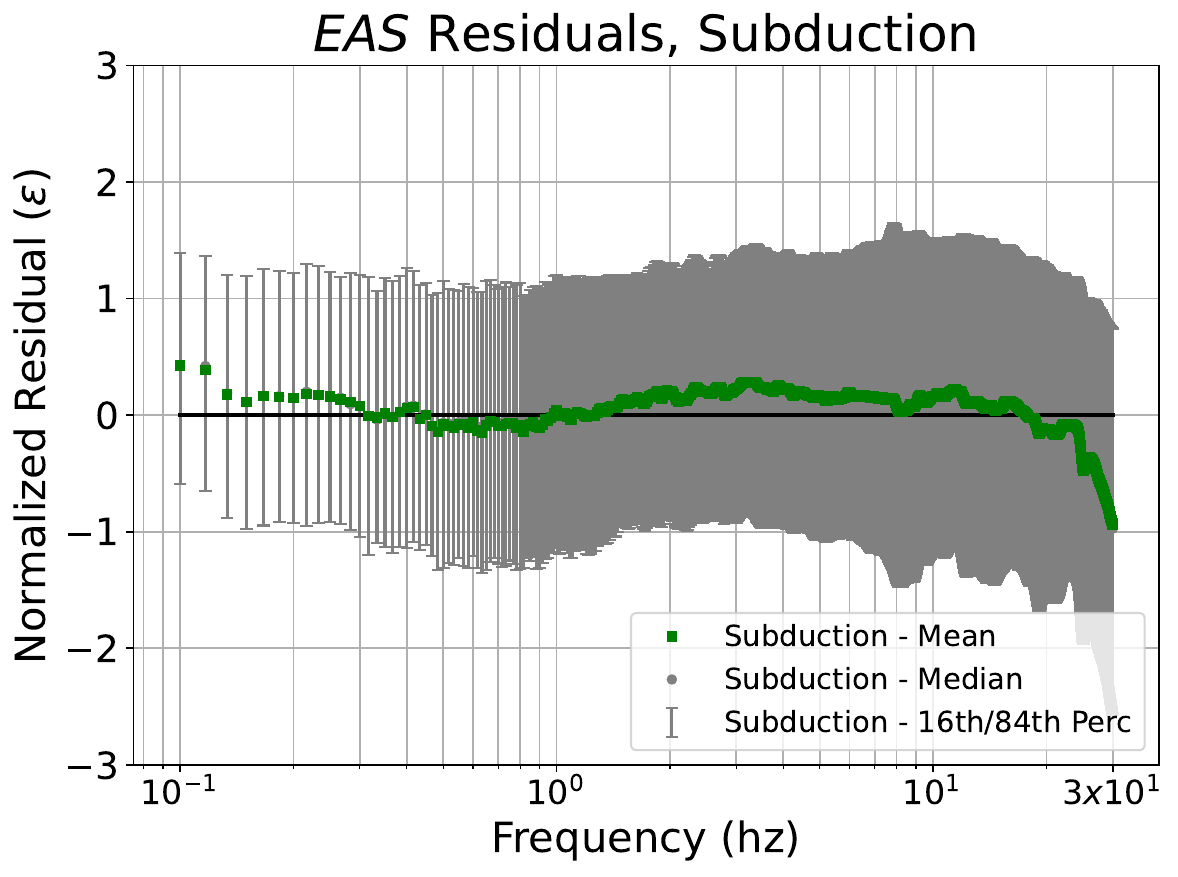}}
    \end{subfigure}
    \quad
    \begin{subfigure}{0.35\textwidth}
        \caption{}
        {\includegraphics[width=\textwidth]{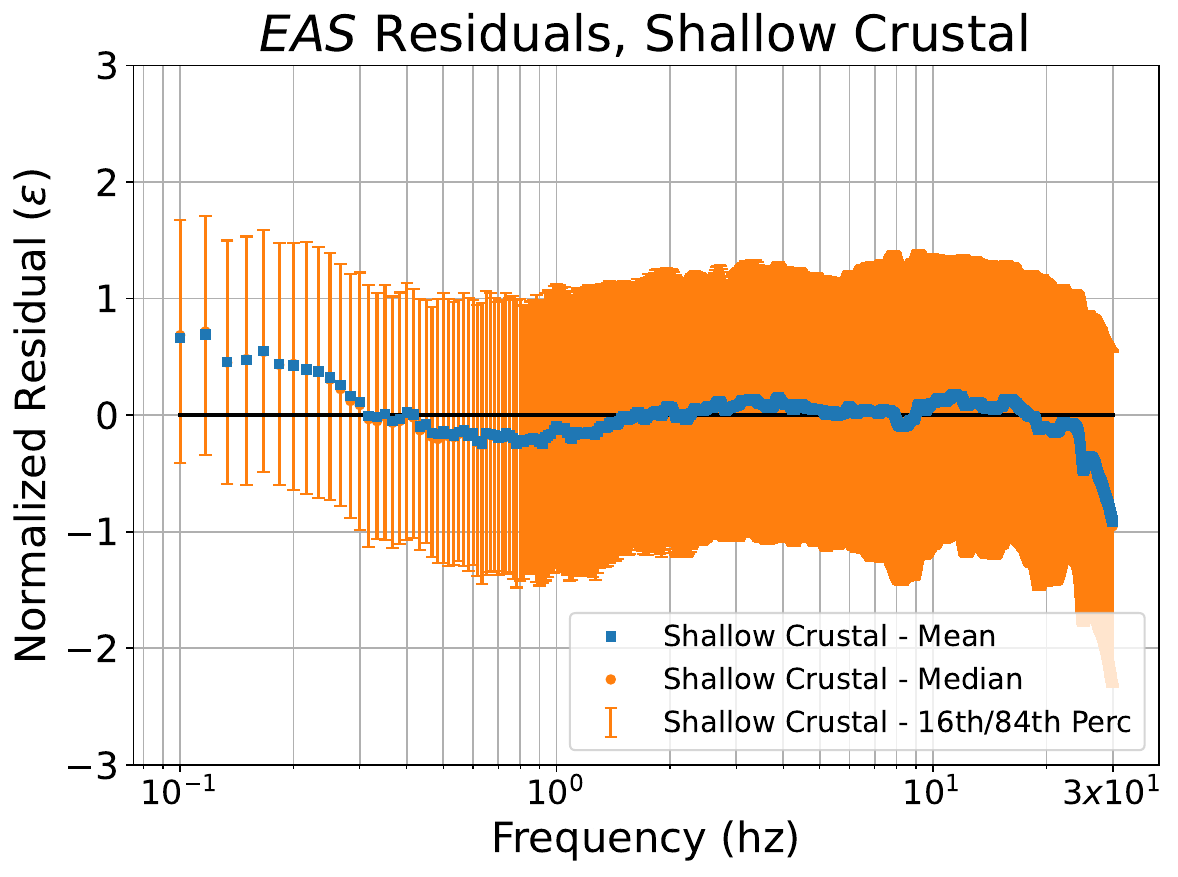}}
    \end{subfigure}  
    
    \begin{subfigure}{0.35\textwidth}
    \caption{}
        {\includegraphics[width=\textwidth]{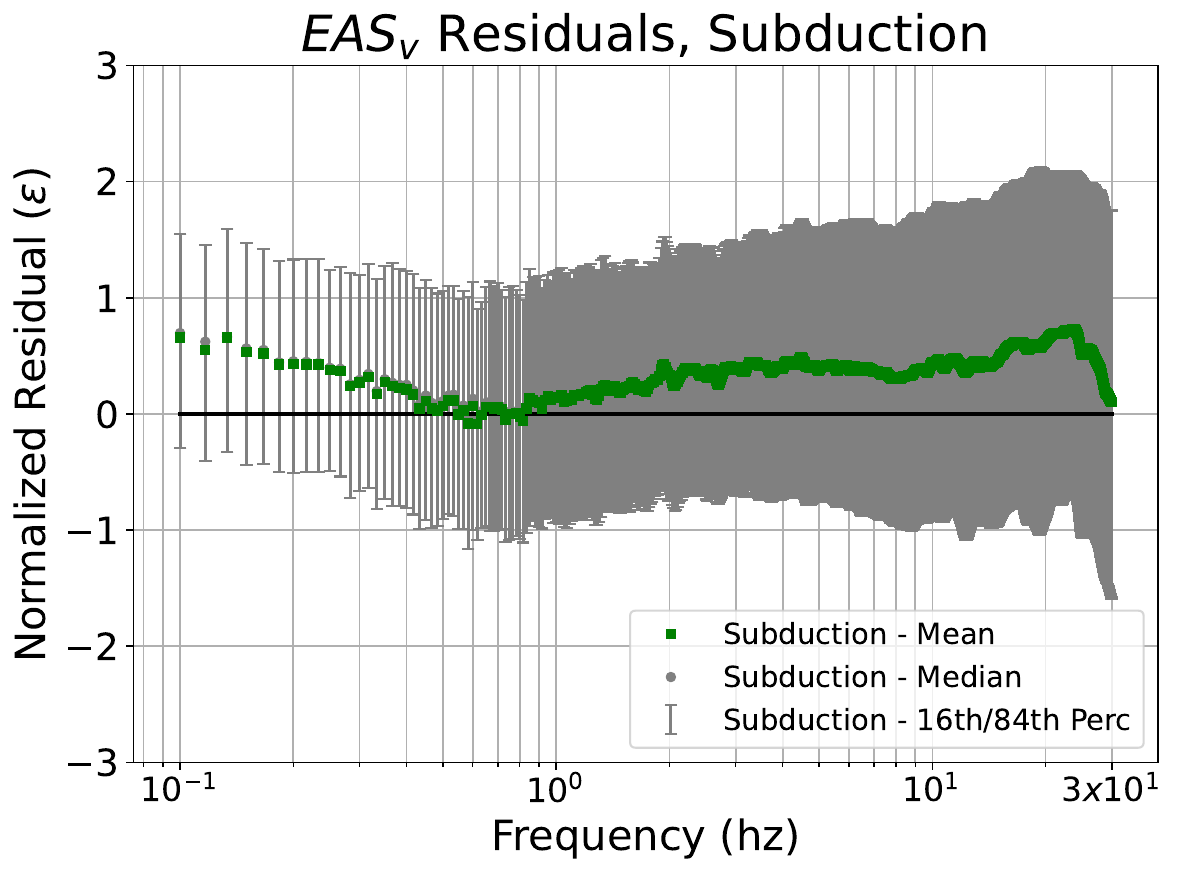}}
    \end{subfigure}
    \quad
    \begin{subfigure}{0.35\textwidth}
        \caption{}
        {\includegraphics[width=\textwidth]{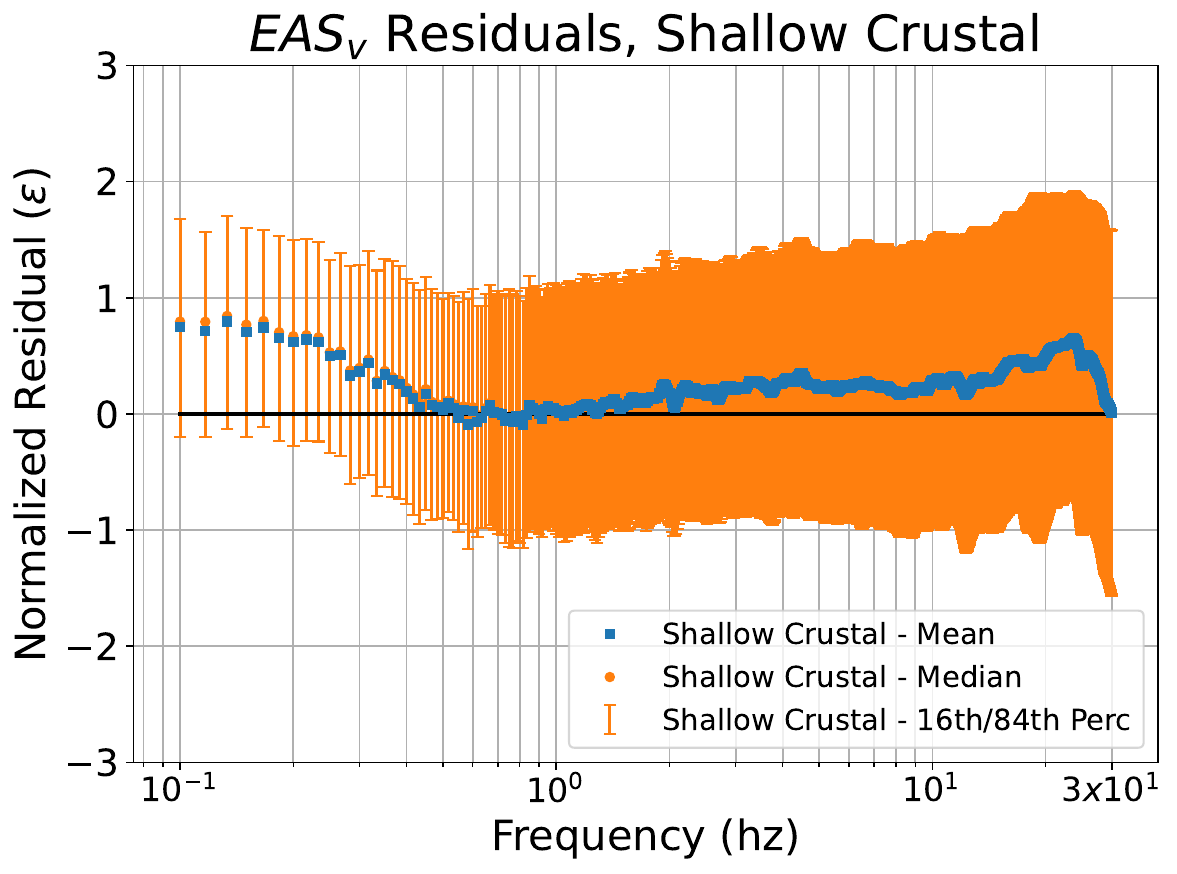}}
    \end{subfigure}  
    
    \medskip
    \begin{subfigure}{0.35\textwidth}
        \caption{}
        {\includegraphics[width=\textwidth]{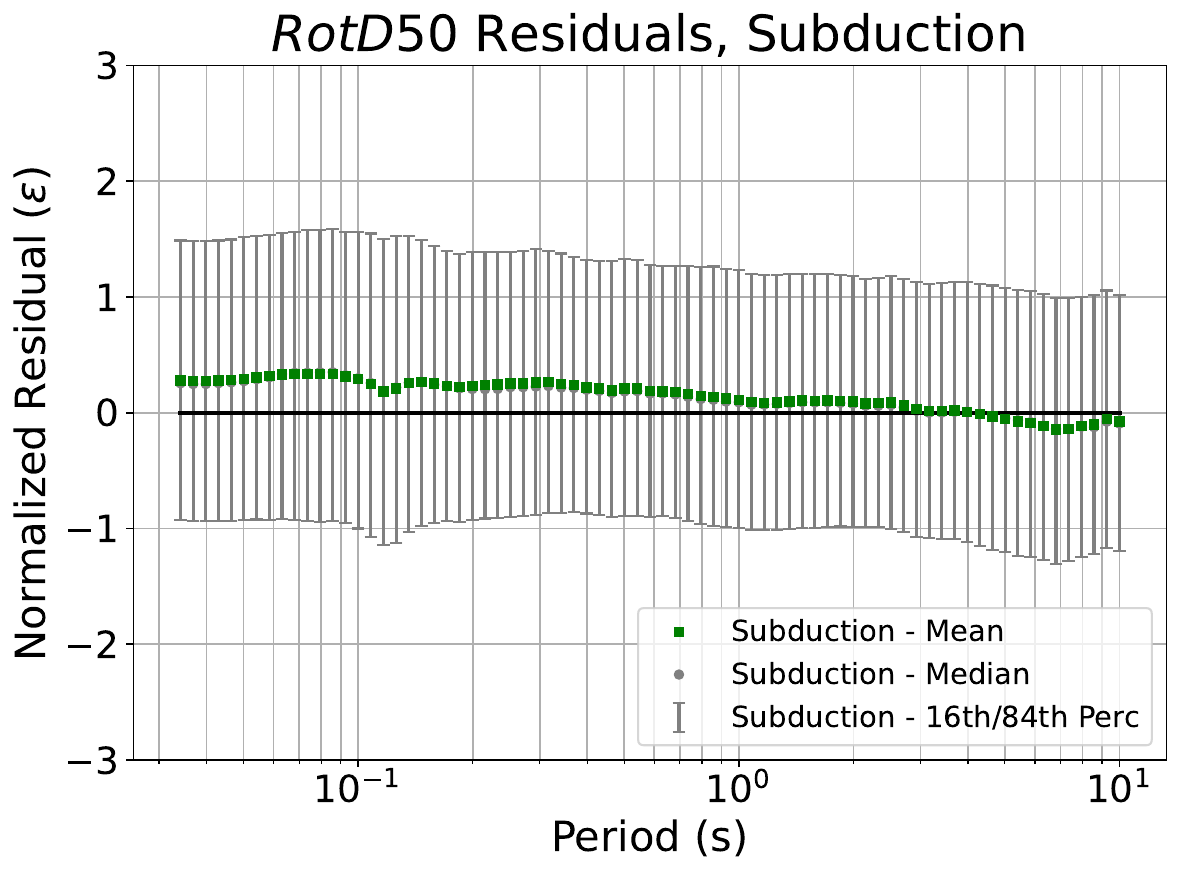}}
    \end{subfigure}
    \quad
    \begin{subfigure}{0.35\textwidth}
        \caption{}
        {\includegraphics[width=\textwidth]{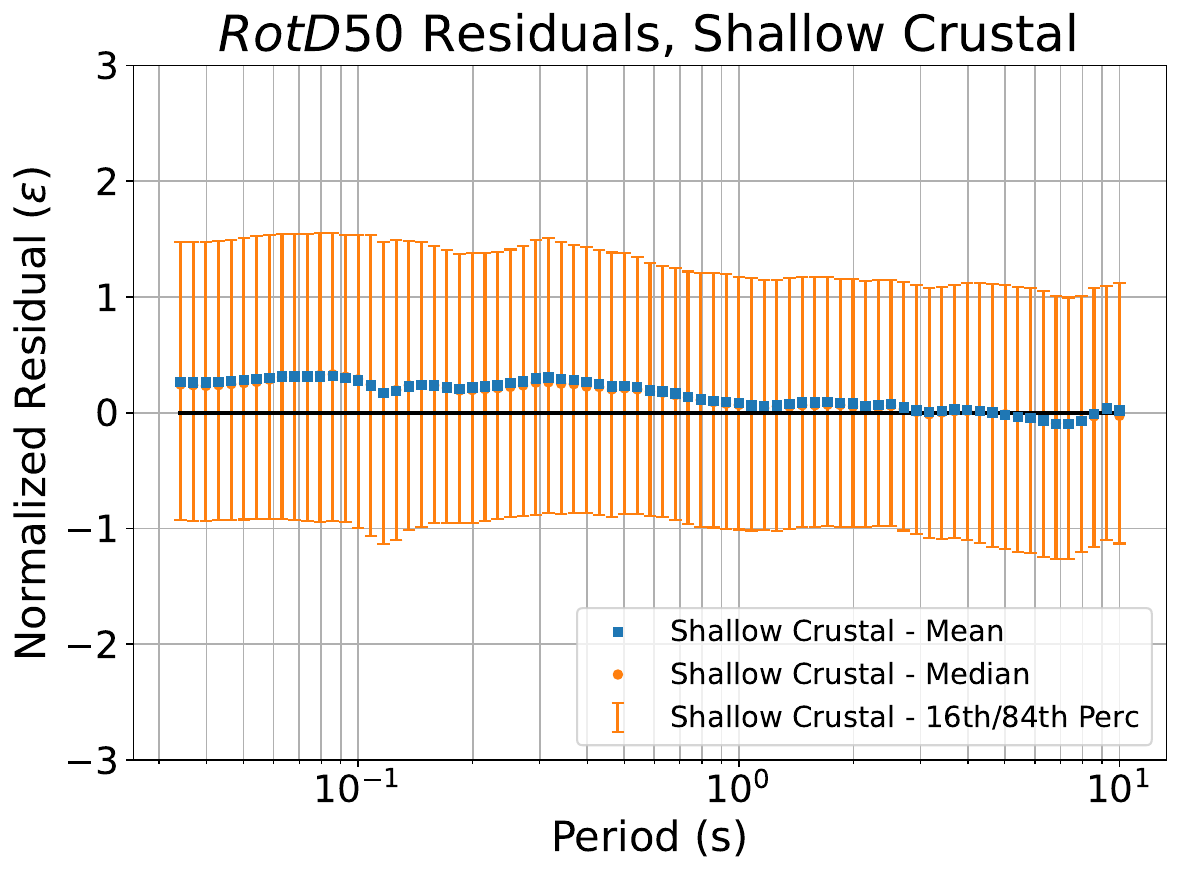}}
    \end{subfigure}  
    \begin{subfigure}{0.35\textwidth}
        \caption{}
        {\includegraphics[width=\textwidth]{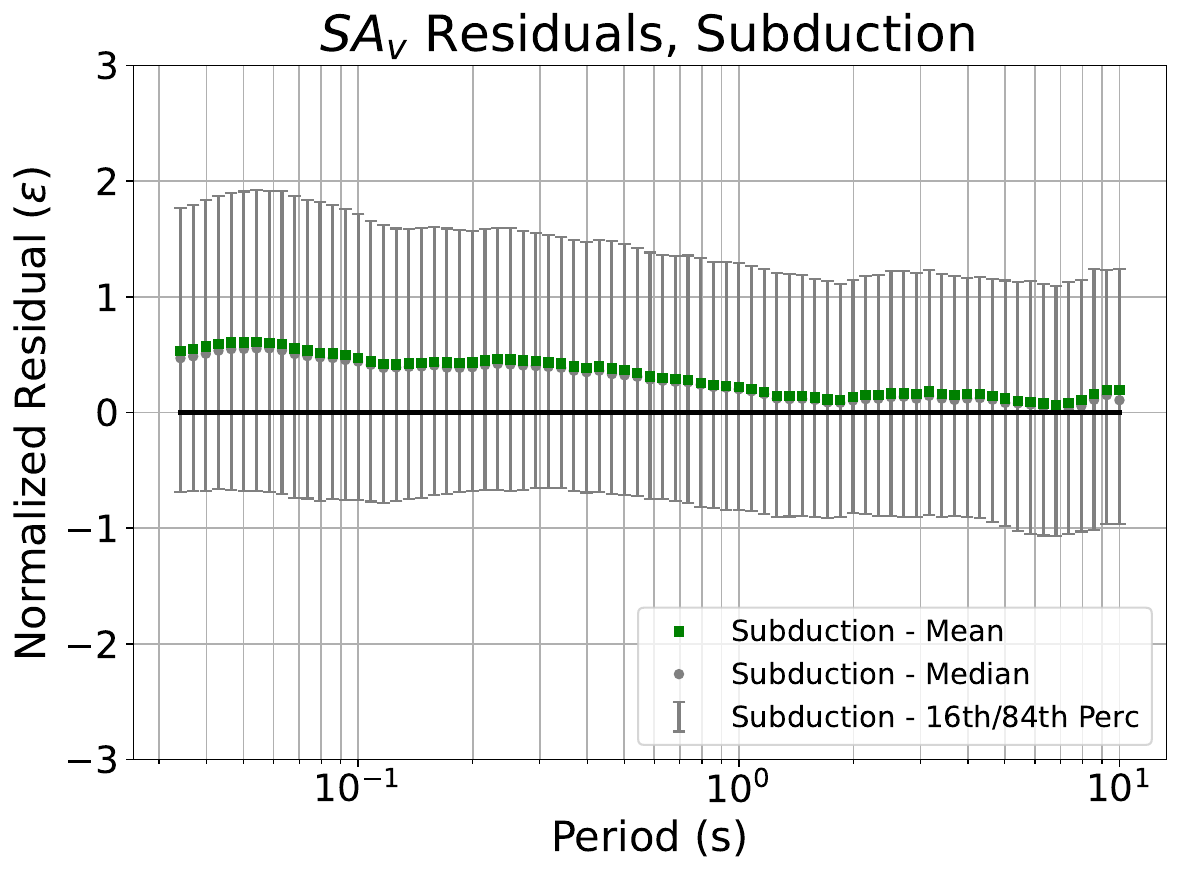}}
    \end{subfigure}
    \quad
    \begin{subfigure}{0.35\textwidth}
        \caption{}
        {\includegraphics[width=\textwidth]{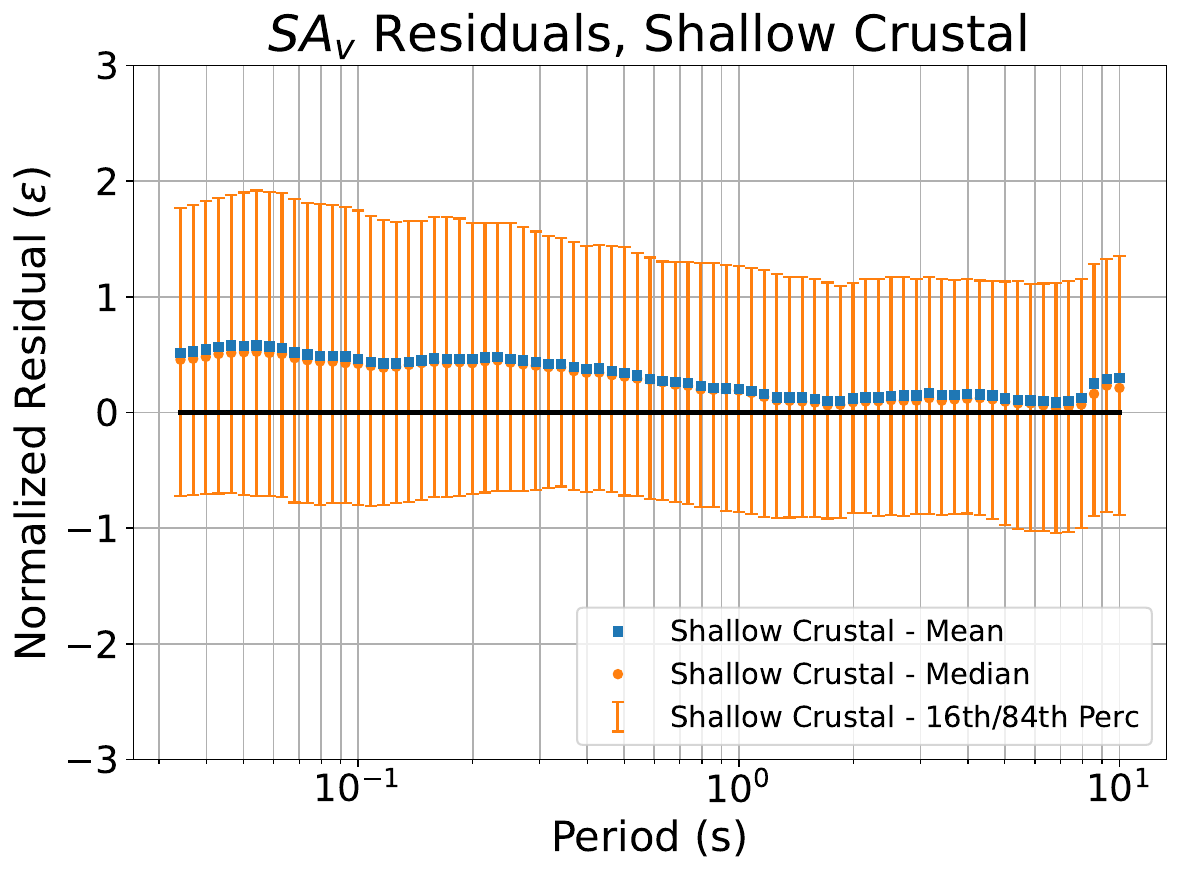}}
    \end{subfigure} 
    \caption{Residual plots of the kik-net trained model. 
    Subfigures (a) and (b) show the horizontal EAS residuals for subduction and shallow crustal events, respectively.
    Subfigures (c) and (d) show the vertical EAS residuals for subduction and shallow crustal events, respectively.
    Subfigures (e) and (f) show the horizontal RotD50 PSA residuals for subduction and shallow crustal events, respectively.
    Subfigures (g) and (h) show the vertical PSA residuals for subduction and shallow crustal events, respectively.}
    \label{fig:residual_plots_fas_sa}
\end{figure*}

We successively examined the extent to which cGM-GANO can recover the $\mathbf{M}$, $R_{rup}$, $V_{S30}$ scaling of the ground motion observations for a given tectonic environment by means of residual analysis (for more information, see Section on Residual Analysis above). Results for the horizontal and vertical components of the EAS and RotD50 residuals are shown in Fig \ref{fig:residual_plots_fas_sa}, separately for the subduction and the shallow crustal events. 
The cGM-GANO results of the horizontal EAS components for both tectonic environments are unbiased in the frequency range $\sim 0.3-20$Hz.
The vertical components show similar behavior with a small increasing bias starting at $0.5$Hz.
The horizontal spectral accelerations are well captured in the $\sim 1-10$sec period range and show small, nearly constant positive bias for shorter periods $T \le 1$sec. The vertical spectral accelerations for both tectonic environments are well captured in the $T \ge 2$sec period range, while for shorter periods, cGM-GANO shows significant bias. While encouraging, these results leave open questions about the frequency range where cGM-GANO can reliably recover the mean and aleatory variability of recorded ground motions and the factors that affect the reliability of the GANO model, which are further investigated by the authors.

\begin{figure*}
\centering
    \begin{subfigure}{0.25\textwidth}
        \caption{}
        {\includegraphics[height=\textwidth]{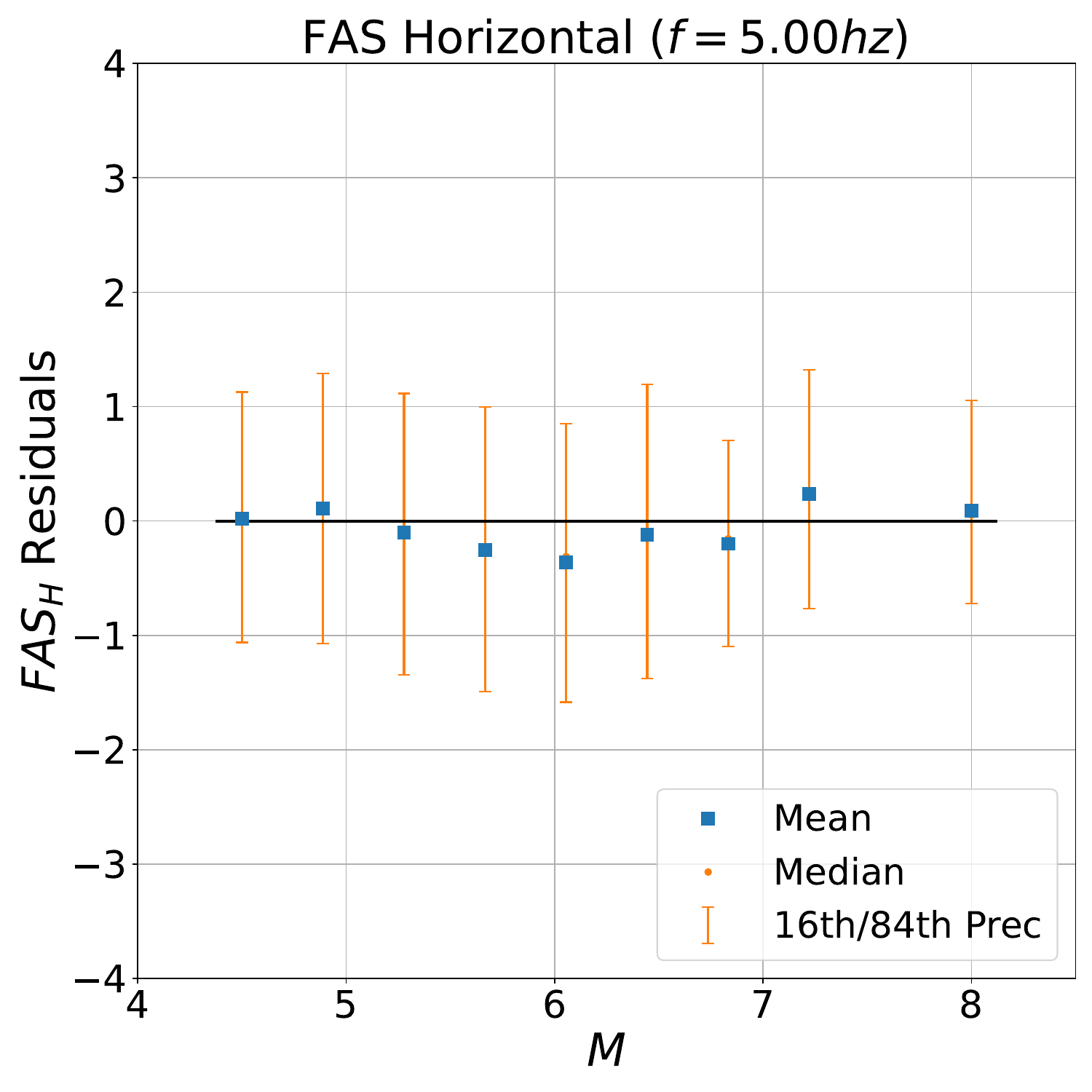}}
    \end{subfigure}
    \quad
    \begin{subfigure}{0.25\textwidth}
        \caption{}
        {\includegraphics[height=\textwidth]{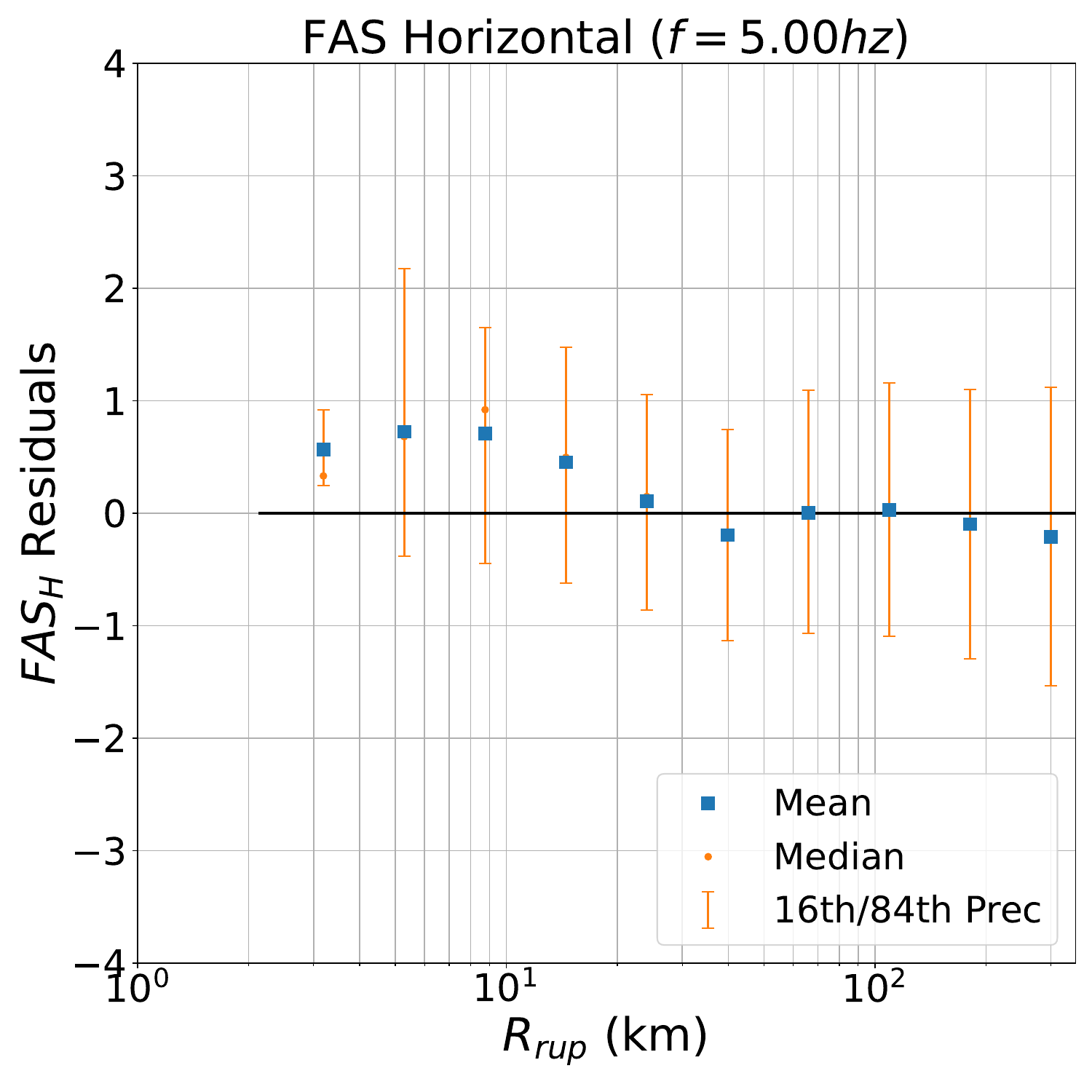}}
    \end{subfigure}
    \quad
    \begin{subfigure}{0.25\textwidth}
        \caption{}
        {\includegraphics[height=\textwidth]{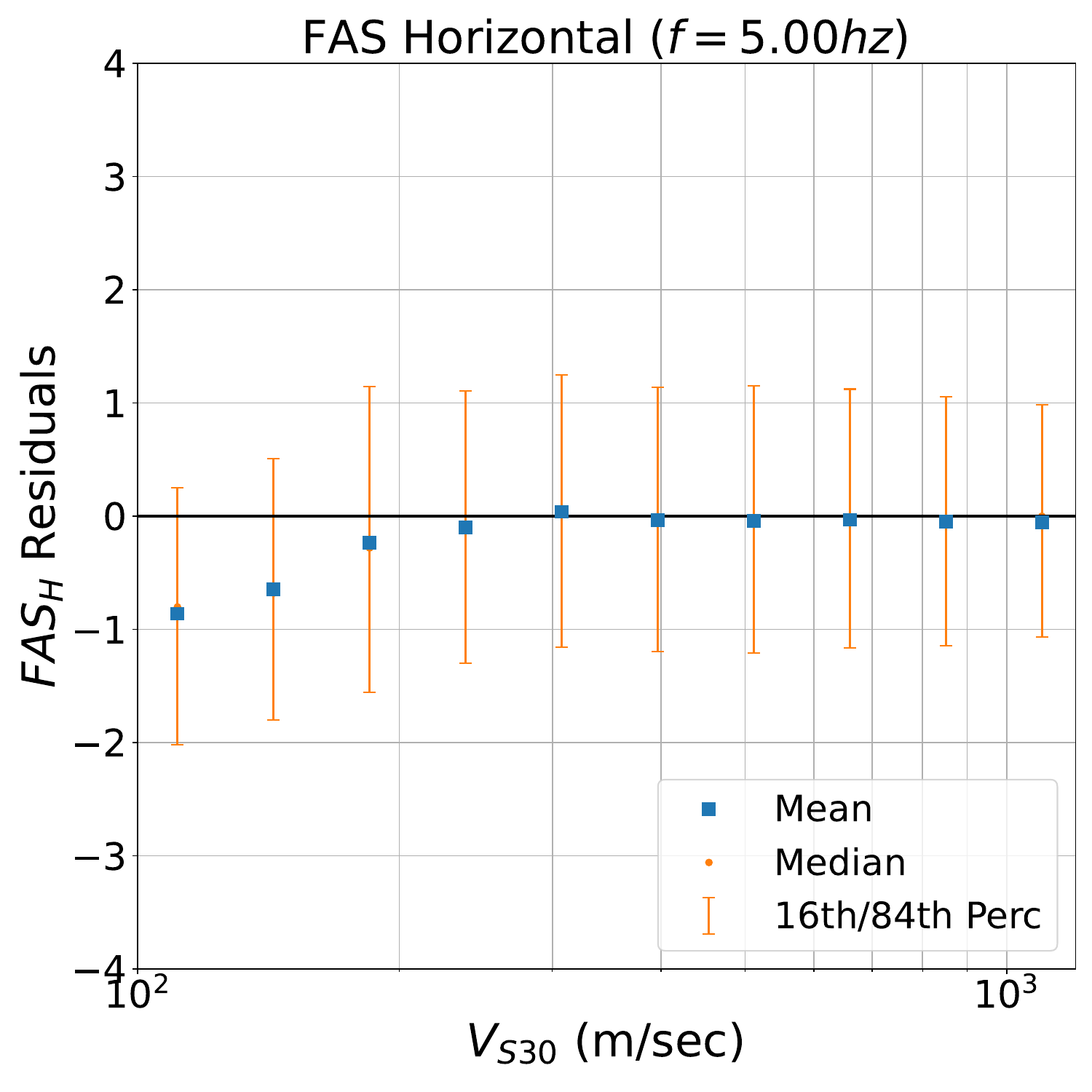}}
    \end{subfigure}
    \medskip

    \begin{subfigure}{0.25\textwidth}
        \caption{}
        {\includegraphics[height=\textwidth]{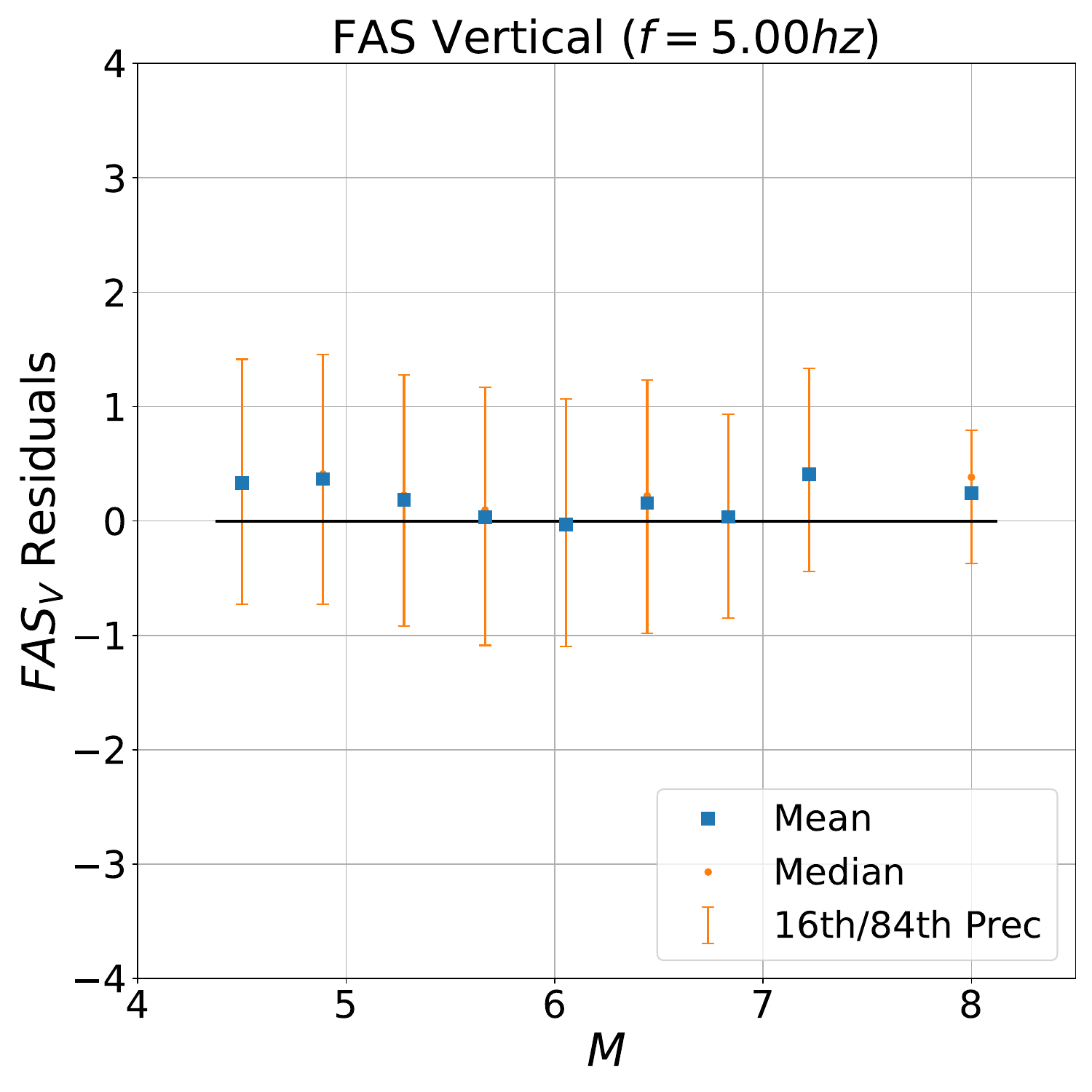}}
    \end{subfigure}
    \quad
    \begin{subfigure}{0.25\textwidth}
        \caption{}
        {\includegraphics[height=\textwidth]{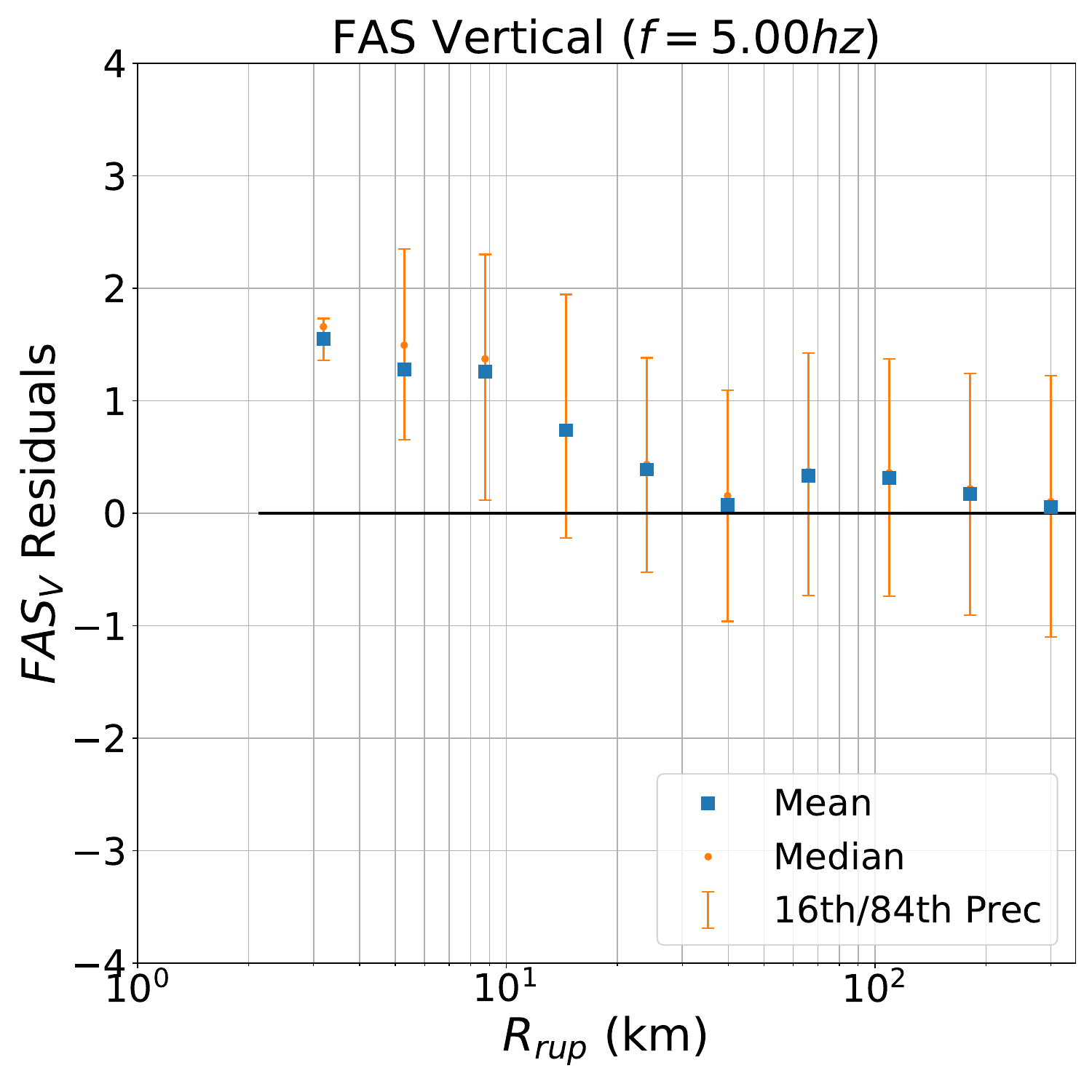}}
    \end{subfigure}
    \quad
    \begin{subfigure}{0.25\textwidth}
        \caption{}
        {\includegraphics[height=\textwidth]{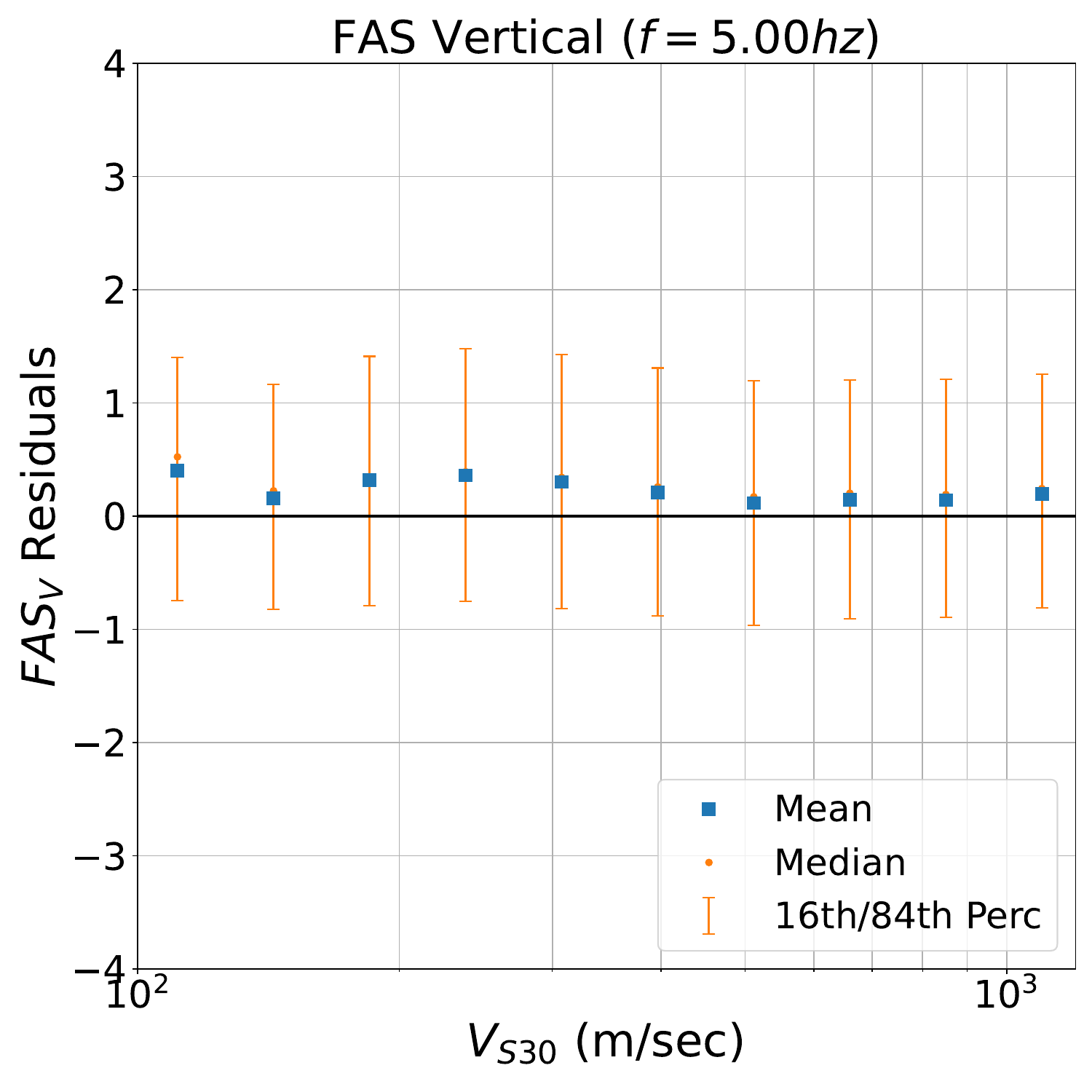}}
    \end{subfigure}
    \caption{Comparison of horizontal and vertical FAS residuals for f=5.0 Hz against magnitude (subfigure (a) and (d), $R_{rup}$ (subfigure (b) and (e)), and $V_{S30}$ (subfigure (c) and (f))}
    \label{fig:scaling_variable_5hz}
\end{figure*}

\begin{figure*}
\centering
    \begin{subfigure}{0.25\textwidth}
        \caption{}
        {\includegraphics[height=\textwidth]{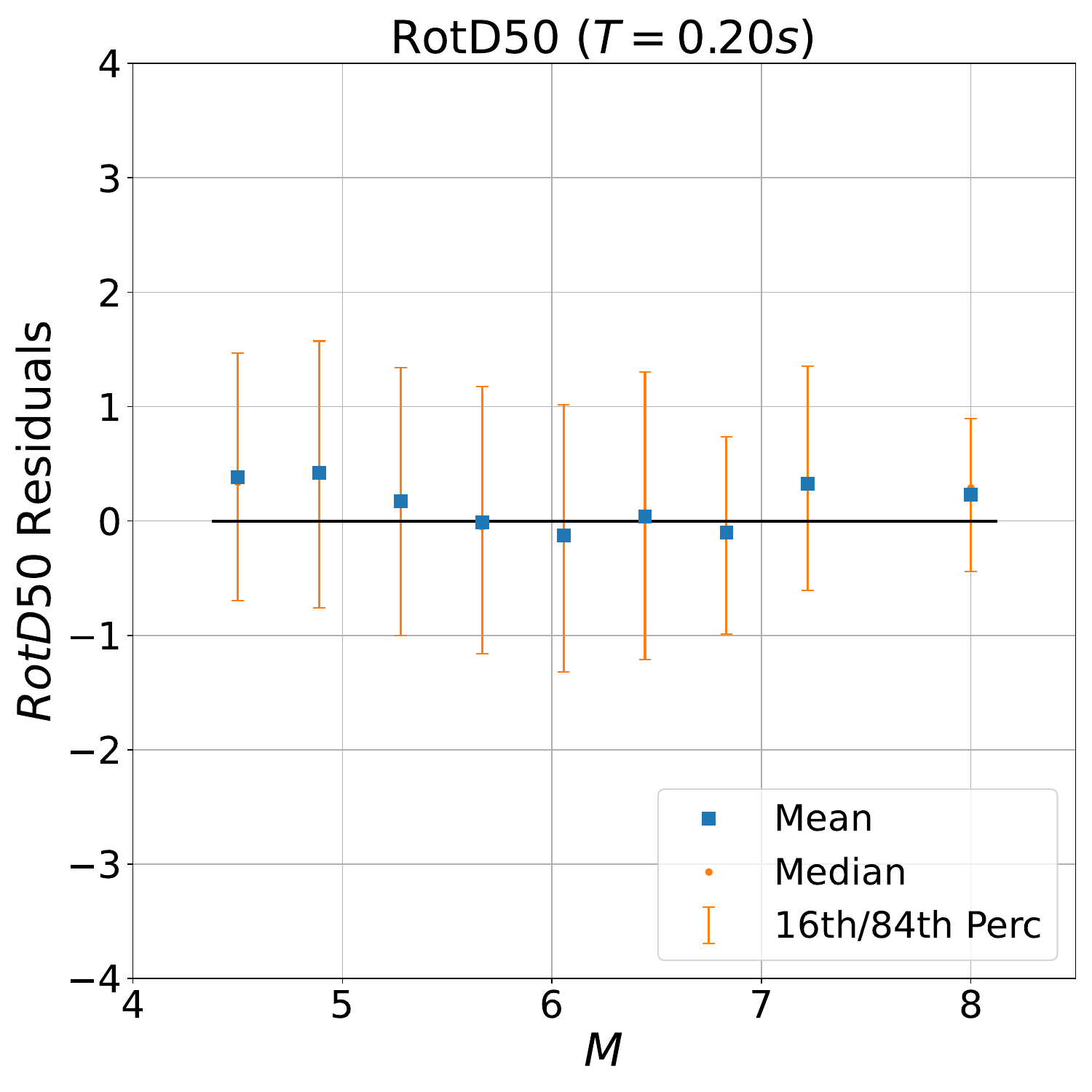}}
    \end{subfigure}
    \quad
    \begin{subfigure}{0.25\textwidth}
        \caption{}
        {\includegraphics[height=\textwidth]{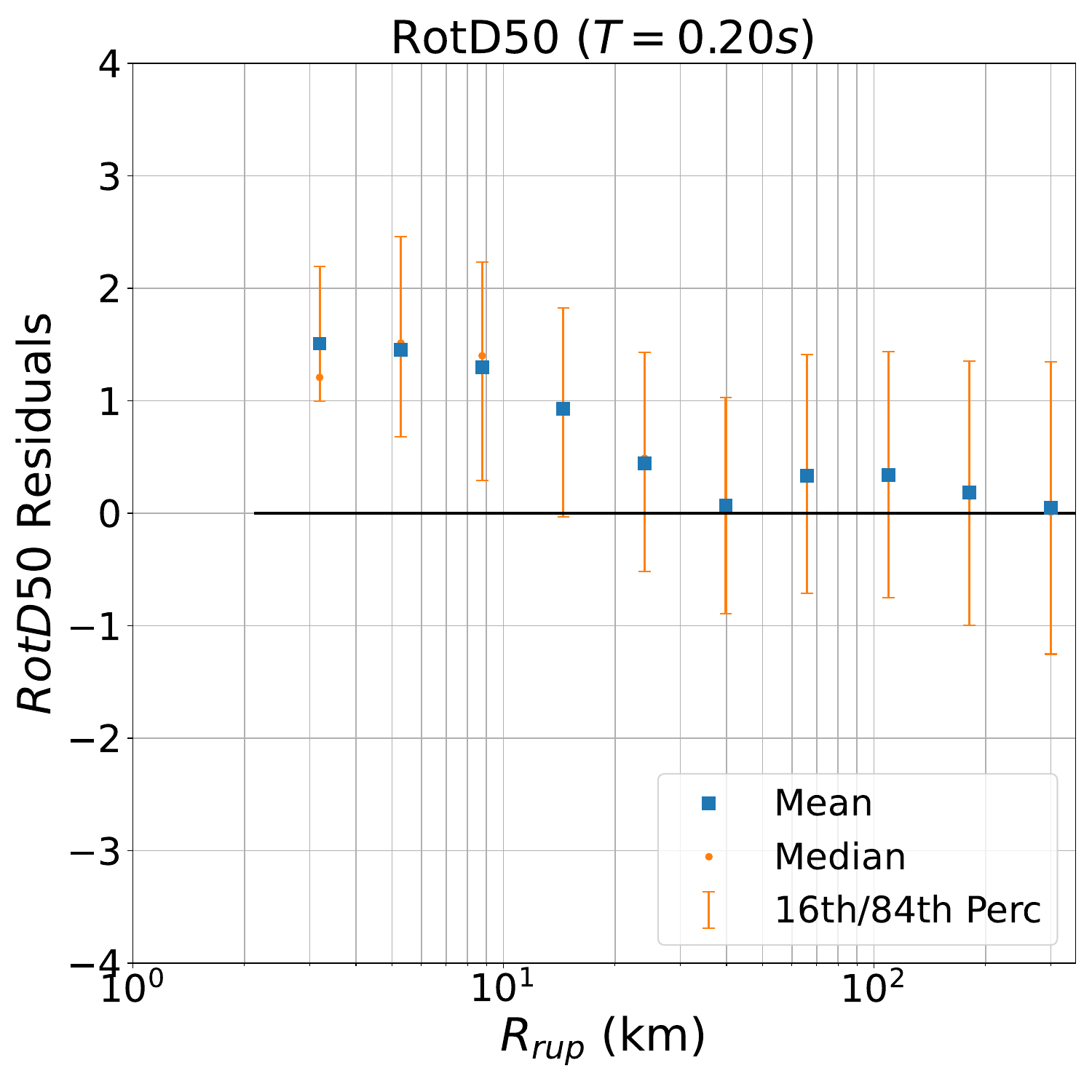}}
    \end{subfigure}
    \quad
    \begin{subfigure}{0.25\textwidth}
        \caption{}
        {\includegraphics[height=\textwidth]{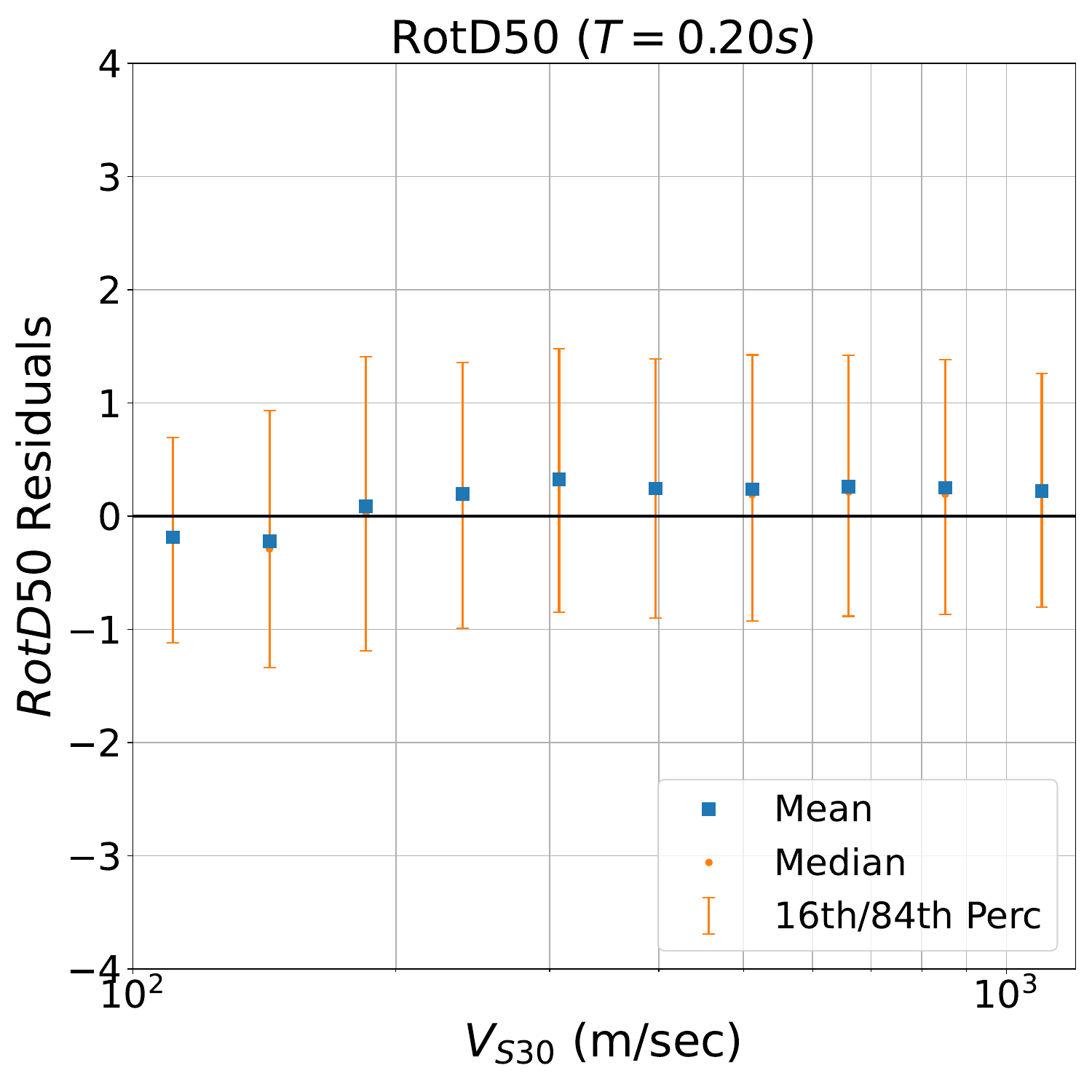}}
    \end{subfigure}
    \medskip

    \begin{subfigure}{0.25\textwidth}
        \caption{}
        {\includegraphics[height=\textwidth]{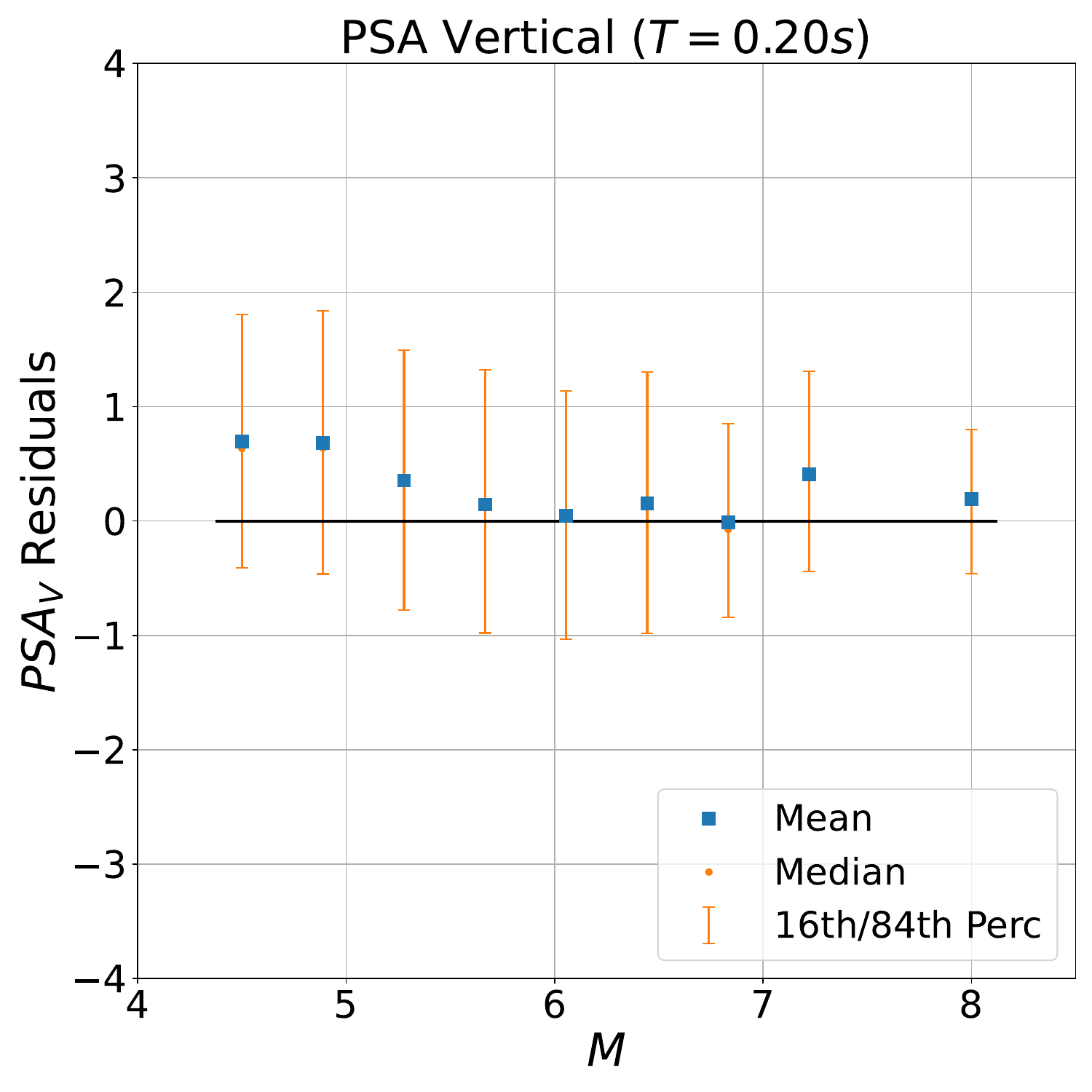}}
    \end{subfigure}
    \quad
    \begin{subfigure}{0.25\textwidth}
        \caption{}
        {\includegraphics[height=\textwidth]{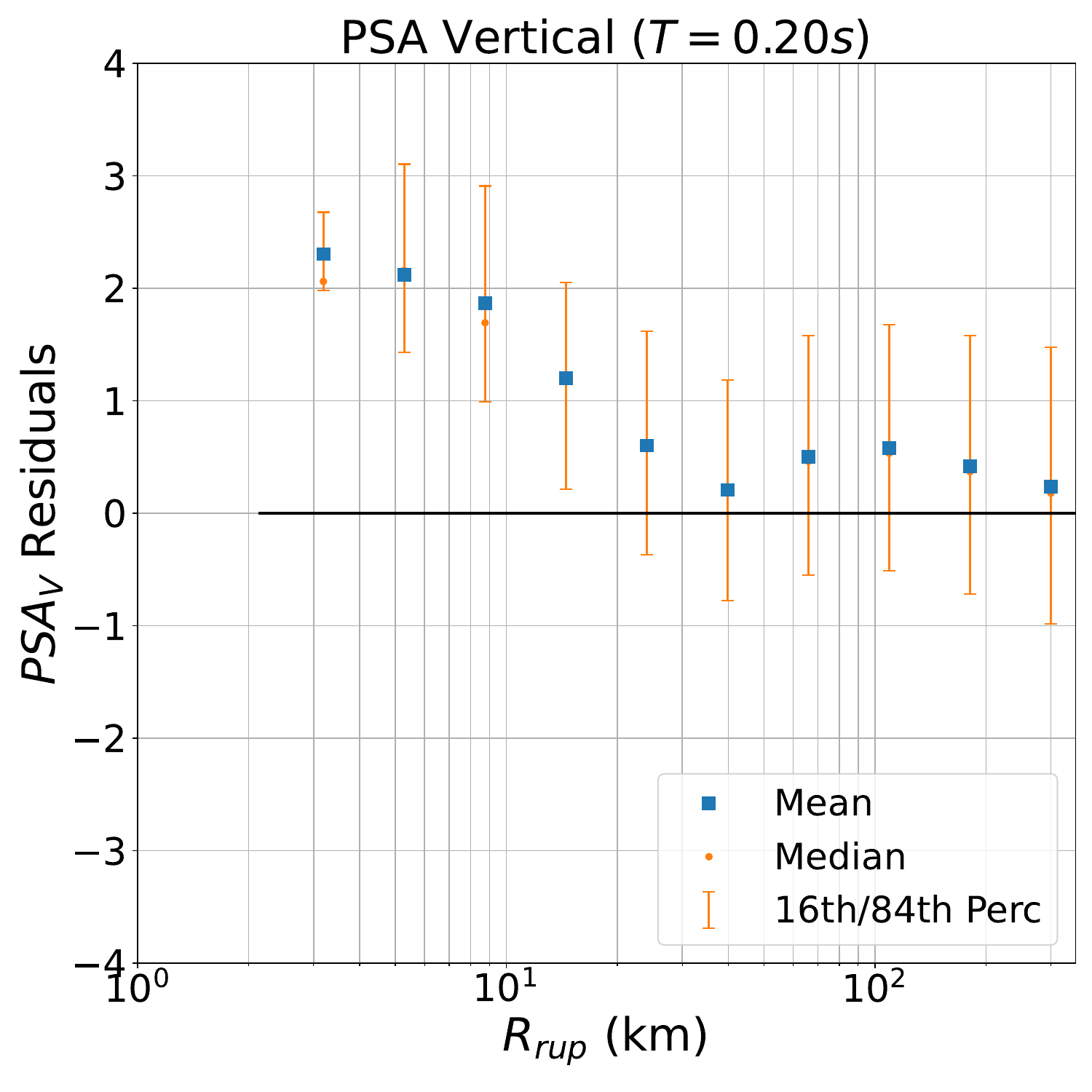}}
    \end{subfigure}
    \quad
    \begin{subfigure}{0.25\textwidth}
        \caption{}
        {\includegraphics[height=\textwidth]{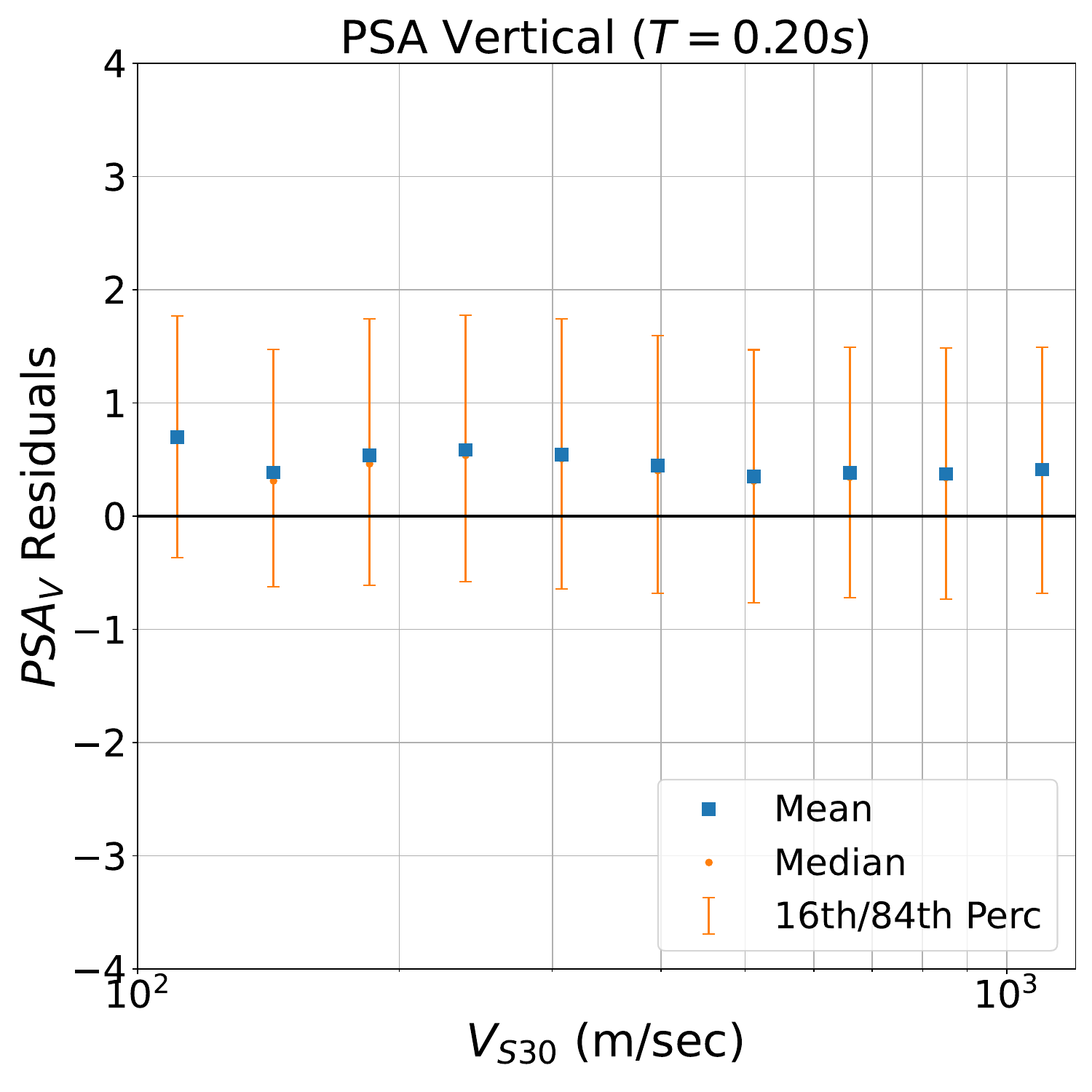}}
    \end{subfigure}    
    \caption{Comparison of horizontal and vertical PSA residuals for T=0.2 s (f=5.0 Hz) against magnitude (subfigure (a) and (d), $R_{rup}$ (subfigure (b) and (e)), and $V_{S30}$ (subfigure (c) and (f))}
    \label{fig:scaling_variable_5hz_sa}
\end{figure*}

The extent to which cGM-GANO recovers the ground motion scaling versus $\mathbf{M}$, $R_{rup}$, $V_{S30}$ is presented in Figs \ref{fig:scaling_variable_5hz} and \ref{fig:scaling_variable_5hz_sa}. 
The horizontal and vertical FAS residuals in subfigures \ref{fig:scaling_variable_5hz} (a) and (c), and horizotnal and vertical PSA residuals in subfigures \ref{fig:scaling_variable_5hz_sa} (a) and (c) for $f= 5Hz$ ($T=0.2 s$)  suggest that cGM-GANO can recover the median magnitude scaling of the ground motions.
By contrast, cGM-GANO shows systematic positive bias when called to evaluate the median scaling against rupture distance, for $R_{rup} \le 20$ km (see Figs \ref{fig:scaling_variable_5hz} and \ref{fig:scaling_variable_5hz_sa} (b) and (e)). Our working hypothesis is that the bias reflects the very small number of near-field observations (0.4\% of the entire training dataset recorded at distance less than 20km), which may not be enough for the model to learn the ground motion characteristics in this regime. The cGM-GANO generated ground motion scaling with $V_{s30}$ recovered in Figs \ref{fig:scaling_variable_5hz} and \ref{fig:scaling_variable_5hz_sa} (c) and (f) shows negative bias for both FAS and RotD50 versus $V_{S30}$ for sites with $V_{S30} \le 200$ m/s. The bias in this $V_{S30}$ range could be associated with the limited number of ground motion observations for soft soil sites within the KiK-net strong motion network (2.7\% of the entire training dataset was recorded at sites with $V_{S30} \le 200$ m/s.) Although this is not a negligible dataset of training ground motions, we hypothesize that training cGM-GANO to generate strong motions at soft sites would be associated with learning ground motion characteristics controlled by nonlinear effects, which are likely less pronounced for stiffer sites and thus less likely to be learned from the remainder of the dataset. This hypothesis is supported by the fact that the vertical component scaling with $V_{S30}$ shows a small, nearly constant positive bias across the same $V_{S30}$ range. The capability of cGM-GANO to learn the effects of nonlinear site response, given the appropriate size of training data and conditional variables, is the focus of a separate study undertaken by the authors. 
Considered strategies for this objective include augmenting the dataset with synthetic datasets that include non-linear effects and employing a weighting sampling, increasing the frequency of soft sites being sampled during training.

\begin{figure*}
    \centering
    \begin{subfigure}{.35\textwidth}
        \caption{}
        \includegraphics[width=\textwidth]{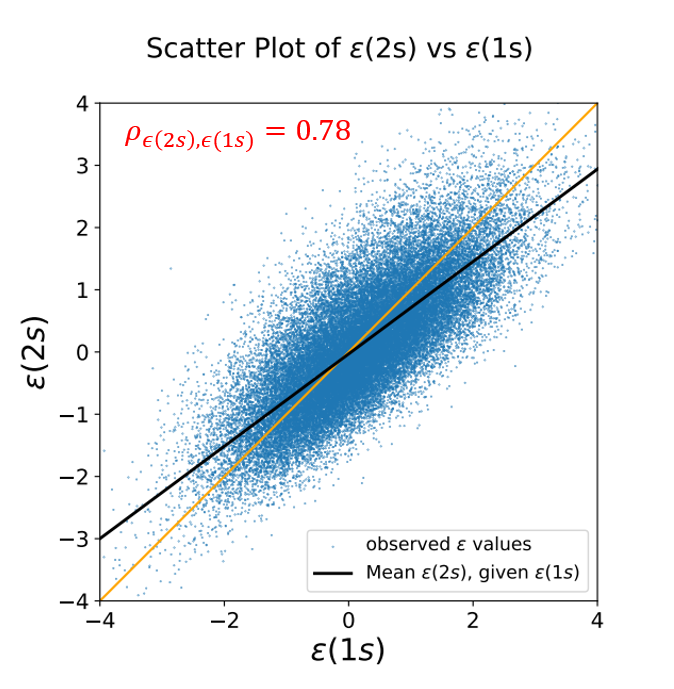}
    \end{subfigure}
    \quad
    \begin{subfigure}{.35\textwidth}
        \caption{}
        \includegraphics[width=\textwidth]{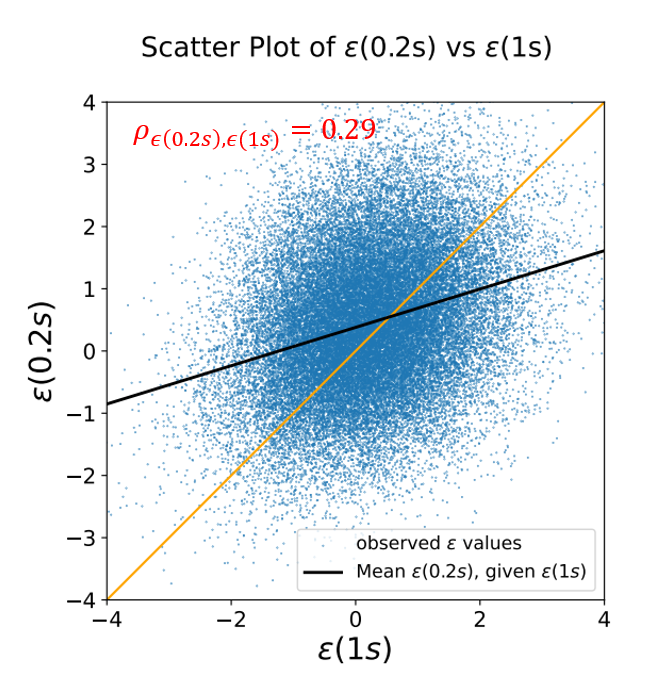}
    \end{subfigure}
    \caption{Inter-period correlation of horizontal PSA residuals. 
    Between T = 1 s and T = 2 s in subfigure (a), and between T = 1 s and T = 0.2 s in subfigure (b). }
    \label{fig:inter-frequency correlation}
\end{figure*}

We also evaluated the inter-frequency correlation of the cGM-GANO-generated ground motion time series. Fig \ref{fig:inter-frequency correlation} shows the correlation of the  RotD50 residuals between T=1s and T=0.2s or T=2.0s for the entire KiK-net dataset.
The Pearson correlation coefficient of residuals ($\rho$) at the period pairs $T_1$ and $T_2$ is also calculated as follows:

\begin{equation}
    \rho_{\epsilon_{T_1}, \epsilon_{T_2}} = \frac{cov(\epsilon_{T_1}, \epsilon_{T_2})}{\sigma_{\epsilon_{T_1}}\sigma_{\epsilon_{T_2}}}
\end{equation}

\noindent where, $cov(\epsilon_{T_1}, \epsilon_{T_2})$ is the covariance between the two periods, and $\sigma_{\epsilon_{T_1}}$ and $\sigma_{\epsilon_{T_2}}$ are the standard deviations at $T_1$ and $T_2$, respectively, as determined empirically form the synthetic ground motions.
In this example, we see a stronger correlation between the T=1.0 s and T=2.0 s residuals ($\rho$=0.78) than the T=1.0 s and T=0.2 s residuals ($\rho$=0.29), which are further apart in log space.  
This observation is in agreement with \cite{baker_conditional_2011}, who found the correlation coefficient between these pairs to be 0.75 and 0.44, respectively, for shallow crustal events. 
This example suggests that cGM-GANO is capable of learning the intricate characteristics of ground motion databases, such as the RotD50 inter-period correlation, that can be important in the seismic response of multi-modal structures.

\begin{figure*}
    \centering
    \begin{subfigure}{.45\textwidth}
        \caption{}
        \includegraphics[width=\textwidth]{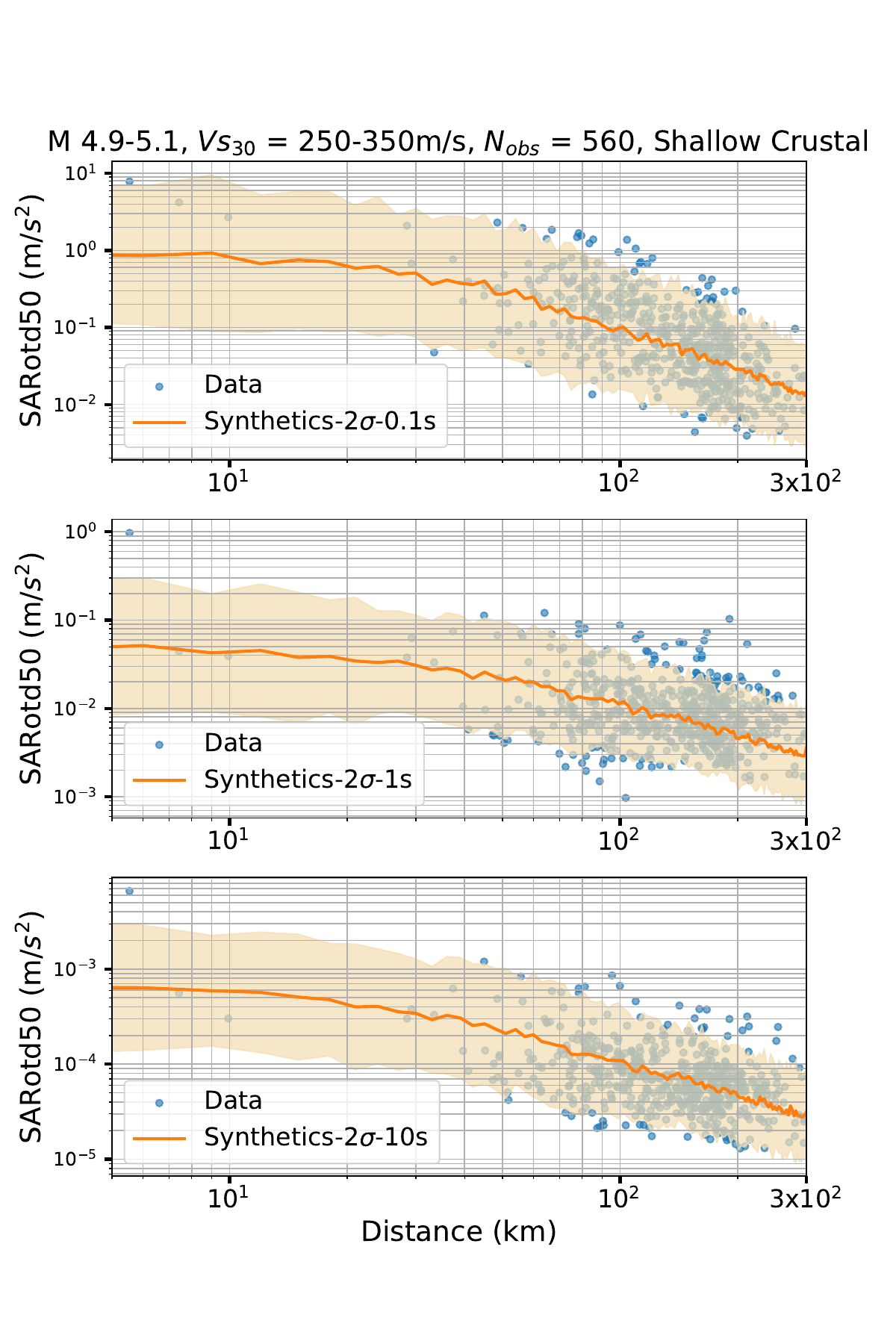}
    \end{subfigure}
    \quad
    \begin{subfigure}{.45\textwidth}
        \caption{}
        \includegraphics[width=\textwidth]{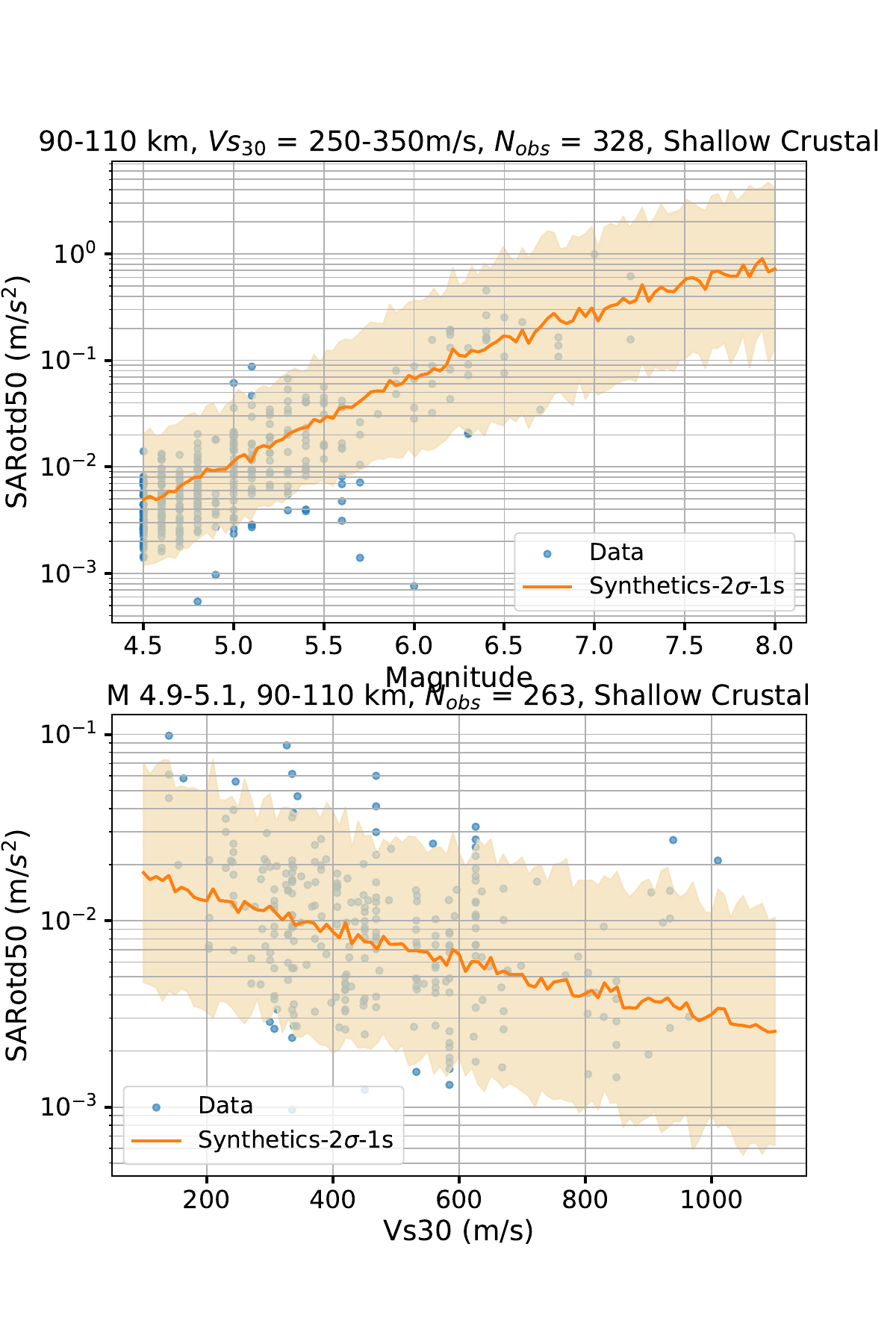}  
    \end{subfigure}
    \caption{Comparison PSA distance, magnitude, and $V_{S30}$ scaling of cmGM-GANO with Kik-net dataset.
    Left panels, from top to bottom, show the distance scaling for T = 0.1, 1.0, and 10 s.
    Top right panel shows the magnitude scaling for  T = 1.0 s.
    Bottom right panel shows the $V_{S30}$ scaling for  T = 1.0 s.
    Solid lines represent the mean scaling in log space. The shaded region shows the $2^{nd}$ to $98^{th}$ percentile of aleatory variability. 
    Solid dots correspond to the kik-net data. }
    \label{figure:fill_the_gap}
\end{figure*}

Next, as we did with the verification SCEC BBP dataset, we tested the capability of cGM-GANO to learn the magnitude and distance scaling in sparsely sampled regions of the training dataset. In Fig \ref{figure:fill_the_gap}, we show scenarios of RotD50 scaling versus magnitude, distance, and $V_{S30}$ generated by cGM-GANO. 
The cGM-GANO-generated ground motions generally follow the mean and aleatory variability of the observations.
In particular, they are able to capture the non-linear distance scaling at periods $T = 0.1, 1.0, 10.0 s$, and the magnitude and $V_{S30}$ scaling at periods  $T = 1.0 s$.

\subsection{Validation against empirical GMMs}

\begin{figure*}
    \centering
    \begin{subfigure}{0.22\textwidth}
        \caption{}
        \includegraphics[width=1.1\textwidth]{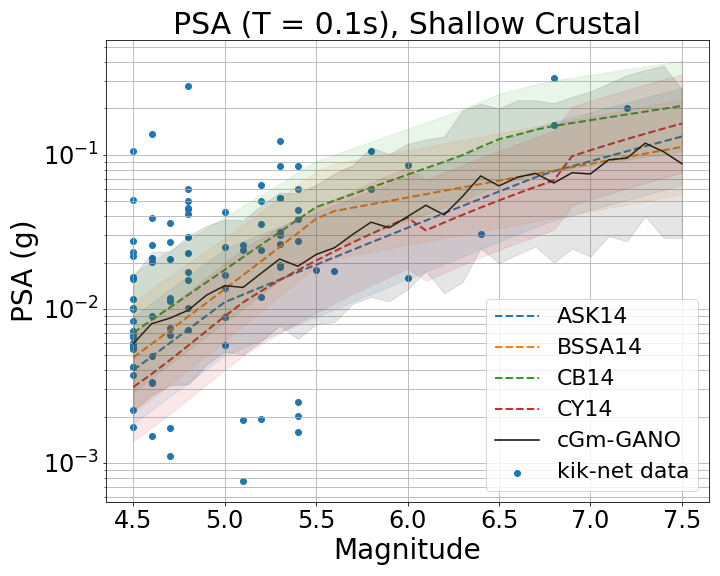}
    \end{subfigure}
    \quad
    \begin{subfigure}{0.22\textwidth}
        \caption{}
        \includegraphics[width=1.1\textwidth]{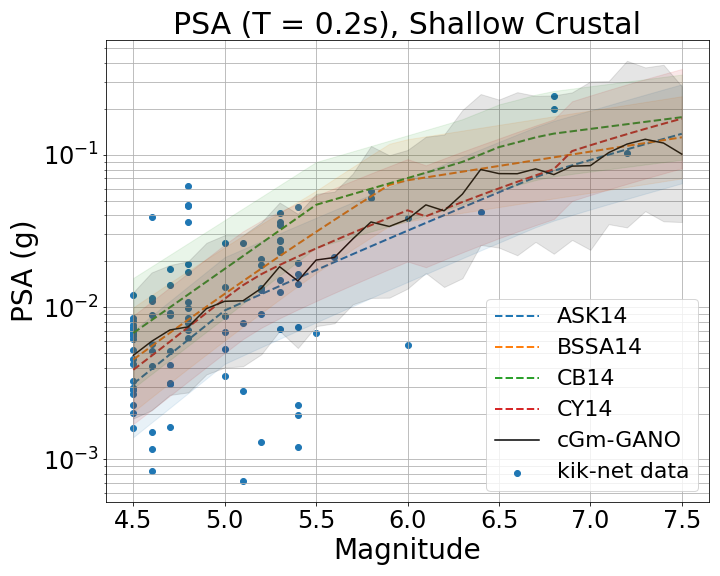}
    \end{subfigure}
    \quad
    \begin{subfigure}{0.22\textwidth}
        \caption{}
        \includegraphics[width=1.1\textwidth]{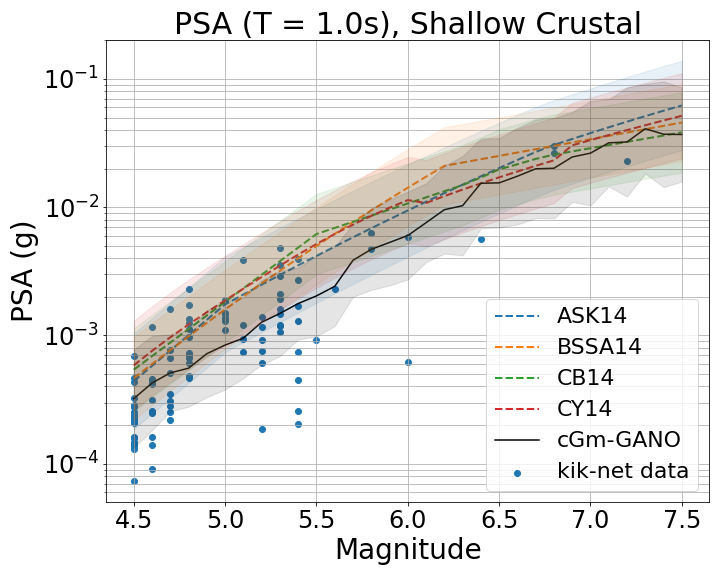}
    \end{subfigure}
    \quad
    \begin{subfigure}{0.22\textwidth}
        \caption{}
        \includegraphics[width=1.1\textwidth]{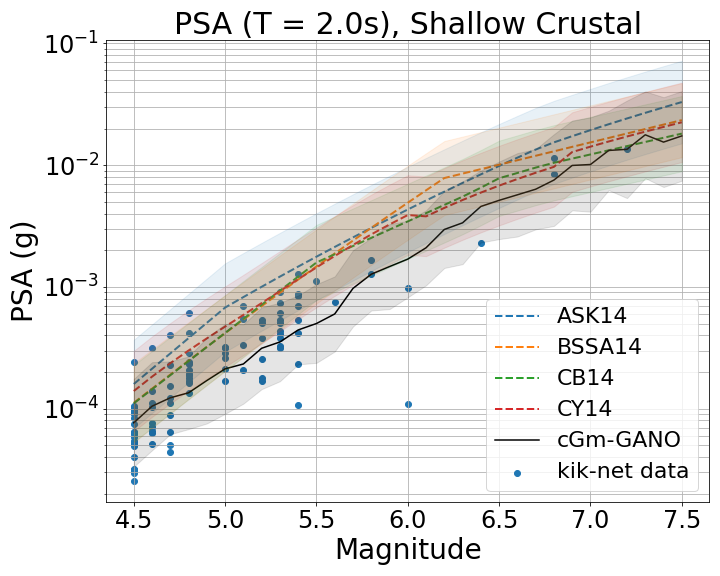}
    \end{subfigure}
        
    \medskip
    \begin{subfigure}{0.22\textwidth}
        \caption{}
        \includegraphics[width=1.1\textwidth]{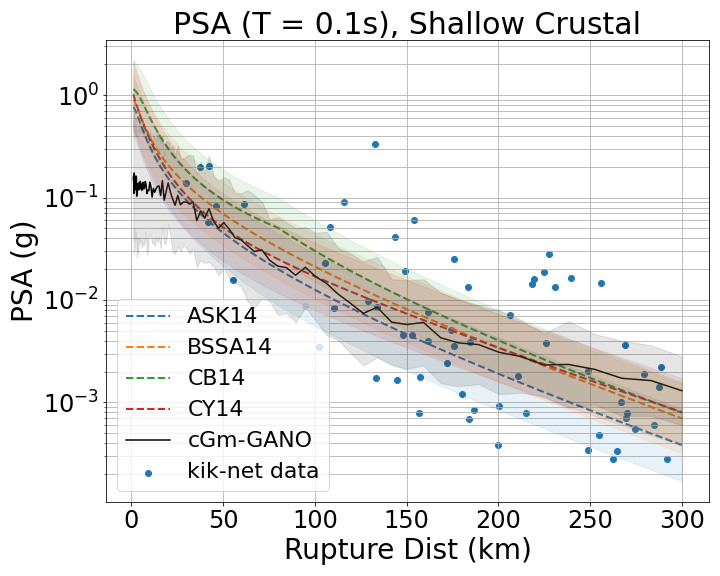}
    \end{subfigure}
    \quad
    \begin{subfigure}{0.22\textwidth}
        \caption{}
        \includegraphics[width=1.1\textwidth]{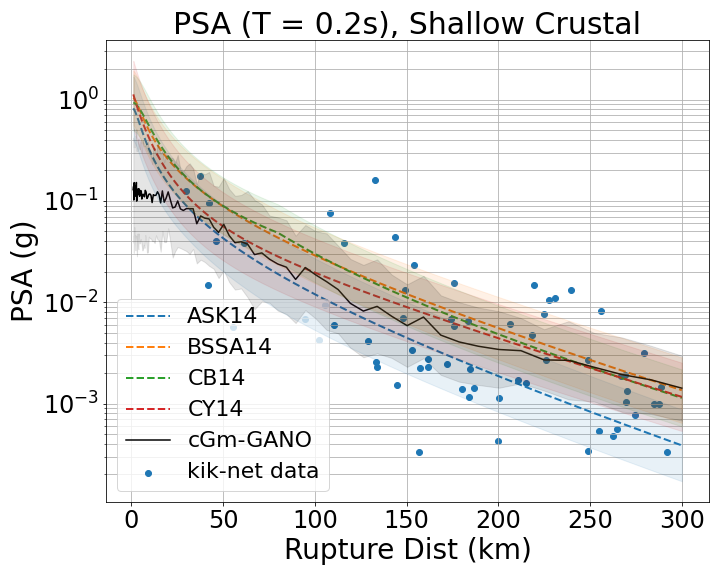}
    \end{subfigure}
    \quad
    \begin{subfigure}{0.22\textwidth}
        \caption{}
        \includegraphics[width=1.1\textwidth]{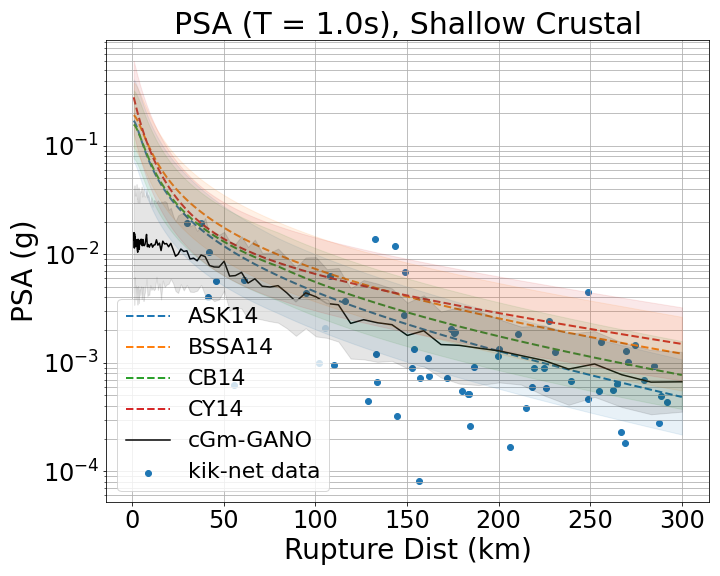}
    \end{subfigure}
    \quad
    \begin{subfigure}{0.22\textwidth}
        \caption{}
        \includegraphics[width=1.1\textwidth]{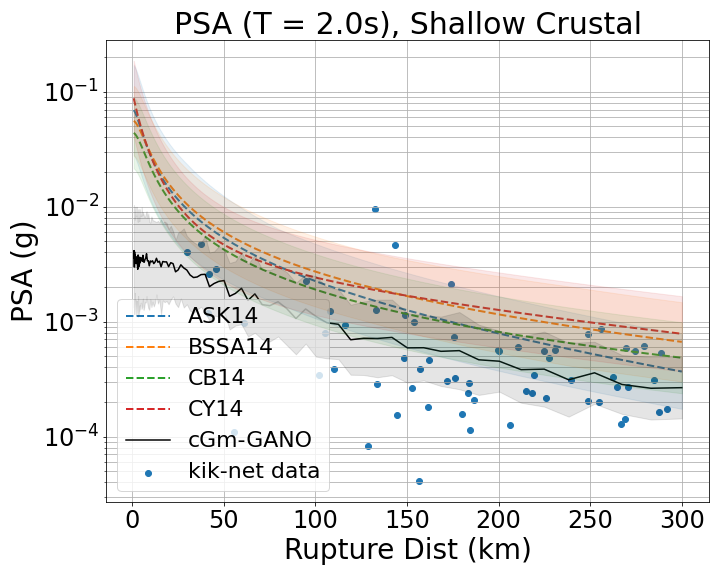}
    \end{subfigure}
    \caption{Comparison of cGM-GANO PSA scaling for shallow crustal events against the NGA-West2 GMMs. 
    Top panel shows the magnitude comparison for $R_{rup} = 60$ km and $V_{s30}$ = 700 m/s, at T = 0.1, 0.2, 1, 2 s (f = 10, 5, 1, 0.5 Hz), respectively.
    Bottom panel shows the distance comparison for $M = 6$ km and $V_{s30}$ = 700 m/s, km, for the same frequencies.
    Solid lines represent the median scaling, the shaded region corresponds to the $16^{th}$ to $84^{th}$ percentile.}
    \label{fig:GMM_psa_crustal}
\end{figure*}

\begin{figure*}
    \centering
    \begin{subfigure}{0.22\textwidth}
        \caption{}
        \includegraphics[width=1.1\textwidth]{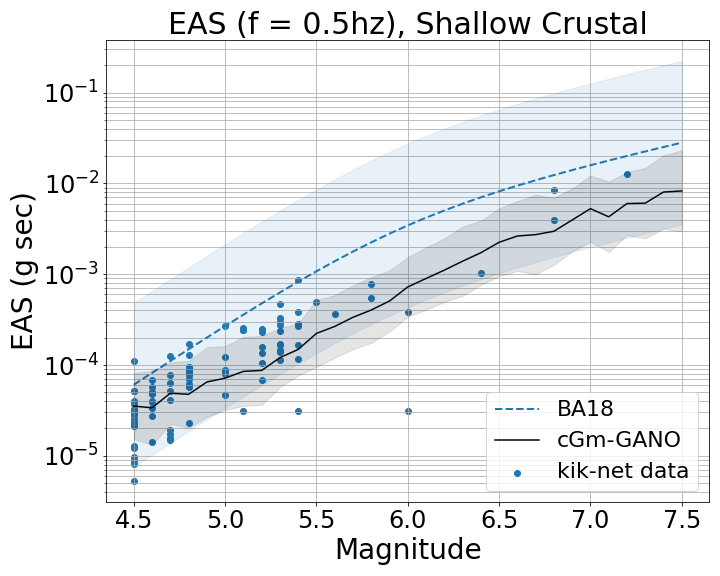}
    \end{subfigure}
    \quad
    \begin{subfigure}{0.22\textwidth}
        \caption{}
        \includegraphics[width=1.1\textwidth]{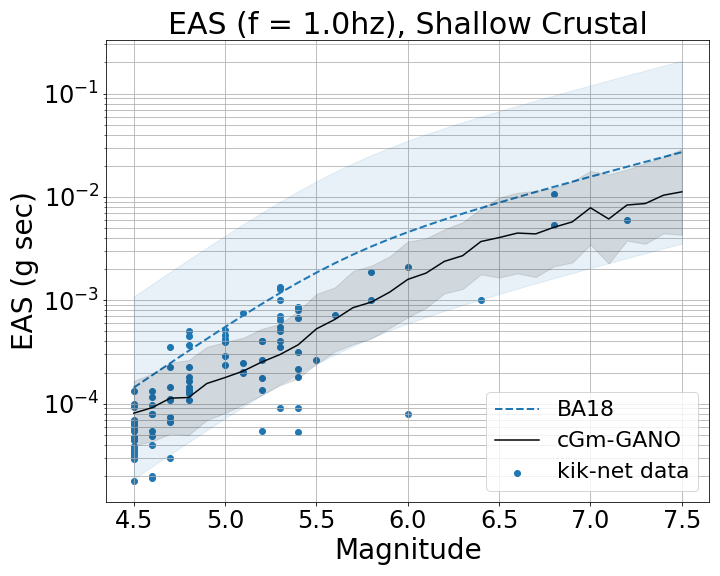}
    \end{subfigure}
    \quad
    \begin{subfigure}{0.22\textwidth}
        \caption{}
        \includegraphics[width=1.1\textwidth]{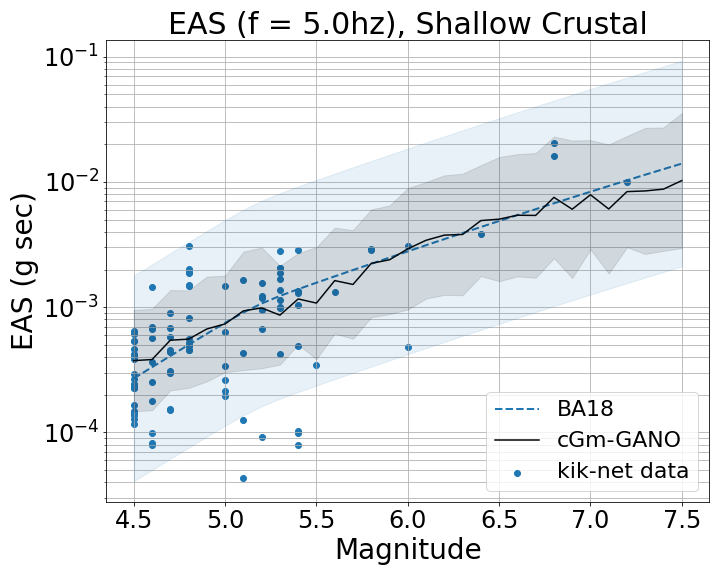}
    \end{subfigure}
    \quad
    \begin{subfigure}{0.22\textwidth}
        \caption{}
        \includegraphics[width=1.1\textwidth]{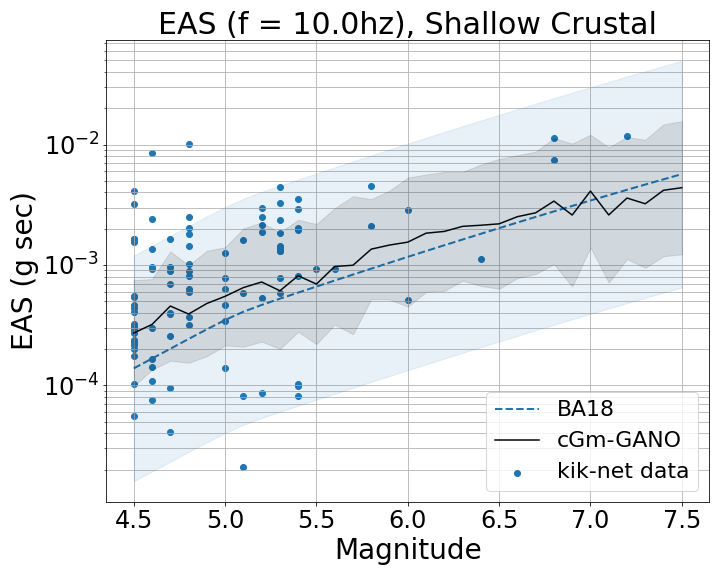}
    \end{subfigure}
    
    \medskip
    \begin{subfigure}{0.22\textwidth}
        \caption{}
        \includegraphics[width=1.1\textwidth]{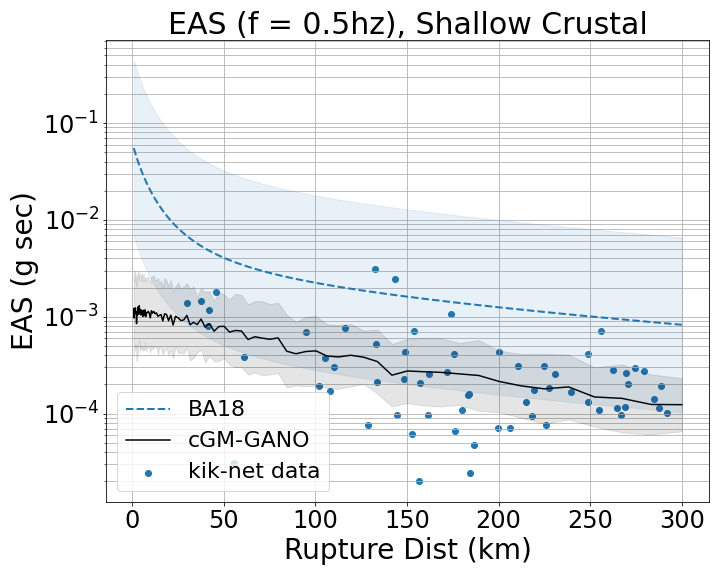}
    \end{subfigure}
    \quad
    \begin{subfigure}{0.22\textwidth}
        \caption{}
        \includegraphics[width=1.1\textwidth]{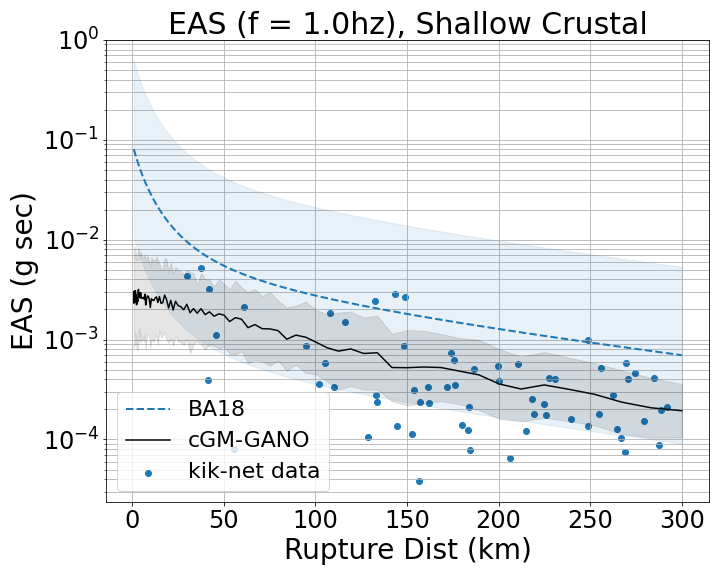}
    \end{subfigure}
    \quad
    \begin{subfigure}{0.22\textwidth}
        \caption{}
        \includegraphics[width=1.1\textwidth]{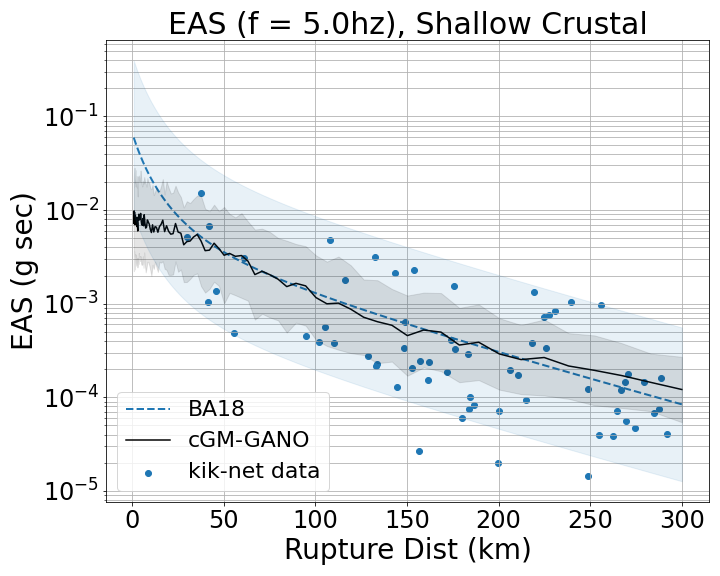}
    \end{subfigure}
    \quad
    \begin{subfigure}{0.22\textwidth}
        \caption{}
        \includegraphics[width=1.1\textwidth]{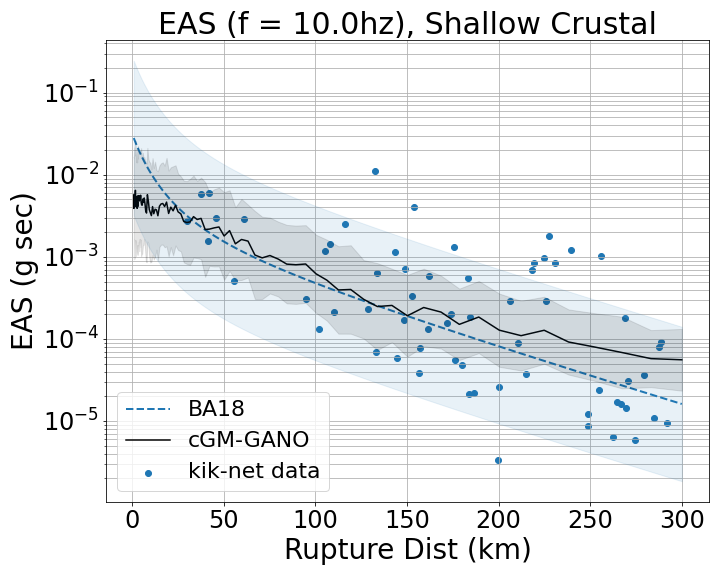}
    \end{subfigure}
    \caption{Comparison of cGM-GANO EAS scaling for shallow crustal events against the BA18 EAS GMMs. 
    Top panel shows the magnitude comparison for $R_{rup} = 60$ km and $V_{s30}$ = 700 m/s, at f = 0.5, 1, 5, and 10 Hz, respectively.
    Bottom panel shows the distance comparison for $M = 6$ km and $V_{s30}$ = 700 m/s, km, for the same frequencies. Solid lines represent the median scaling, the shaded region corresponds to the $16^{th}$ to $84^{th}$ percentile.}
    \label{fig:GMM_eas_crustal}
\end{figure*}

Lastly, we performed a series of comparisons of the cGM-GANO generated ground motions, trained on the KiK-net data set, to selected published ground motion models (GMMs) described in Section Residual Analysis and Comparison to Empirical Ground Motion Models. 
Fig. \ref{fig:GMM_psa_crustal} compares the PSA magnitude and distance scaling of the cGM-GANO generated ground motions for shallow crustal events with the NGA-West 2 GMMs.
Overall, cGM-GANO magnitude and distance scaling show similar trends with the published models in terms of median ground motion and aleatory variability. 
The biggest difference is at short distances which as explained above is believed to be a result of sparse data to constrain the model over that distance range. 
Similar comparisons are made in Fig. \ref{fig:GMM_eas_crustal} against the BA18 EAS GMM. 
Here, a systematic misfit is observed at f<1 Hz, which is introduced by the misfit between BA18 and the training data set; cGM-GANO scaling is consistent with the training dataset.
Otherwise, similar conclusions are drawn for the median magnitude and distance scaling. 
The underestimation of EAS aleatory variability by cGM-GANO compared to BA18 and the cause of the misfit between the BA18 and the data will be further investigated by the authors. 

\begin{figure*}
    \centering
    \begin{subfigure}{0.22\textwidth}
        \caption{}
        \includegraphics[width=1.1\textwidth]{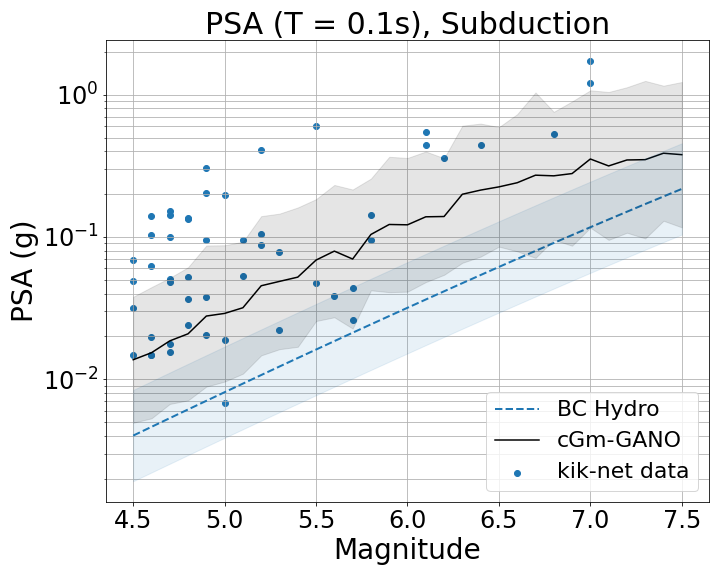}
    \end{subfigure}
    \quad
    \begin{subfigure}{0.22\textwidth}
        \caption{}
        \includegraphics[width=1.1\textwidth]{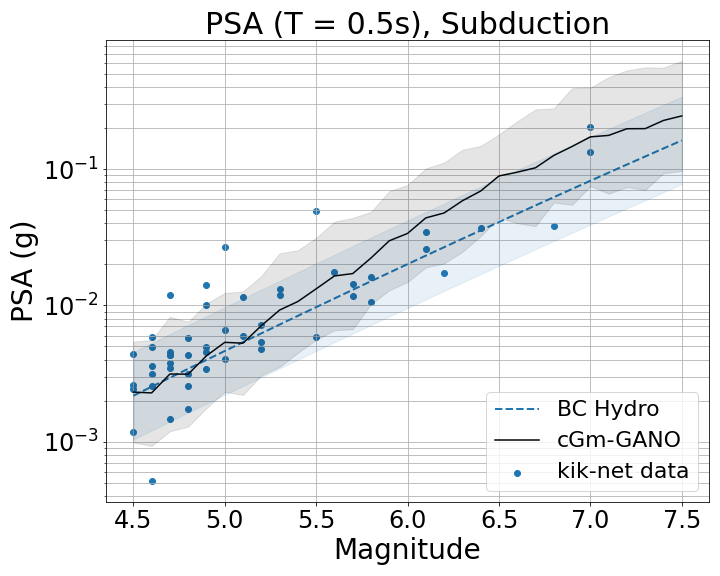}
    \end{subfigure}
    \quad
    \begin{subfigure}{0.22\textwidth}
        \caption{}
        \includegraphics[width=1.1\textwidth]{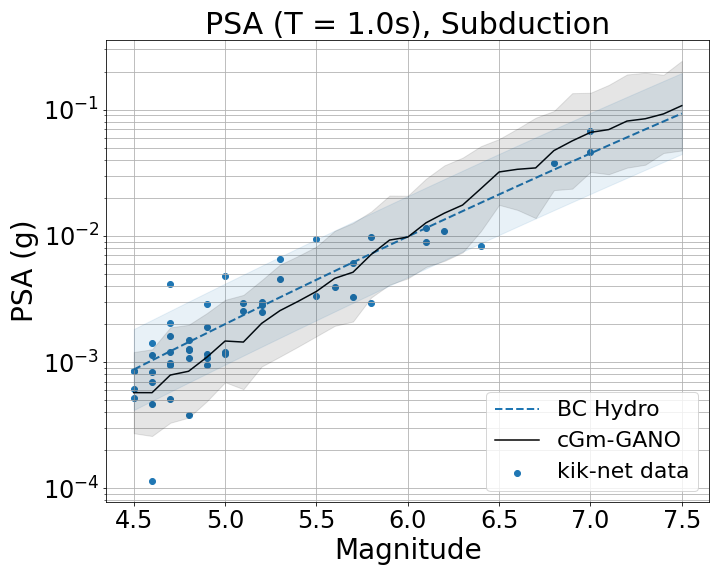}
    \end{subfigure}
    \quad
    \begin{subfigure}{0.22\textwidth}
        \caption{}
        \includegraphics[width=1.1\textwidth]{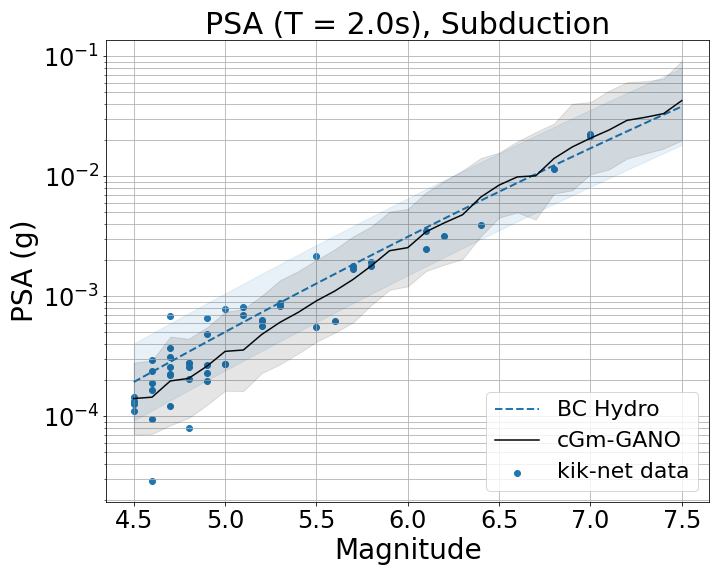}
    \end{subfigure}
    
    \medskip
    \begin{subfigure}{0.22\textwidth}
        \caption{}
        \includegraphics[width=1.1\textwidth]{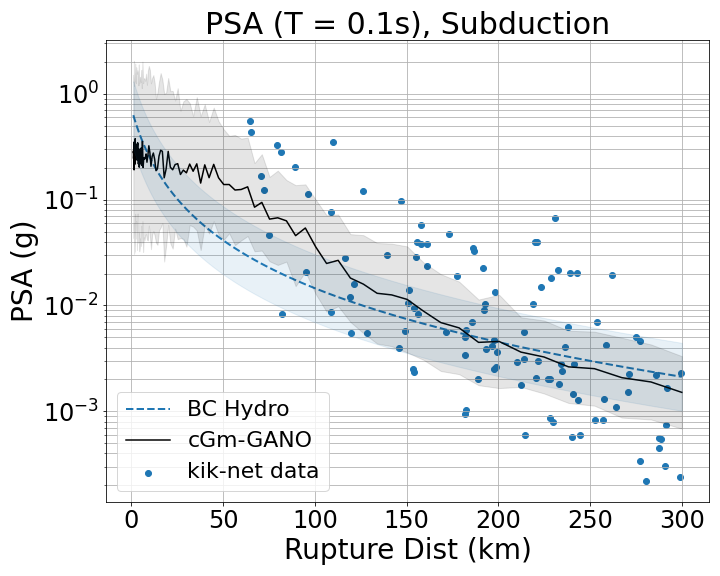}
    \end{subfigure}
    \quad
    \begin{subfigure}{0.22\textwidth}
        \caption{}
        \includegraphics[width=1.1\textwidth]{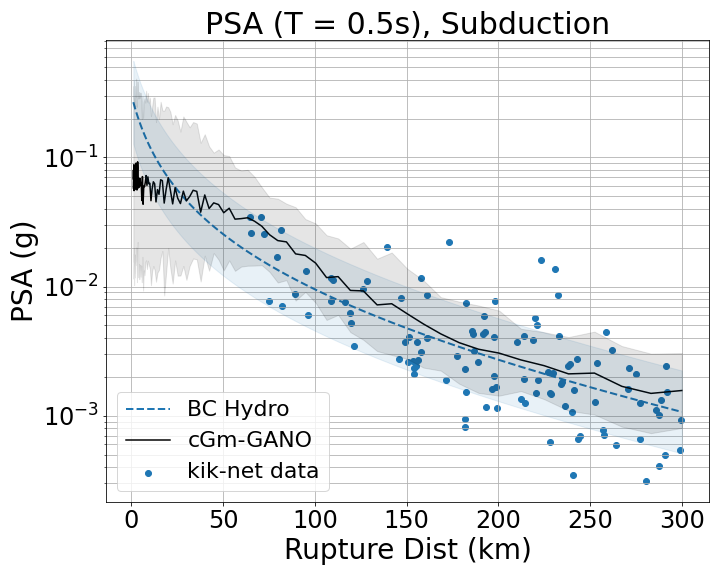}
    \end{subfigure}
    \quad
    \begin{subfigure}{0.22\textwidth}
        \caption{}
        \includegraphics[width=1.1\textwidth]{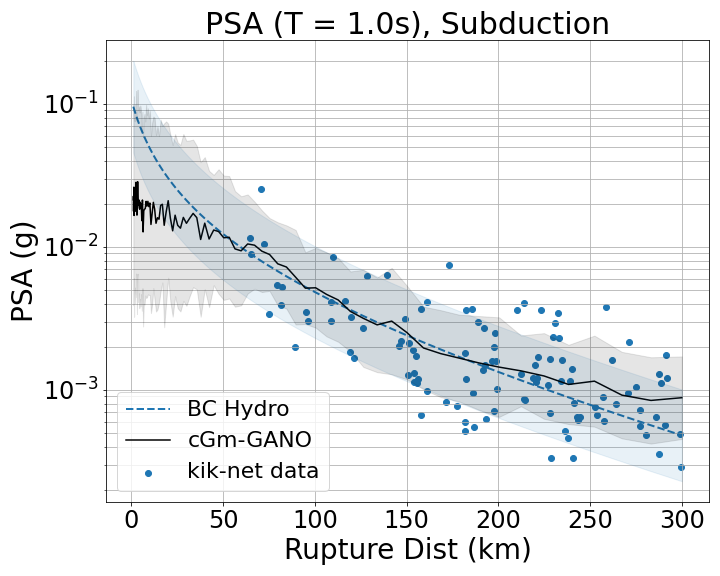}
    \end{subfigure}
    \quad
    \begin{subfigure}{0.22\textwidth}
        \caption{}
        \includegraphics[width=1.1\textwidth]{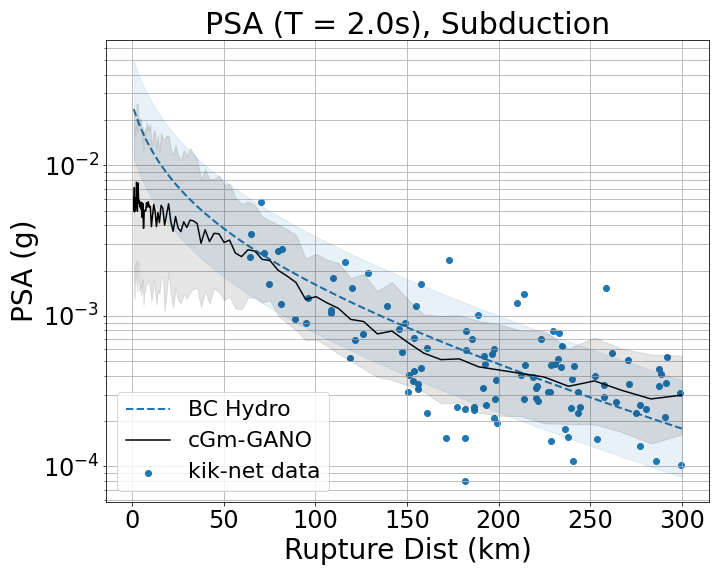}
    \end{subfigure}
    \caption{Comparison of cGM-GANO PSA scaling for subduction events against BC-Hydro GMM. 
    Top panel shows the magnitude comparison for $R_{rup} = 60$ km and $V_{s30}$ = 700 m/s, at T = 0.1, 0.5, 1, and 2 s (f = 10, 2, 1, and 0.5 Hz), respectively.
    Bottom panel shows the distance comparison for $M = 6$ km and $V_{s30}$ = 700 m/s, km, for the same frequencies. Solid lines represent the median scaling, the shaded region corresponds to the $16^{th}$ to $84^{th}$ percentile.}
    \label{fig:GMM_subduction}
\end{figure*}

For the subduction events, PSA scaling is compared in Fig. \ref{fig:GMM_subduction}.
The synthetic ground motions have consistent scaling with BCHydro both in terms of median ground motion and aleatory variability over the entire frequency range. 
The main misfit is observed at short distances. 
cGM-GANO is believed to have better performance for subduction events compared to shallow crustal due to the larger dataset size. 

\section{Summary and Discussion}

We presented a data-driven model for conditional ground motion synthesis that can be used to generate 3-component ground motion time series covering a broadband range of frequencies. The model is based on the architecture of Neural Operators, a resolution-invariant generalization of Neural Networks. The ground motion synthesis is conditioned on moment magnitude ($\mathbf{M}$), rupture distance ($R_{rup}$), time-average shear-wave velocity at the top $30m$ of the crust ($V_{S30}$), and tectonic environment or style of faulting ($f_{type}$). 

Technical advancements of this work compared to previous efforts, including of the authors', are the use of Neural Operators and training of the model on velocity time series. Using Neural Operators guarantees both discretization-invariance of the model from the resolution of the training data set and universal approximation. We specifically employed the U-shaped Neural Operator (U-NO) architecture, which alleviates some issues associated with the memory requirements of fully connected Neural Operators. As well as, the training on velocity time series (as opposed to the target acceleration ground motions) also led to performance improvements due to the flatter Fourier spectral shape of the training signals.

The model was verified against a dataset of simulated ground motion time series from the SCEC BBP. 
The comparison of scenarios in terms of Fourier amplitude in the low-frequency regime $f < 1 Hz$ ($T > 1$ sec) systematically showed the mean and standard deviation of the simulated and cGM-GANO-generated time series to be in agreement. At the same time, for higher frequencies, the cGM-GANO-generated time series of the same scenarios have systematically higher variability and a positive bias for reverse events. Our working hypothesis for the discontinuity that occurs at $f = 1 Hz$, the cross-over period between the 
low- and high-frequency methods in the BBP hybrid GP module, is that cGM-GANO cannot learn the stochastic high-frequency method, and resorts to carrying over the learned PDE solution of $f < 1 Hz$ into the higher frequency range. 
The comparison of the response spectral ordinates showed a small constant positive bias across the period range for both styles of faulting. 
This bias is attributed to an unsuitable inter-period correlation of the synthetic ground motions that are influenced by the stochastic component of the training dataset; however, further investigation is needed.

The model was next trained and compared to a dataset curated and processed from observations recorded by the Japanese strong motion network KiK-net, which contains both shallow crustal and subduction events. Bias was observed in the distance scaling of near field ground motions and $V_{S30}$ scaling for soft soil sites, which we attributed to a lack of adequate training data in the corresponding regimes: in the first case, the training dataset contains less than 0.5\% of ground motions recorded at stations with $R_{rup} \le 20$ km; in the second case, the training dataset was larger but we hypothesize that cGM-GANO is also called to learn the scaling that includes nonlinear site response, and as such, requires an even larger amount of soft site data. This last hypothesis is being tested by the authors as part of a separate project.
% Potential strategies to address this issue include developing a hybrid dataset composed of both empirical and simulation data and using weighting sampling during training towards short distance and low $V_{S30}$ ground motion records.  

Comparisons with traditional ground motion models (GMMs) led to similar conclusions: the cGM-GANO generated ground motions showed consistent median scaling with the GMMs for the corresponding tectonic environments. The largest misfit was observed at short distances due to the scarcity of training data, as mentioned above. A systematic misfit in EAS in the low-frequency regime ($f < 1 Hz$) between cGM-GANO and BA18 is likely caused by the inconsistency of the available data and the GMM. The EAS variability, on the other hand, is underestimated across the entire frequency range. The source of the variability inconsistency between data, cGM-GANO, and GMMs is currently under investigation by the authors. The median scaling of PSA is better captured by cGM-GANO, except at short distances, and the aleatory variability is more accurately recovered, especially for subduction events, due to the availability of the data. 

Ongoing work by the authors focuses on addressing the limitations of the model training and on expanding the overall framework. Until recently, strong motions recorded by the National Research Institute for Earth Science and Disaster Resilience (NIED) have been the only available open-access datasets that could be used in the training of cGM-GANO. As publicly available, larger and higher quality recorded datasets from regions such as California are slowly emerging, we plan to use them to improve the distance scaling over short distances and the $V_{S30}$ scaling at soft soils of cGM-GANO, as well as to resolve the bias observed in the low-frequency range of the empirical dataset presented here. Further validations with traditional GMMs will also be performed, including duration and intensity of ground shaking intensity measures. For conditions not well constrained in the empirical data, we plan to employ additional techniques, such as reinforcement learning with physics-based simulations. 

Framework enhancements that we are investigating include the effects of ground motion filtering and instrument direction, as well as the partitioning of the stochastic component into within and between event variability. More specifically, in its current architecture, cGM-GANO does not recognize the effect of filtering in recorded ground motions, and thus, any differences in the filtering of ground motions are mapped onto the stochastic component of cGM-GANO output. Similarly, the horizontal components are currently being provided in the North and South direction, irrespective of the fault's azimuth, but, as past studies \citep[e.g.,][]{wald_1994, archuleta_1984} have demonstrated, radiation patterns and thus the fault's azimuth can have a significant impact in the ground motion amplitude. Last, in this study's framework, each ground motion realization is produced for a random event, which may be adequate for characterizing the seismic risk of individual structures, but it is not suitable for evaluating the seismic risk at a system's level. These improvements are developed as part of several ongoing projects led by the authors, and the results thereof will be submitted for publication consideration in forthcoming manuscripts. 

Data-driven models for ground motion synthesis may still be at the pilot stage, but with further refinements, they have the potential to bridge the gap between physics-based simulations and empirical models. Already in its current form, cGM-GANO can generate hundreds of site-specific realistic 3-component coherent ground motions with clear P-S arrivals and surface waves --currently only feasible with physics-based simulations-- with the simplicity of empirical methods in less than a second. One should also consider this: data-driven generated ground motions can span a broad range of frequencies (up to 30 Hz in this study), all of which are learned from real recorded ground motion data. By contrast, physics-based simulations may have broken the computational barrier of high-frequency deterministic wave propagation with EXASCALE computing, but they will continue to lag behind the empirical ground motion models in the high-frequency regime until the propagation medium reaches the input resolution necessary for these frequencies to realistically propagate through and be modified appropriately by the heterogeneous, nonlinear shallow crust. 

\section{Data and Resources}

KiK-net time histories can be downloaded through National Research Institute for Earth Science and Disaster Resilience \url{https://www.bosai.go.jp/e/index.html}, while the kik-net flatfile of the selected events is summarized in \cite{bahrampouri_updated_2021} and can be downloaded through DesignSafe \url{https://www.designsafe-ci.org/data/browser/public}. 
The code for cGM-GANO is published on \url{https://github.com/yzshi5/GM-GANO}. The supplementary material contains a graphical example supporting the selection of velocity time series as our training data (Fig. S1); a realization of fault and station locations for the BBP simulations, depicting the sparsity of the recorded ground motions  (Fig. S2); two scenarios of M7.0 events at 10km simulated using the BBP for reverse and strike-slip faults in the time and frequency domains, and their comparison to cGM-GANO generated ground motions for the same conditional variables (Fig. S3); the inter-frequency correlation of the cGM-GANO ground motions for the model trained on BBP, clearly depicting the separation between the deterministic and stochastic regimes of the algorithm (Fig. S4); residual plots of the KiK-net trained model based on the validation (hold-off) dataset (Fig. S5); velocity time histories from observations and cGM-GANO generated ground motions using the model trained on the KiK-net data (Fig. S6); and displacement time histories from observations and cGM-GANO generated ground motions using the model trained on the KiK-net data (Fig. S7).

\section{Declaration of Competing Interests}
The authors declare no competing interests

\section{Acknowledgments}
We would like to thank Dr. Weiqiang Zhu and Dr. Mahdi Bahrampouri for their help during the curation and preprocessing of the KiK-net dataset.
We would also like to thank Dr. Jian Shi and Dr. Robert W Graves for providing the BBP dataset. 
This material is based upon work supported in part by the U.S. Geological Survey under Grant No. G22AP00231. The views and conclusions contained in this document are those of the authors and should not be interpreted as representing the opinions or policies of the U.S. Geological Survey. Mention of trade names or commercial products does not constitute their endorsement by the U.S. Geological Survey.
We would also like to thank Dr. Kim Olsen, an anonymous reviewer and associate editor, for the review and constructive comments that helped to improve the final article.

%\bibliographystyle{unsrt}  
%\bibliography{references_draft}  

\end{document}